\documentclass[preprint,preprintnumbers,amsmath,amssymb,floatfix,nofootinbib,superscriptaddress,longbibliography]{revtex4-1}
\usepackage[dvips]{graphicx}
\usepackage{url}
\usepackage{subfigure}
\usepackage{xcolor}
\usepackage{hyperref}
\usepackage[export]{adjustbox}
\usepackage{setspace}
\def\tb{\bar{t}}
\def\bb{\bar{b}}
\def\GeV{\mathrm{GeV}}
\def\TeV{\mathrm{TeV}}
\def\nlo{\mathrm{NLO}}
\def\nlops{\mathrm{NLO+PS}}
\def\pTmin{p_T^\textrm{min}}
\def\mur{\mu_\mathrm{R}}
\def\muf{\mu_\mathrm{F}}
\def\muq{\mu_\mathrm{Q}}
\def\powhegbox{\texttt{POWHEG-BOX}}
\def\nlox{\texttt{NLOX}}
\def\oneloop{\texttt{OneLOop}}
\def\sherpa{\texttt{Sherpa}}
\def\blackhat{\texttt{Blackhat}}
\def\madspin{\texttt{MadSpin}}
\def\rivet{\textsc{Rivet}}
\def\openloops{\texttt{OpenLoops}}

\def\hepmc{\texttt{HepMC}}
\def\pythia{\texttt{PYTHIA8}}
\def\powhel{\texttt{PowHel}}
\def\recola{\texttt{Recola}}
\def\mgfive{\texttt{MG5\_aMC@NLO}}
\def\mg5{\texttt{MadGraph5}}
\def\powheg{POWHEG}
\def\mcnlo{MC@NLO}
\def\SSL{$2\ell SS$}
\begin{document}          
\title{Top-quark pair production in association with a $W^\pm$ gauge boson in the
  \powhegbox{}}
\author{F.~Febres Cordero}
\email{ffebres@hep.fsu.edu}
\author{M.~Kraus}
\email{mkraus@hep.fsu.edu}
\author{L.~Reina}
\email{reina@hep.fsu.edu}
\affiliation{Physics Department, Florida State University,
Tallahassee, FL 32306-4350, U.S.A.}

\date{\today}

\begin{abstract}
  We present a new Monte Carlo event generator for the production of a top-quark 
  pair in association with a $W^\pm$ boson at hadron colliders in the 
  \powhegbox{} framework. We consider the next-to-leading-order QCD corrections 
  to the $pp\to t\tb W^\pm$ cross section, corresponding to the 
  $\mathcal{O}(\alpha_s^3\alpha)$ and $\mathcal{O}(\alpha_s\alpha^3)$ terms in 
  the perturbative expansion of the parton-level cross section, and model the 
  decays of $W$ and top quarks at leading order retaining spin correlations. 
  The fixed-order QCD calculation is further interfaced with the \pythia{} 
  parton-shower event generator via the \powheg{} method as implemented in the 
  \powhegbox{}. The corresponding code is now part of the public repository of 
  the \powhegbox{}.
  We perform a comparison of different event generators for both the case of 
  inclusive production and the case of the two same-sign leptons signature at the 
  Large Hadron Collider operating at a center-of-mass energy of $13~\TeV$. We 
  investigate theoretical uncertainties in the modelling of the fiducial volume 
  stemming from missing higher-order corrections, the different parton shower
  matching schemes, and the modelling of decays.
  We find that the subleading contribution at $\mathcal{O}(\alpha_s\alpha^3)$ is 
  particularly sensitive to differences in the matching scheme and higher-order 
  parton shower effects. We observe that in particular jet observables can differ
  quite visibly although these differences play only a subordinate role in the 
  description of physical observables once all contributions are combined.
\end{abstract}
\maketitle

\section{Introduction}
\label{sec:introduction}
The production of top-quark pairs in association with electroweak gauge bosons 
($W,Z,\gamma$) can be measured at the Large Hadron Collider (LHC) and future 
hadron colliders (HL-LHC, FCC-hh, CppC) in a multitude of decay channels and 
provides new avenues to test the consistency of the Standard Model (SM) in 
processes that have been beyond the energy reach of existing colliders or 
statistically limited until recently. At the same time, these processes represent 
some of the most important backgrounds for Higgs-boson precision measurements and 
searches of new physics beyond the Standard Model (BSM). In this context, the 
hadronic production of $W^\pm$ bosons in association with top-quark pairs is 
particularly interesting both from a phenomenological and a theoretical point of 
view.

On top of its intrinsic interest as a SM process, $t\tb W^\pm$ provides very 
distinctively polarized top quarks that can be used to unveil the imprint of new 
physics interactions. Indeed, in contrast to top-quark pair production, the top 
quarks originating from the $t\tb W^\pm$ production process are highly polarized 
and give rise to a large $t\tb$ charge 
asymmetry~\cite{Maltoni:2014zpa,Bevilacqua:2020srb}. A measurement of the $t\tb$ 
charge asymmetry in $t\tb W^\pm$ production can then be sensitive to the chiral 
nature of new physics contributing to the process and can become a unique 
indirect probe of BSM physics. 

The $t\tb W^\pm$ processes also represent a very important background to the 
production of a Higgs boson in association with top quarks in the multi-lepton
decay channels~\cite{Maltoni:2015ena,Sirunyan:2018hoz,Aaboud:2018urx,
ATLAS:2019nvo,Sirunyan:2020icl}, where it limits the accuracy of the direct 
measurement of the top-quark Yukawa coupling. In addition, $t\tb W^\pm$ is the 
dominant background in searches for the SM production of four top 
quarks~\cite{Sirunyan:2019wxt,Aad:2020klt}. In general, $t\tb W^\pm$ is a 
background to any search of new physics in signatures with same-sign leptons, 
missing energy, and $b$ jets, common in many BSM models.

Due to its phenomenological relevance, the process has been studied extensively 
on the theory side, starting from the first calculation of next-to-leading order 
(NLO) QCD corrections to the production and decay process in 
Ref.~\cite{Campbell:2012dh}. Further studies at fixed order include the 
calculation of the leading NLO electroweak (EW) 
corrections~\cite{Frixione:2015zaa} and the assessment of the impact of formally 
subleading mixed QCD and EW corrections~\cite{Frederix:2017wme}. Beyond fixed 
order the pure NLO QCD $t\tb W^\pm$ calculation has been matched to parton 
showers using the \powheg{} method~\cite{Nason:2004rx,Frixione:2007vw} as
implemented in the \powhel{} framework~\cite{Garzelli:2012bn} as well as using 
the \mcnlo{} method~\cite{Frixione:2002ik,Frixione:2003ei} in the \mgfive{} 
framework~\cite{Maltoni:2014zpa}\footnote{A comparison of $t\tb W^\pm$ 
differential distributions obtained from \mgfive{}~\cite{Alwall:2014hca}, 
\powhel{}~\cite{Garzelli:2012bn} and \sherpa{}~\cite{Gleisberg:2008ta,
Bothmann:2019yzt}+\openloops{}~\cite{Cascioli:2011va,Buccioni:2017yxi,
Buccioni:2019sur} including $\mathcal{O}(\alpha_s^3\alpha)$ NLO QCD corrections 
has been presented in Ref.~\cite{deFlorian:2016spz} as a validation of the  
corresponding Monte Carlo tools. Ref.~\cite{deFlorian:2016spz} also provided LHC 
$t\tb W^\pm$ cross sections for $\sqrt{s}=13$ and $14$ TeV including 
$\mathcal{O}(\alpha_s^3\alpha)$ and $\mathcal{O}(\alpha_s^2\alpha^2)$ NLO 
corrections as obtained from both \mgfive{} and \sherpa{}+\openloops.} 
Separately, the resummation of soft gluon emission effects have been studied at 
the next-to-next-to-leading logarithmic (NNLL) accuracy~\cite{Li:2014ula,
Broggio:2016zgg,Kulesza:2018tqz,Broggio:2019ewu,Kulesza:2020nfh}.

On the experimental side, the associate $t\tb W^\pm$ production has been measured 
by the LHC experiments both as inclusive cross section~\cite{Sirunyan:2017uzs,
Aaboud:2019njj} and as a background in searches for $t\tb H$ and $t\tb t\tb$
signals~\cite{ATLAS:2019nvo,Sirunyan:2020icl,Sirunyan:2019wxt,Aad:2020klt}
in multi-lepton decay channels. Some of these measurements have resulted in
larger values with respect to SM predictions. Because of this, the modelling of 
the $t\tb W^\pm$ processes has come under more thorough scrutiny, with the
recent inclusion of off-shell and non-resonant effects at fixed-order NLO QCD
\cite{Bevilacqua:2020pzy,Denner:2020hgg} and by studying the effect of
spin-correlations and formally subleading EW corrections in the fiducial volume 
of specific $t\tb W^\pm$ signatures~\cite{Frederix:2020jzp}. Also, estimates of 
the impact of higher-order QCD corrections beyond NLO corrections, estimated via
multi-jet merging, on inclusive $t\tb W^\pm$ samples, has been presented in 
Ref.~\cite{vonBuddenbrock:2020ter}. Furthermore, the ATLAS collaboration 
recently performed a dedicated comparison of the implementation of $t\tb W^\pm$
production in existing NLO parton-shower Monte Carlo event generators, including 
both QCD and EW effects and allowing for multi-jet merging~\cite{ATLAS:2020esn}.

In this paper we continue the investigation of modelling uncertainties of the 
$t\tb W^\pm$ process, by presenting a study based on a new implementation of the 
$pp\rightarrow t\tb W^\pm$ production in the \powhegbox{} 
framework~\cite{Alioli:2010xd}. We consider next-to-leading-order QCD corrections 
to the $pp\rightarrow t\tb W^\pm$ cross section, corresponding to the 
$\mathcal{O}(\alpha_s^3\alpha)$ and $\mathcal{O}(\alpha_s\alpha^3)$ terms in the 
perturbative expansion of the parton-level cross section, and model the decays of
$W$ and top quarks at leading order (LO) retaining spin correlations.

As part of our study, we perform a detailed comparison between different NLO 
parton-shower Monte Carlo event generators at both the inclusive and the fiducial 
level in order to address modelling uncertainties.  We compare results obtained
with our \powhegbox{} implementation (interfaced to the 
\pythia{}~\cite{Sjostrand:2006za,Sjostrand:2014zea} parton shower), with 
\sherpa{} (using its parton shower based on Catani-Seymour 
dipoles~\cite{Schumann:2007mg}), and with \mgfive{} (also interfaced to 
\pythia{}) and present a first study of the consistency between these different 
frameworks. Our comparison provides a solid basis on which to develop a more 
robust estimate of the residual theoretical uncertainty, and suggests which 
aspects of the theoretical prediction for hadronic $t\tb W^\pm$ production still 
need improvement. 

The paper is organized as follows. In section~\ref{sec:powheg} we review the 
\powheg{} method to the extent of allowing us to establish our notation. In 
section~\ref{sec:details} we provide details of our implementation of the 
$t\tb W^\pm$ processes in the \powhegbox{} framework. 
In section~\ref{sec:results} we present theoretical predictions for both 
inclusive $t\tb W^\pm$ production and for a two same-sign leptons signature, 
comparing results from different Monte Carlo event generators. Finally, we give 
our summary and outlook in section~\ref{sec:conclusions}.

\section{Review of the \powheg{} framework}
\label{sec:powheg}
Matching  parton-shower Monte Carlo event generators with fixed-order
perturbative calculations to achieve NLO accuracy for inclusive observables, has
proven fundamental to describe LHC data. Two major strategies for this
\textit{matching} are commonly employed, the
\mcnlo{}~\cite{Frixione:2002ik,Frixione:2003ei} and
\powheg{}~\cite{Nason:2004rx,Frixione:2007vw} methods. In this article we employ 
the latter using the \powhegbox{} framework~\cite{Alioli:2010xd} to study event 
simulation associated to $t\tb W^\pm$ production at hadron colliders.

Within the \powheg{} method the matching of NLO matrix elements to parton showers 
is achieved by generating the hardest emission first with NLO QCD accuracy, while 
subsequent emissions are modelled by the parton shower. Starting with the NLO 
fixed-order cross section 
\begin{equation}
  \sigma^\nlo = \int d\Phi_n~\Big[B(\Phi_n) + V(\Phi_n)\Big] + 
  \int d\Phi_{n+1}~R(\Phi_{n+1})\;,
\end{equation}
where $d\Phi_n$ denotes the $n$-particle Lorentz-invariant phase space measure
and $B(\Phi_n)$, $V(\Phi_n)$, and $R(\Phi_{n+1})$ are the differential Born, 
virtual, and real cross sections, one introduces a \textit{jet function} 
$F(\Phi_{n+1})$ to split the real radiation contribution as:
\begin{equation}
 R(\Phi_{n+1}) = F(\Phi_{n+1})R(\Phi_{n+1}) + 
 \left[1-F(\Phi_{n+1})\right]R(\Phi_{n+1})\;,
\end{equation}
and uses it to define the \textit{soft} $R_s(\Phi_{n+1})$ and \textit{hard} 
$R_h(\Phi_{n+1})$ real contributions according to:
\begin{equation}
 R_s(\Phi_{n+1}) \equiv F(\Phi_{n+1})~R(\Phi_{n+1})\;, \qquad
 R_h(\Phi_{n+1}) \equiv \Big[1-F(\Phi_{n+1})\Big]~R(\Phi_{n+1})\;.
\label{eqn:shregions}
\end{equation}
The jet function $F(\Phi_{n+1})$ is a real function which takes values between
$0$ and $1$, and should approach smoothly $1$ in the infrared (soft and 
collinear) limits of the $(n+1)$-particle phase space $\Phi_{n+1}$. The precise 
functional form of $F(\Phi_{n+1})$ is in principle arbitrary, but a judicious 
choice is in certain cases necessary to avoid large matching corrections, which 
are formally subleading. Below we discuss standard choices made in the 
\powhegbox{}, and in section~\ref{sec:results} we will study their impact on the 
process at hand.

To generate the first (hardest) emission while keeping the NLO accuracy for  
inclusive observables, a one-step parton shower is introduced according to
\begin{equation}
\begin{split}
 \sigma^\nlops &= \int d\Phi_n~\overline{B}(\Phi_n)\left[ \Delta(\Phi_n,\pTmin) + 
 \int d\Phi_r \frac{R_s(\Phi_{n+1})}{B(\Phi_n)}\Delta(\Phi_n,p_T) \right] \\
 &+ \int d\Phi_{n+1}~R_h(\Phi_{n+1})\;,
 \label{eqn:pwg_master}
\end{split}
\end{equation}
where the real emission phase space is factorized as $d\Phi_{n+1} = 
d\Phi_n d\Phi_r$ in terms of the underlying Born phase space with $n$ final-state 
particles ($d\Phi_n$) and the phase space of the radiated particle ($d\Phi_r$),
the $\overline{B}(\Phi_n)$ function is defined by
\begin{equation}
 \overline{B}(\Phi_n) \equiv B(\Phi_n) + V(\Phi_n) + 
 \int d\Phi_r~R_s(\Phi_{n+1})\;,
\end{equation}
and we have introduced the modified Sudakov form factor $\Delta(\Phi_n,p_T)$,
which is defined as:
\begin{equation}
 \Delta(\Phi_n,p_T) = \exp\left( - \int d\Phi_r 
 \frac{R_s(\Phi_n,\Phi_r)}{B(\Phi_n)} \Theta(k_T(\Phi_n,\Phi_r)-p_T)\right)\;.
 \label{eqn:sudakov} 
\end{equation}
In Eq.~\eqref{eqn:pwg_master} the parton shower infrared cutoff scale is denoted 
with $\pTmin$ and $p_T=p_T(\Phi_r)$ is the transverse momentum of the emitted 
particle.

The modified Sudakov form factor ensures that no double counting of real 
radiation in $R_s$ is produced, while radiation from $R_h$ is treated as an 
independent contribution. This highlights the importance of the choice of 
$F(\Phi_{n+1})$ for the \powheg{} method. As already mentioned, after the first 
(hardest) emission, subsequent splittings can be generated by a standard parton 
shower without affecting the NLO accuracy for inclusive observables.

In the \powhegbox{} the function $F(\Phi_{n+1})$ is written as the product of
two functions~\cite{Alioli:2010xd,Alioli:2008tz} according to
\begin{equation}
\label{eq:F-function}
 F(\Phi_{n+1}) = F_\mathrm{damp}(\Phi_{n+1})~F_\mathrm{bornzero}(\Phi_{n+1})\;,
\end{equation}
where by default $F_\mathrm{damp}(\Phi_{n+1})$ and 
$F_\mathrm{bornzero}(\Phi_{n+1})$ are set identical to $1$, such that 
$F(\Phi_{n+1})=1$ and consequently $R_s=R$ and $R_h=0$.
When enabling a non-trivial $F_\mathrm{damp}$ in the \powhegbox{}, the 
corresponding function takes the form
\begin{equation}
 F_\mathrm{damp}(\Phi_{n+1}) = 
 \frac{h_\mathrm{damp}^2}{h_\mathrm{damp}^2+p_T^2}\;,
 \label{eqn:hdamp}
\end{equation}
where $h_\mathrm{damp}$ is a dimensionful parameter that can either be set to a
constant value or to a function of the underlying Born kinematics $\Phi_n$.
The function $F_\mathrm{damp}$ then ensures that events with large $p_T$ are 
treated as part of the hard real contributions $R_h$, while events in 
soft/collinear regions (where $p_T \to 0$) are associated to $R_s$.

On the other hand, when enabling a non-trivial $F_\mathrm{bornzero}$,
the corresponding function takes the form~\cite{Alioli:2008gx,Alioli:2010xd}
\begin{equation}
 F_\mathrm{bornzero}(\Phi_{n+1}) = \Theta\left(h_\mathrm{bornzero} - 
 \frac{R(\Phi_{n+1})}{P_{ij}(\Phi_r)\otimes B(\Phi_n)} \right)\;,
 \label{eqn:bornzero}
\end{equation}
where $P_{ij}(\Phi_r)\otimes B(\Phi_n)$ is an approximation to $R(\Phi_{n+1})$ 
based on the factorization properties of the real amplitudes in the soft and 
collinear limits. The dimensionless parameter $h_\mathrm{bornzero}$ controls how 
much phase space outside of the singular limits is associated to 
$R_s$\footnote{The default value of $h_\mathrm{bornzero}$ is $5$ in the 
\powhegbox{}.}. In particular, it ensures that if $B(\Phi_n)$ vanishes in certain 
regions of phase space, no large contributions from the first term on the 
right-hand side of Eq.~\eqref{eqn:pwg_master} will be produced. While 
historically this was the main reason to introduce $F_\mathrm{bornzero}$, this 
function also helps to identify and distinguish different enhancement mechanisms 
of the real matrix elements, which should not be interpreted as due to QCD 
splittings. As we will see, this plays a crucial role for the
$\mathcal{O}(\alpha_s\alpha^3)$ contributions in our present study.

As a final remark, we would like to highlight that the role played by the 
\powheg{} damping functions is different from the role played by the initial 
shower scale $\muq$ in the \mcnlo{} method. In the \powheg{} method the damping 
functions are used to control the degree of resummation for non-singular 
contributions, while the initial shower scale is always assigned by the \powheg{} 
framework to be the transverse momentum of the parton splitting independent of 
whether that splitting is associated to $R_s$, the \textit{soft} or to $R_h$ the
\textit{hard} contribution. Thus, the function
$F(\Phi_{n+1})=F_\mathrm{damp}(\Phi_{n+1})~F_\mathrm{bornzero}
(\Phi_{n+1})$ can have a substantial impact on the generated event sample but 
only a mild impact on the subsequent parton shower evolution. In contrary, the 
initial shower scale $\muq$ of the \mcnlo{} method directly controls the 
available phase space for subsequent parton shower emissions and therefore can 
have a strong impact on the shower evolution. The impact of the damping function
and the initial shower scale is formally of higher order and does not spoil the 
NLO accuracy of the predictions. Nonetheless, these higher-order corrections can 
become sizable.

\section{Details of the calculation}
\label{sec:details}
In this section we present our implementation of the $pp \to t\tb W^\pm$ process 
in the \powhegbox{} including NLO corrections. We start by discussing the 
perturbative orders in $\alpha_s^n\alpha^k$ that we are considering.

\begin{figure}[h!]
\begin{center}
 \includegraphics[scale=0.9]{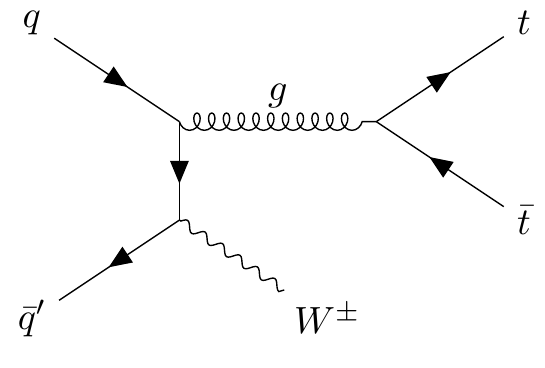}
 \includegraphics[scale=0.9]{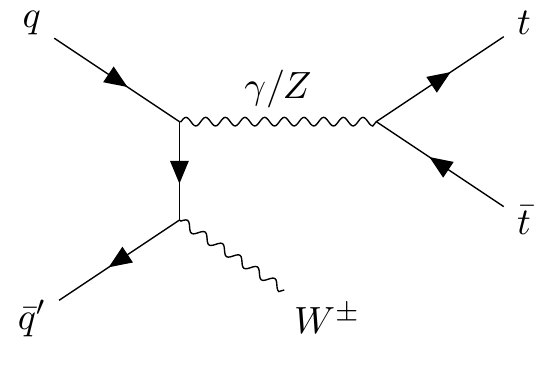}
 \includegraphics[scale=0.9]{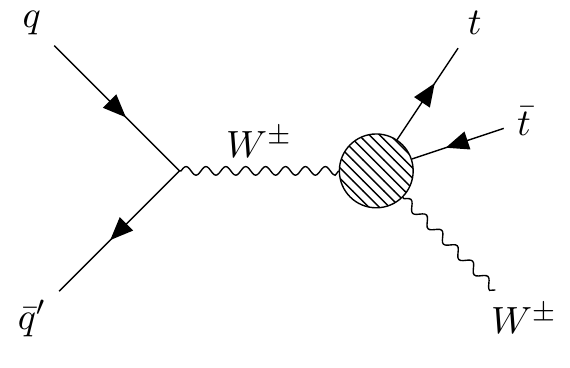}
\end{center}
\caption{Representative Feynman diagrams for QCD (left) and EW (middle, right) 
tree-level contributions to the $pp\rightarrow t\tb W^\pm$ process. Contributions 
to the $WWtt$ blob on the right are shown in Fig.~\ref{fig:WWtt-blob}.}
\label{fig:tree-level}
\end{figure}
\begin{figure}[h!]
\begin{equation*}
  \includegraphics[scale=0.75]{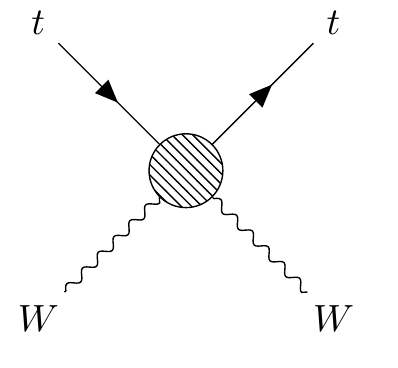}
   ~\raisebox{1.4cm}{=}~
  \includegraphics[scale=0.75]{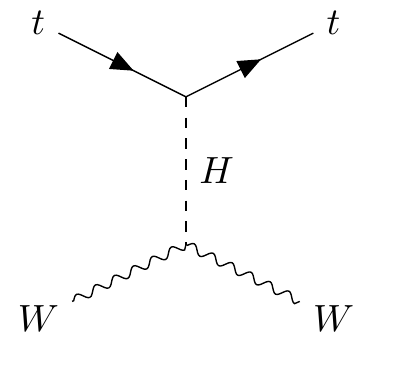}
   ~\raisebox{1.4cm}{+}~
  \includegraphics[scale=0.75]{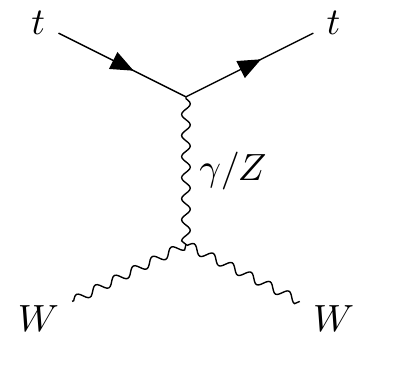} 
   ~\raisebox{1.4cm}{+}~
  \includegraphics[scale=0.75]{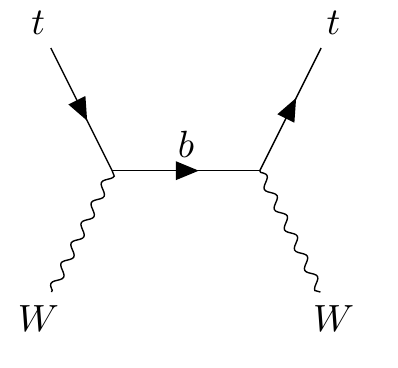} 
   ~\raisebox{1.4cm}{+}~
  \includegraphics[scale=0.75]{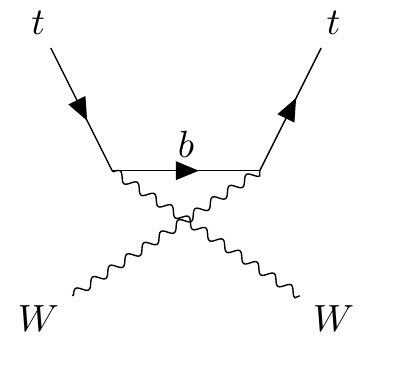} 
\end{equation*}
\caption{The diagrams contributing to the $WWtt$ tree-level amplitude in unitary 
gauge.}
\label{fig:WWtt-blob}
\end{figure}
\begin{figure}[h!]
\begin{center}
 \includegraphics[width=\textwidth]{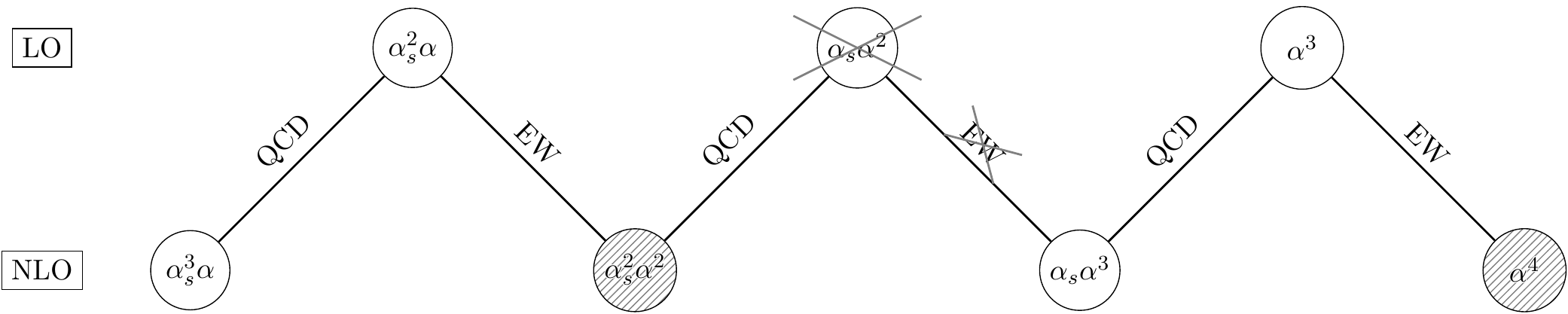}
\end{center}
\caption{The possible coupling combinations contributing to the LO and NLO 
$t\tb W^\pm$ cross section. The links indicate how a given NLO order originates 
from a corresponding LO order via QCD or EW corrections. The terms crossed out 
vanish by color structure. In this study the orders corresponding to the shaded 
bubbles are neglected.}
\label{fig:couplings}
\end{figure}

At tree level the $t\tb W^\pm$ final state can be generated in hadron collisions 
via $q\bar{q}^\prime\rightarrow t\tb W^\pm$ subprocesses, as illustrated in 
Fig.~\ref{fig:tree-level}. As illustrated in Fig.~\ref{fig:couplings} the 
corresponding LO cross section receives contributions from three different 
coupling combinations, namely $\mathcal{O}(\alpha_s^2\alpha)$, 
$\mathcal{O}(\alpha_s\alpha^2)$, and $\mathcal{O}(\alpha^3)$, although the 
QCD-EW interference term of $\mathcal{O}(\alpha_s\alpha^2)$ vanishes by color
structure (as indicated in Fig.~\ref{fig:couplings} by having it crossed out).
At the next order in the QCD or EW couplings all four orders illustrated in the 
second line of Fig.~\ref{fig:couplings} are generated. At this level, the 
classification into pure QCD or EW corrections generally breaks down since the 
two kinds of corrections start to mix.  
For example terms of order $\mathcal{O}(\alpha_s^2\alpha^2)$ can be interpreted 
as QCD corrections to the $\mathcal{O}(\alpha_s\alpha^2)$ or as EW corrections to 
the $\mathcal{O}(\alpha_s^2\alpha)$ LO cross section. In general, in the 
presence of mixed NLO corrections the complexity of these computations increases 
substantially, as infrared QED and QCD singularities have to be subtracted 
simultaneously. At the same time, a proper matching to a parton shower Monte 
Carlo event generator would have to consider both QCD and QED radiation in the 
parton shower.

In the case of $t\tb W^\pm$, the identification in terms of QCD and EW 
corrections can be restored to a very good approximation by considering the 
hierarchy of the different leading and subleading orders of the NLO cross 
section. As shown in Refs.~\cite{Dror:2015nkp,Frederix:2017wme}, apart from the 
dominant $\mathcal{O}(\alpha_s^3\alpha)$ term which amounts to about $50\%$ of 
the LO cross section (at $\sqrt{s}=13$ TeV), among all other subleading 
contributions the $\mathcal{O}(\alpha_s\alpha^3)$ is surprisingly the most 
relevant, amounting to about $10\%$ of the LO cross section. 
In comparison the $\mathcal{O}(\alpha_s^2\alpha^2)$ represents only about $4\%$ 
of the LO cross sections and the $\mathcal{O}(\alpha^4$) is below permille 
level.\footnote{Incidentally, notice that the $\mathcal{O}(\alpha_s^2\alpha^2)$ 
and $\mathcal{O}(\alpha^4)$ NLO corrections include photon-initiated 
contributions of the form $q\gamma\rightarrow t\tb W^\pm q^\prime$ which however 
amount to a small fraction of these already subleading 
corrections~\cite{Frixione:2015zaa}.}

\begin{figure}[h!]
\begin{center}
 \includegraphics[scale=1.0]{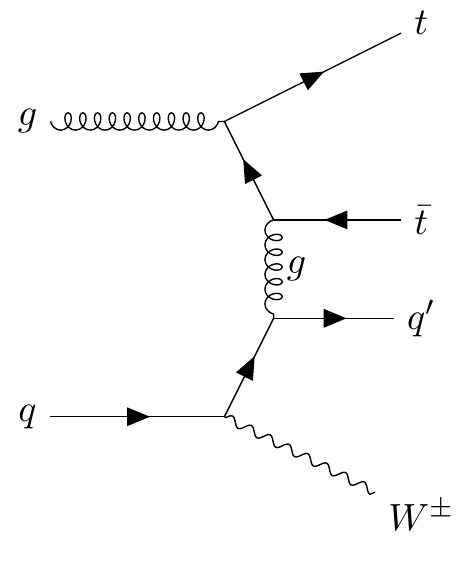} 
 \hspace{2cm}
 \includegraphics[scale=1.0]{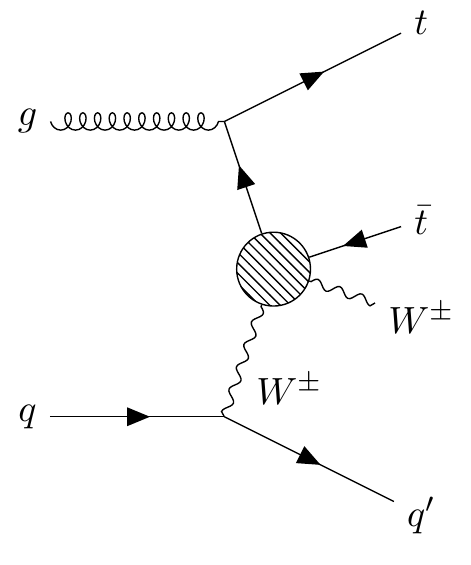}
\end{center}
\caption{Representative Feynman diagrams for the new $gq$ channel that
  opens up at NLO and that contributes at $\mathcal{O}(\alpha_s^3\alpha)$ 
  (l.h.s.) and $\mathcal{O}(\alpha_s\alpha^3)$ (r.h.s.). Contributions to the 
  $WWtt$ blob on the right are shown in Fig.~\ref{fig:WWtt-blob}.}
\label{fig:qg-t-channel}
\end{figure}

The enhancement of the $\mathcal{O}(\alpha_s\alpha^3)$ terms originates from NLO 
QCD real corrections from the $qg$ channel that opens at NLO QCD, and more 
specifically from the kind represented by the right-hand side diagram in 
Fig.~\ref{fig:qg-t-channel}. In these particular contributions, the parton 
density enhancement of $qg$ versus $q\bar{q}$ is largely amplified by the 
combined effect of several factors, from the $t$-channel kinematic, to the 
rescattering of $W$ and $Z$ longitudinal components, and the presence of a large
top-quark Yukawa coupling.

Being able to consider only the \textit{leading} $\mathcal{O}(\alpha_s^3\alpha)$ 
and $\mathcal{O}(\alpha_s\alpha^3)$ NLO contributions allows to unambiguously 
interpret them as NLO QCD corrections to the $\mathcal{O}(\alpha_s^2\alpha)$ and 
the $\mathcal{O}(\alpha^3)$ LO cross sections respectively.
Indeed, the $\mathcal{O}(\alpha_s\alpha^3)$ terms can only arise from QCD
corrections to the $\mathcal{O}(\alpha^3)$ LO since EW corrections to the 
$\mathcal{O}(\alpha_s\alpha^2)$ LO interference term vanish by color structure 
(see Fig.~\ref{fig:couplings}). As such, the fixed-order NLO QCD calculation can 
be consistently interfaced with a QCD parton-shower event generator.

In the following we will then focus solely on the contributions at the
perturbative orders  $\mathcal{O}(\alpha_s^2\alpha)$ and 
$\mathcal{O}(\alpha_s^3\alpha)$, which we will denote from now on as 
`$t\tb W^\pm$ QCD', and the orders $\mathcal{O}(\alpha^3)$ and 
$\mathcal{O}(\alpha_s\alpha^3)$, which we will denote as `$t\tb W^\pm$ EW'. 
Results from our implementation of $pp\to t\tb W^\pm$ in the \powhegbox{} have 
the same level of theoretical accuracy as the ones presented in
Ref.~\cite{Frederix:2020jzp}, and can be directly compared to the ones that can 
be obtained via analogous NLO QCD parton-shower Monte Carlo event generators such 
as \sherpa{} or \mgfive{}~+~\pythia{}. Indeed, we will show a corresponding 
comparison among these tools in section~\ref{sec:results}.

Next, we will discuss in section~\ref{subsec:ttW-nlo-powheg-box} the 
implementation of $pp\to t\tb W^\pm$ including the aforementioned orders of NLO 
QCD corrections in the \powhegbox{} framework. Furthermore, in 
section~\ref{subsec:ttW-decays} we will give some further details on the 
modelling of a fully realistic final state by including decays of unstable 
particles while keeping spin correlations at LO accuracy.

\subsection{NLO corrections to the production of $t\tb W^\pm$}
\label{subsec:ttW-nlo-powheg-box}
The implementation of a new process in the \powhegbox{} requires to provide 
process-specific ingredients such as the LO and NLO virtual and real matrix 
elements as well as the parametrization of the Born-level phase space, while all 
process-independent parts, such as the subtraction of infrared singularities, are 
automated. In our implementation of $t\tb W^\pm$, all tree-level matrix elements
including spin- and color-correlated Born matrix elements are taken from the 
\texttt{MadGraph 4}~\cite{Stelzer:1994ta,Alwall:2007st} interface that is 
provided within the \powhegbox{}. The finite remainders of the virtual loop 
corrections interfered with the Born matrix elements are computed by the fairly 
new one-loop provider for QCD and EW corrections 
\nlox{}~\cite{Honeywell:2018fcl,Figueroa:2021txg} that uses
\oneloop{}~\cite{vanHameren:2010cp} for the evaluation of scalar Feynman 
integrals. The parametrization of the $t\tb W^\pm$ phase space has been directly 
modelled on the \powhegbox{} implementation of $t\tb H$~\cite{Hartanto:2015uka}.

We performed several checks to validate our implementation. For example, virtual 
amplitudes have been successfully compared at a few phase-space points against 
\recola{}~\cite{Actis:2012qn,Actis:2016mpe} and 
\texttt{MadGraph5}~\cite{Alwall:2011uj}. At the same time, total inclusive cross 
sections at fixed order have been cross checked with 
\mgfive{}~\cite{Alwall:2014hca}, while a full validation at the differential 
level has been performed by comparing to an independent calculation obtained from 
\sherpa{}~\cite{Gleisberg:2008ta,Bothmann:2019yzt} in conjunction with either the 
version of the \blackhat{} library of Ref.~\cite{Anger:2017glm} or with the 
\openloops{} program~\cite{Cascioli:2011va,Buccioni:2017yxi,Buccioni:2019sur}.

On the other hand, as explained in section~\ref{sec:powheg}, in matching the 
fixed-order calculation with parton-shower event generator within the 
\powhegbox{} framework, we have implemented the following choices for the jet 
function $F(\Phi_{n+1})$ of Eq.~\eqref{eq:F-function}. First, similar to the case 
of $b\bb W^\pm$~\cite{Oleari:2011ey}, the leading-order matrix element at 
$\mathcal{O}(\alpha_s^2\alpha)$ has vanishing Born-level configurations.
Therefore, the usage of $F_\mathrm{bornzero}(\Phi_{n+1})$ 
(see Eq.~\eqref{eqn:bornzero}) is mandatory for $t\tb W^\pm$ QCD production. 
This is also applied to $t\tb W^\pm$ EW production, where it plays a critical 
role given the presence of the strongly enhanced real matrix element discussed 
before (due to the diagrams of the right of Fig.~\ref{fig:qg-t-channel}). Finally 
we have enabled the damping function $F_\mathrm{damp}(\Phi_{n+1})$ of 
Eq.~\eqref{eqn:hdamp} to further suppress hard radiation using a dynamic value of 
$h_{damp}$. More specifically, our default choices for the \textit{damping} 
parameters will be:
\begin{equation}
 h_\mathrm{damp} = \frac{H_T}{2}\;, \qquad h_\mathrm{bornzero} = 5\;,
\label{eqn:dampdefault}
\end{equation}
where
\begin{equation}
 H_T = \sum_{i \in \{t,\bar{t},W\}} \sqrt{m_i^2 + p_{T,i}^2}\;,
\end{equation}
is evaluated on the underlying Born kinematics. 

\begin{figure}
\begin{center}
 \includegraphics[width=0.48\textwidth]{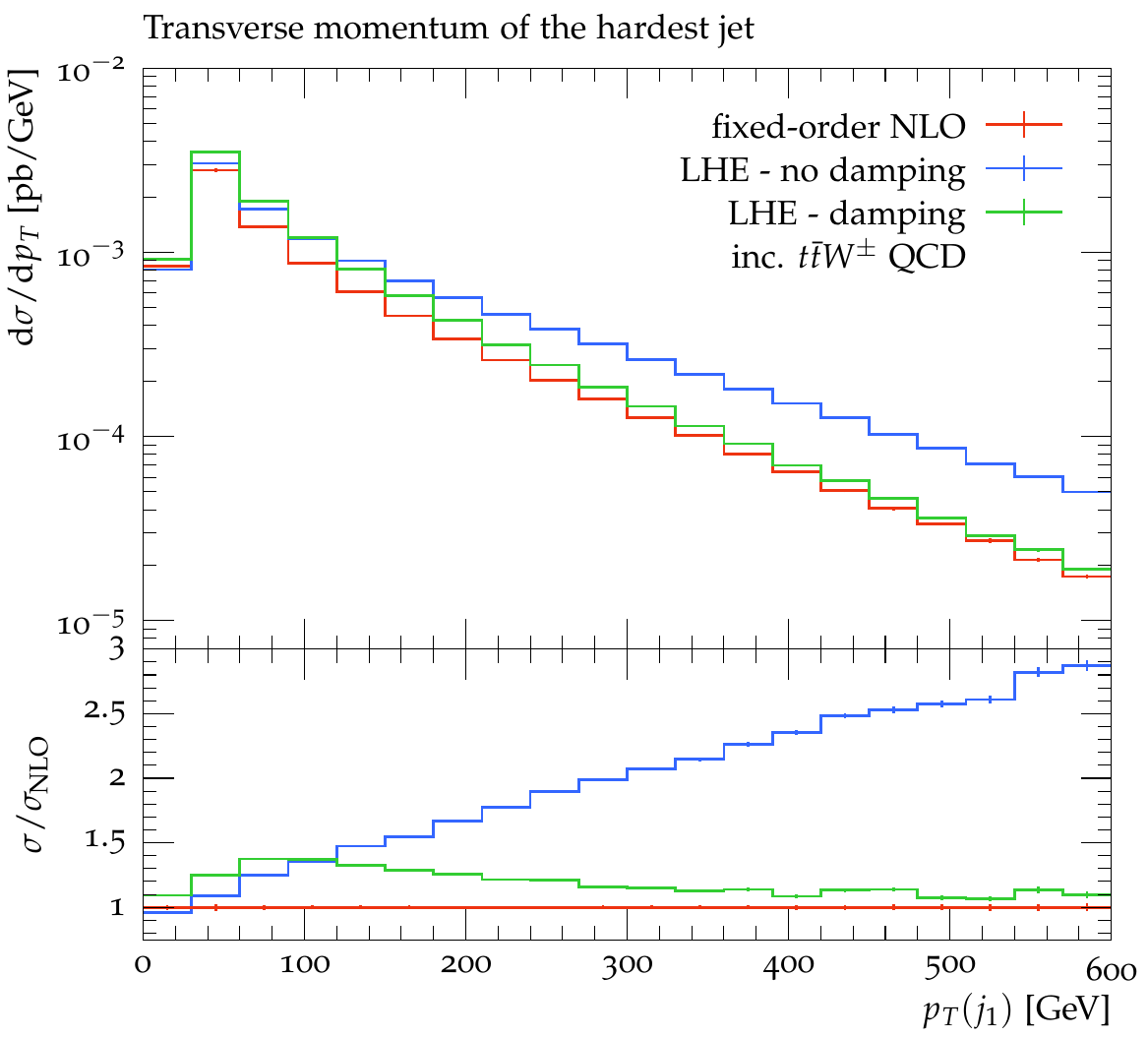}
 \includegraphics[width=0.48\textwidth]{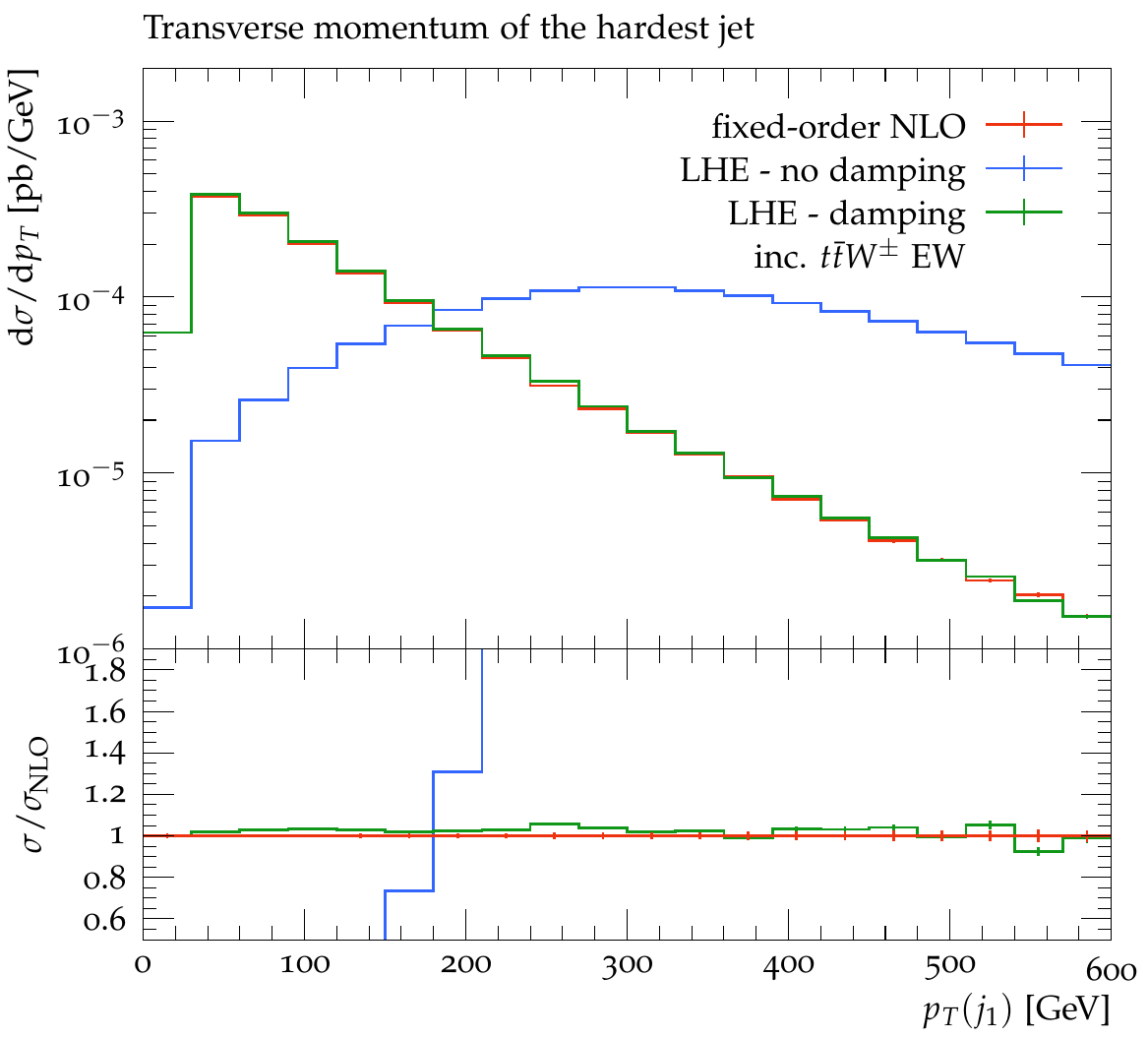} \\
 \includegraphics[width=0.48\textwidth]{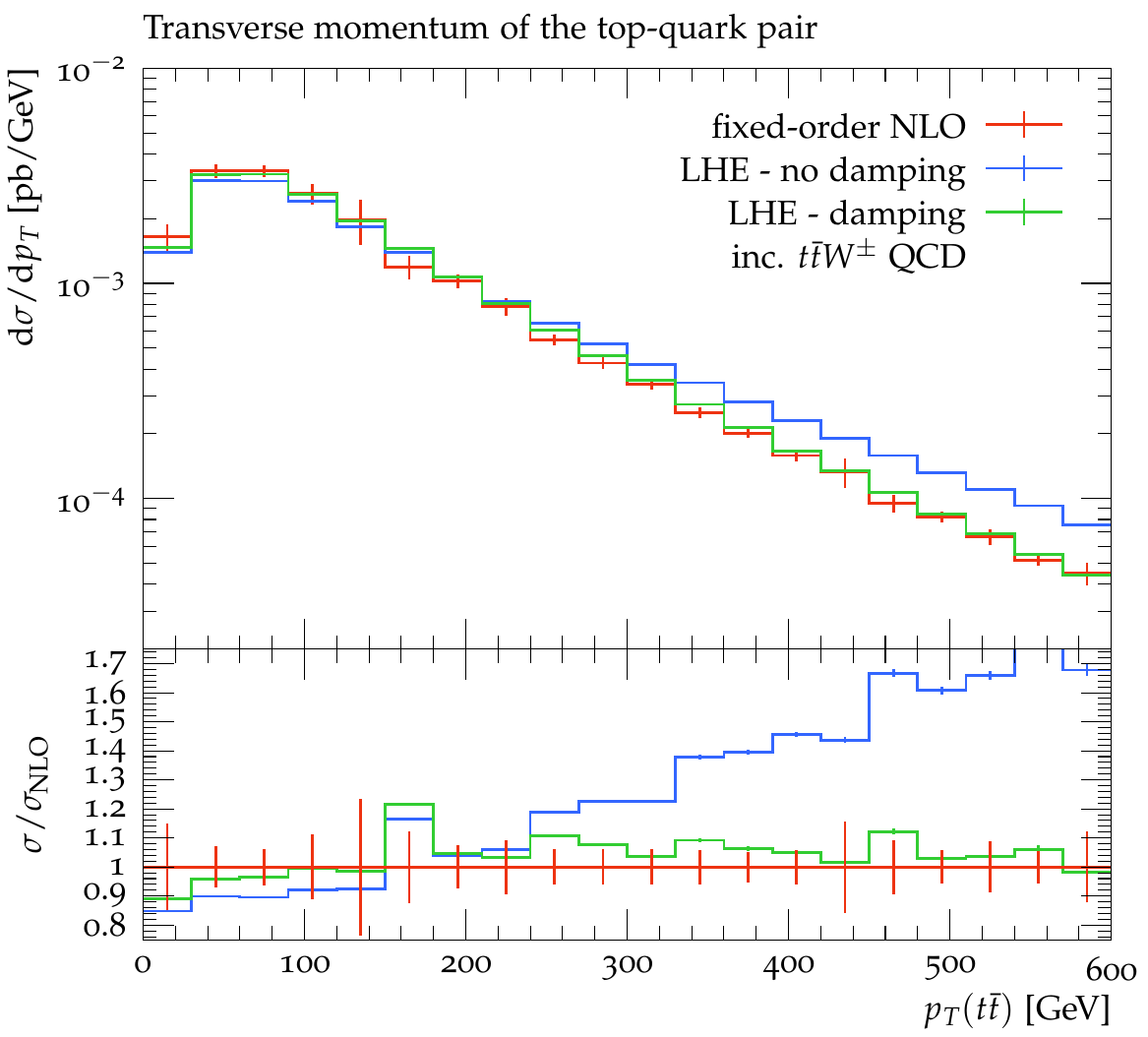}
 \includegraphics[width=0.48\textwidth]{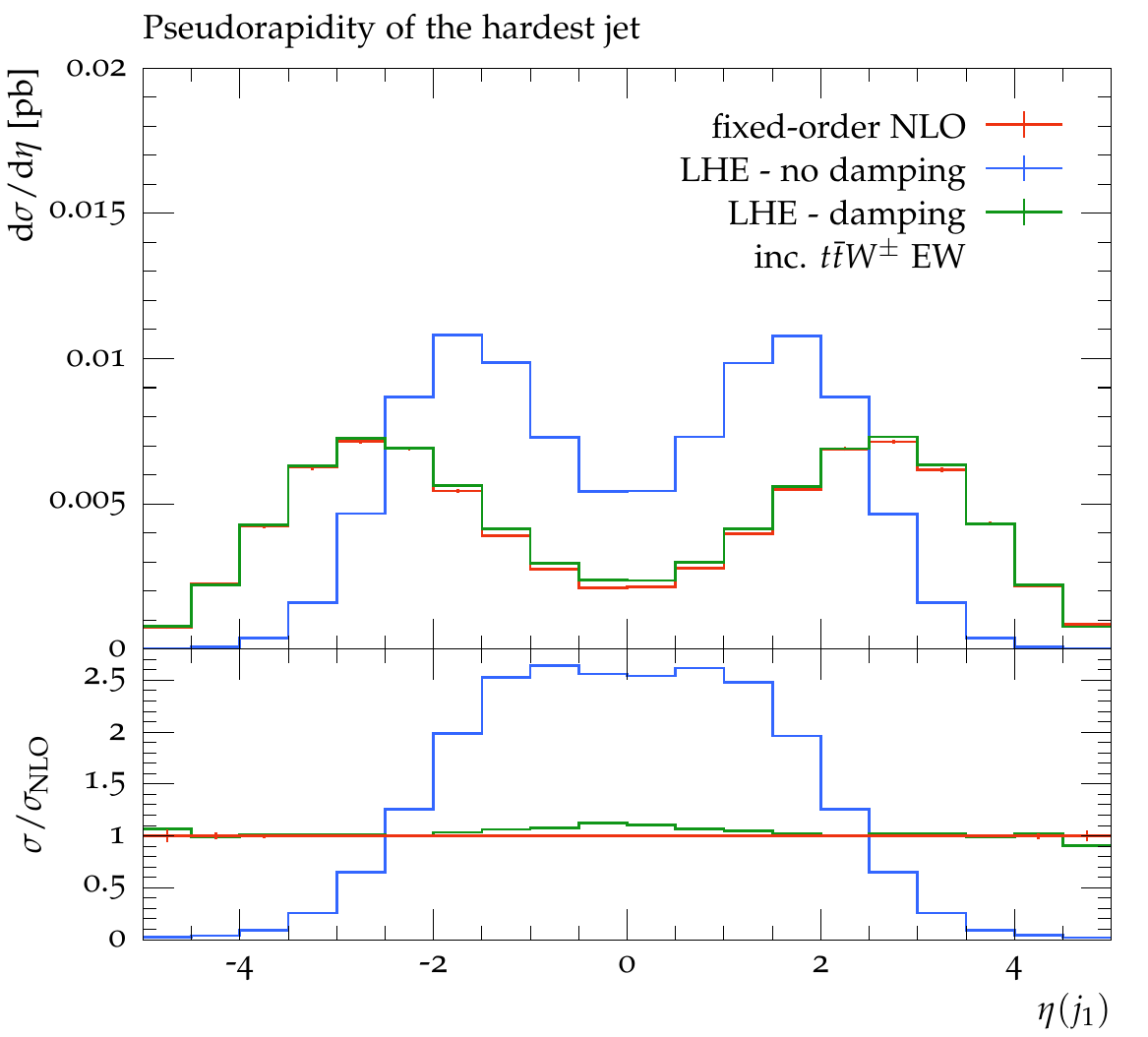} \\
 \includegraphics[width=0.48\textwidth]{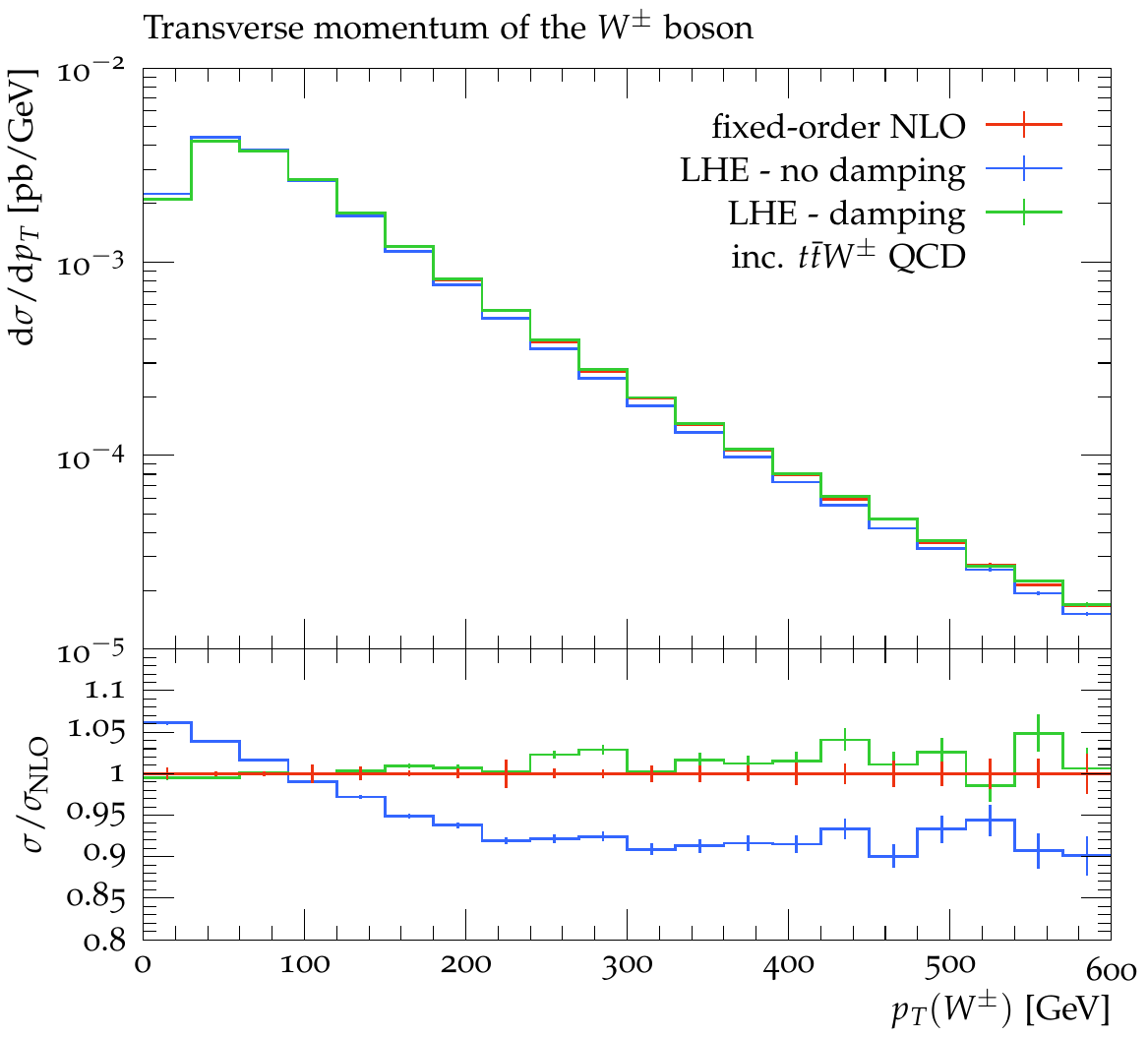}
 \includegraphics[width=0.48\textwidth]{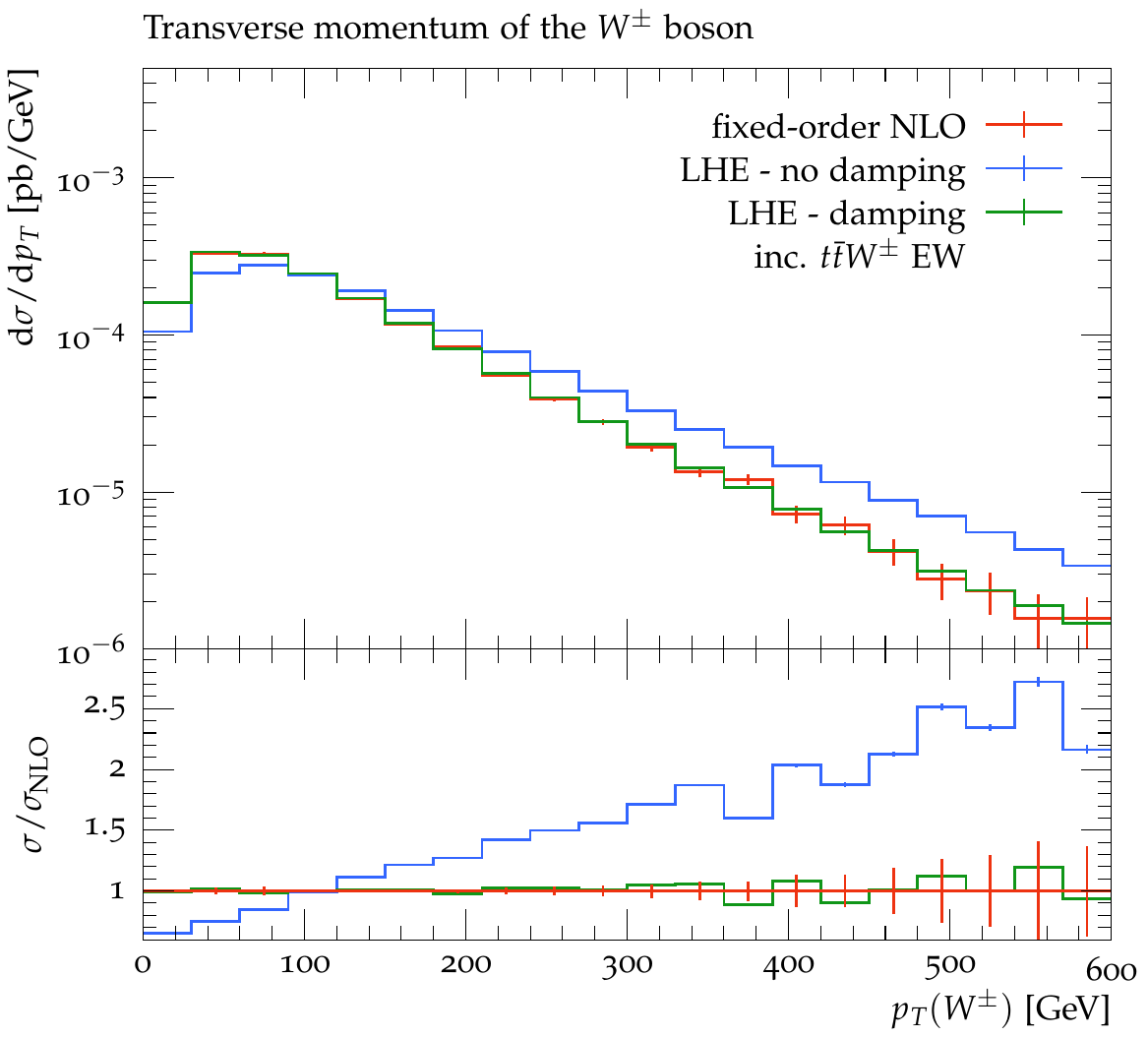} 
\end{center}
\caption{Impact of the jet function $F(\Phi_{n+1})$ (see 
Eq.~\eqref{eq:F-function}) on differential distributions. The \textit{LHE - 
damping} curves correspond to our default choice of parameters according to
Eq.\eqref{eqn:dampdefault}}
\label{fig:damp}
\end{figure}

In order to disentangle the impact of the jet function $F(\Phi_{n+1})$ from 
possible parton-shower corrections, we investigate the differential distributions 
of \powhegbox{} events without taking into account the parton-shower evolution. 
We show a few representative observables in Fig.~\ref{fig:damp} at the level of 
Les Houches events (LHE)~\cite{Boos:2001cv,Alwall:2006yp}. On the left we show 
transverse momentum distributions for the leading jet, the top-quark pair, and 
the $W^\pm$ boson for inclusive $t\tb W^\pm$ QCD production, while on the right 
we show the transverse momentum and pseudorapidity distributions of the leading 
jet, as well as the transverse momentum distribution of the $W^\pm$ boson for 
inclusive $t\tb W^\pm$ EW production. In all cases, we compare \powhegbox{} 
predictions with and without the effect of the jet function $F(\Phi_{n+1})$ 
(labeled as `LHE - damping' and  `LHE - no damping' respectively) to the 
corresponding fixed-order differential distributions.

For inclusive $t\tb W^\pm$ QCD production we observe large shape differences 
between the fixed-order NLO prediction and the \powhegbox{} results without the 
jet function. In hadronic observables such as the transverse momentum of the 
hardest jet the differences reach up to a factor of 3 at a transverse momentum of 
around $600~\GeV$. A similar behavior can be seen in the transverse momentum of 
the top-quark pair, where the deviations are up to $+70\%$ at the end of the 
plotted range. Even in non-hadronic observables such as the transverse momentum 
of the $W^\pm$ boson moderate differences at the level of $10\%$ are visible. 
However once the damping mechanism is taken into account the predictions recover 
the tails of the NLO fixed-order distributions, which are reliably described by 
fixed-order matrix elements. Though considerable improvement is achieved for the 
transverse momentum distribution of the hardest jet with the inclusion of the jet
function, still a small difference of about $10\%$ is observed for large 
$p_{T}(j_1)$. On the other hand, shape differences for low $p_T$ values such as 
in the case of the transverse momentum of the leading jet are attributed to the 
resummation of soft and collinear QCD splittings via the Sudakov form factor and 
are thus expected to be different from the fixed-order result. In the case of 
inclusive $t\tb W^\pm$ EW production the impact of the jet function is more 
dramatic. Here, the NLO QCD corrections are dominated by real radiative 
contributions that are enhanced by the $t$-channel EW scattering of $tW \to tW$ 
shown in Fig.~\ref{fig:WWtt-blob} and Fig.~\ref{fig:qg-t-channel}. Therefore, the 
resummation of the full real matrix element yields unphysical results as 
non-factorizing contributions are resummed and thus the jet function 
$F(\Phi_{n+1})$ has to be used to restrict the resummation to singular QCD 
splittings that indeed factorize. As a consequence of choosing a trivial jet 
function $F(\Phi_{n+1})=1$, the shapes of the transverse momentum and the 
rapidity distributions of the leading jet are described poorly over the entire 
phase space. Also the transverse momentum of the $W^\pm$ boson shows large 
discrepancies in the tail of the distribution.  Nonetheless, using our default 
jet function $F(\Phi_{n+1})$ parameters (see Eq.~\eqref{eqn:dampdefault}) we 
observe an excellent agreement between the NLO fixed-order computation and the
\powhegbox{} results.  Furthermore, the impressive agreement of the transverse 
momentum distribution of the leading jet down to low values of $p_{T}(j_1)$ just 
highlights how radiation attributed to the \textit{hard} contribution $R_h$ 
(see Eq.~\eqref{eqn:pwg_master}) dominates the production of that leading jet, 
while enhancements from resummation effects are nearly negligible.

\subsection{Decay modelling of the $t\tb W^\pm$ system}
\label{subsec:ttW-decays}
In order to be able to study a broader range of exclusive observables, which 
depend strongly on spin correlations from production to decaying particles, we 
include the decay of the $t\tb W^\pm$ system in our implementation. Typically, 
parton-shower event generators decay unstable particles during the shower 
evolution. This approach is the simplest but it cannot preserve spin correlations 
in the decays, as each particle is decayed independently. In the following we
briefly discuss our implementation of the decays in the \powhegbox{} that 
preserves spin correlations at least to LO accuracy, which can have a sizable 
impact on the top-quark decay products since the emission of the $W^\pm$ boson 
in the initial-state polarizes the top quarks~\cite{Maltoni:2014zpa,
Frederix:2020jzp}.

Our approach follows closely the method of Ref.~\cite{Frixione:2007zp}. This 
method has been adopted in the \powhegbox{} already for several processes (see 
for example Refs.~\cite{Alioli:2009je,Oleari:2011ey,Alioli:2011as,
Hartanto:2015uka}). The basic idea can be summarized as follows. Starting from an 
on-shell phase space point for the $t\tb W^\pm$ momenta one performs a 
reshuffling of the momenta to allow for off-shell virtualities of the unstable 
particles. Afterwards, momenta of decay products are generated uniformly in the 
decay phase space and finally these momenta are unweighted against the fully 
decayed matrix element by constructing a suitable upper bounding function. 

All previous implementations in the \powhegbox{} have in common that only the 
decay of a top-quark pair has been taken into account. In order to allow these 
reshuffles for more than two unstable particles, we introduce a new momentum 
mapping in a process-independent and Lorentz-invariant way.

The \powhegbox{}, after the single-step parton shower process (see 
section~\ref{sec:powheg}), generates on-shell momentum configurations 
$\Phi_n^\mathrm{OS}$ that either refer to a $t\tb W^\pm$ or a $t\tb W^\pm j$ 
final state. We start by generating independent virtualities $v^2$ for each top 
quark and $W^\pm$ boson according to Breit-Wigner distributions
\begin{equation}
 \rho(v^2) = \frac{m_f\Gamma_f}{\pi}~\frac{1}{(v^2-m_f^2)^2 + m_f^2\Gamma_f^2}\;,
\end{equation}
where $m_f$ and $\Gamma_f$ are the particle mass and decay width.
We constrain the generated virtualities to the window $|v-m_f| < 5~\Gamma_f$.
These virtualities will be imprinted onto the momenta of the on-shell phase space
point $\Phi_n^\mathrm{OS}$ by the repeated application of the mapping presented
in Appendix~\ref{app:offshellproj}, where we choose to preserve the momentum 
of the light jet in the case of a real radiation event. This amounts to choosing 
$Q=p_t + p_{\bar{t}} + p_W$ as the total available momentum in the mapping of 
Appendix ~\ref{app:offshellproj} and excluding the jet from the Lorentz boost.
In the following we will call the off-shell phase space configuration simply 
$\Phi_n$.

Afterwards, momenta of the decay products are uniformly generated as a sequence 
of $1 \to 2$ decays. This allows also to include off-shell $W$ bosons in the 
top-quark decays. In the last step we apply the \textit{hit-and-miss} technique 
(see e.g. Ref.~\cite{Alioli:2009je}) on the so obtained final state momenta using 
an upper bounding function 
$U_\textrm{dec}(v_t^2,v_{\tb}^2,v_W^2,\Phi_{t\to b\ell\nu},
\Phi_{\tb \to \bar{b}\ell\nu},\Phi_{W\to \ell\nu})$, constructed according to 
Ref.~\cite{Frixione:2007zp}, such that
\begin{equation}
\frac{\mathcal{M}_\textrm{dec}(\Phi_n,\Phi_{t\to b\ell\nu},
\Phi_{\tb \to \bar{b}\ell\nu},\Phi_{W\to \ell\nu})}{\mathcal{M}_\textrm{undec}
(\Phi_n^\mathrm{OS})}
\le U_\textrm{dec}(v_t^2,v_{\tb}^2,v_W^2,\Phi_{t\to b\ell\nu},
\Phi_{\tb \to \bar{b}\ell\nu},\Phi_{W\to \ell\nu})\;,
\end{equation}
where $\mathcal{M}_\textrm{undec}(\Phi_n^\mathrm{OS})$ is the LO matrix element
for the undecayed process, while $\mathcal{M}_\textrm{dec}(\dots)$ is a so-called 
decay-chain matrix element, which corresponds to a LO matrix element for the 
fully decayed process where only diagrams with the resonance structure of 
interest are kept. These decay-chain matrix elements are taken from 
\mgfive{}~\cite{Alwall:2014hca}. 

The result of this procedure is such that all spin correlations between unstable 
particles and decay particles are kept with LO precision.

For completeness, we briefly discuss the extension to the $t\tb W^\pm$ process 
of the method to choose particular decay signatures, which has been previously 
used in \powhegbox{} implementations. We present details for the $t\tb W^+$ 
process, as the $t\tb W^-$ process is treated in an analogous way. All top quarks 
decay into a $W$ boson and a $b$-quark, and we now have to take into account the 
decay of three $W$ bosons. Based on the branching ratios $\mathrm{Br}(W \to
\ell_i \nu_i)$ and $\mathrm{Br}(W \to q_i \bar{q}^\prime_i)$ we can construct a 
density matrix $\rho$ for the decay probabilities
\begin{equation}
 \rho_{ijk} = \mathrm{Br}(W^+ \to X_i)\mathrm{Br}(W^- \to \overline{X}_j)
 \mathrm{Br}(W^+ \to X_k)\;,
\end{equation}
where $X_i~(\overline{X}_i)$ represent the possible (charge conjugated) final 
states, namely
\begin{equation}
 X_1 = e^+\nu_e\;, \quad X_2 = \mu^+\nu_\mu\;, \quad X_3 = \tau^+\nu_\tau\;, \quad
 X_4 = u\bar{d}\;, \quad X_5 = c\bar{s}\;.
\end{equation}
The total branching ratio is then given by
\begin{equation}
 \mathrm{Br}_\mathrm{tot} = \sum_{i,j,k} \rho_{ijk}\;,
\end{equation}
which can be less than $1$ if, for example, only particular decay channels are 
selected. 

To choose a particular decay channel we simply perform a hit-and-miss procedure 
on the components of the probability density matrix $\rho$. To allow for hadronic 
decays we take into account the CKM mixing of the first two generations, which is 
parametrized by
\begin{equation}
 V_\mathrm{CKM} = \begin{pmatrix} V_{ud} & V_{us} & V_{ub} \\
 V_{cd} & V_{cs} & V_{cb} \\
 V_{td} & V_{ts} & V_{tb} \end{pmatrix} = \begin{pmatrix}
 \cos\theta_c & \sin\theta_c & 0 \\
 -\sin\theta_c & \cos\theta_c & 0 \\
 0 & 0 & 1\end{pmatrix}\;.
\end{equation}
For example, when the decay $W^+ \to u\bar{d}$ has been chosen we simply generate 
a random number $r \in [0,1]$ and if $r \le |V_{us}|^2 = \sin^2\theta_c$ we 
choose the decay $W^+ \to u\bar{s}$, otherwise we keep the decay 
$W^+ \to u\bar{d}$.

At last we would like to mention that with the procedure outlined above the 
symmetry factors for identical final state particles are obtained in the correct 
way. For example, for a calculation employing full off-shell matrix elements it 
is clear that 
\begin{equation}
 \sigma(pp \to b\bar{b}~e^+ \nu_e~\mu^-\bar{\nu}_\mu~e^+ \nu_e) = \frac{1}{2}~ 
 \sigma(pp \to b\bar{b}~e^+ \nu_e~\mu^-\bar{\nu}_\mu~\tau^+ \nu_\tau)\;,
\end{equation}
due to symmetry factors. Since in our methodology we generate decays subsequently
we have to ensure that the probability to generate the $e^+\mu^-\tau^+$ final 
state is twice as large as the one for the $e^+\mu^-e^+$ final state. This is 
trivially ensured in our procedure, since
\begin{equation}
 \mathrm{Br}(t\tb W^\pm\rightarrow e^+\mu^-e^+) = \rho_{121} = \frac{1}{729}\;, 
\end{equation}
and
\begin{equation}
 \mathrm{Br}(t\tb W^\pm\rightarrow e^+\mu^-\tau^+) = \rho_{123} + \rho_{321}=  
 \frac{2}{729}\;,
\end{equation}
where we have used $\mathrm{Br}(W\to \ell_i \nu_i) = 1/9$.

We performed several cross checks on the modelling of the decays. For instance,
we checked that the correct branching ratios are obtained from inclusive event
samples that take into account all possible decay modes. Also, we compared at the 
differential level unshowered leading-order events for a particular decay mode 
with events obtained through the same procedure by \mgfive{} in conjunction with 
\madspin{}~\cite{Artoisenet:2012st} and found perfect agreement.

\section{Phenomenological results}
\label{sec:results}

In this section we present and discuss numerical results for $pp\to t\tb W^\pm$ 
obtained with the new \powhegbox{} implementation described in this paper and 
compare them with analogous results obtained from \mgfive{} and \sherpa. After 
reviewing the general settings for the input parameters of our study in
section~\ref{sec:setup}, we will consider the case of inclusive $t\tb W^\pm$ 
production in sec.~\ref{sec:ttW-inclusive} and the case of a specific signature 
with two same-sign leptons and jets (usually referred to as $2\ell SS$) in 
section~\ref{sec:2lSS}. In the following we will denote by $t\tb W^\pm$ the sum 
of both $t\tb W^+$ and $t\tb W^-$ production.

\subsection{Computational setup}
\label{sec:setup}

In our study we consider $t\tb W^\pm$ production at the LHC with a center-of-mass 
energy of $\sqrt{s}=13$ TeV.  All results presented in this section have been 
obtained using the NNPDF3.0~\cite{Ball:2014uwa} (\texttt{NNPDF30-nlo-as-0118}) 
parton distribution functions as provided by 
\textsc{LHAPDF}~\cite{Buckley:2014ana}. We have not performed a detailed study of 
the PDF uncertainty associated with this production mode since it does not 
directly affect either the comparison between fixed-order and parton-shower 
results or the comparison between different NLO parton-shower Monte Carlo event
generators. Of course such uncertainty should be included in future more 
comprehensive assessments of the overall theoretical uncertainty on this 
production mode.

The necessary Standard Model parameters have been chosen to be:
\begin{equation}
\begin{array}{lll}
 G_F = 1.166378\cdot 10^{-5}~\GeV^{-2}\;, &  m_t = 172.5~\GeV\;, & m_b=0~\GeV\;,\\ 
 M_W = 80.385~\GeV\;, & M_Z = 91.1876~\GeV\;, & M_H = 125~\GeV\;, \\
 \Gamma_t = 1.33247~\GeV\;, & \Gamma_W = 2.09767~\GeV\;, & 
 \Gamma_Z = 2.50775~\GeV\;,
\end{array}
\end{equation}
in terms of which the electromagnetic coupling is defined as:
\begin{equation}
 \alpha = \frac{\sqrt{2}}{\pi} G_F M_W^2\left( 1- \frac{M_W^2}{M_Z^2}\right)\;.
\end{equation}
The central values of the renormalization and factorization scales are set to:
\begin{equation}
 \mur = \muf = \mu_0 = \frac{H_T}{2}\;,
\end{equation}
with 
\begin{equation}
 H_T = \sum_{i~\in~\text{final state}} \sqrt{m_i^2 + p_{T,i}^2}\;.
\end{equation}
The theoretical uncertainties associated with this choice of scales are estimated 
via the $7$-point envelope that corresponds to $\mu_R$ and $\mu_F$ assuming the 
following sets of values:
\begin{equation}
 \left(\frac{\mur}{\mu_0}, \frac{\muf}{\mu_0}\right) = \Big\{ 
 (0.5,0.5), (0.5,1), (1,0.5), (1,1), (1,2), (2,1), (2,2)\Big\}\;.
\end{equation}
We notice that in quoting the dependence of the results presented in this section 
from scale variation, the previously defined scale variation has been applied 
only to the hard matrix elements, while scale choices in the parton shower have 
not been altered.

In order to assess the nature and size of theoretical uncertainties present in 
the modelling of the $t\tb W^\pm$ process, we will compare results from three 
different NLO parton-shower event generators, \powhegbox{}, \mgfive{}, and 
\sherpa{}, using the setups described next. In the following figures 
we will label the results obtained using different tools accordingly.

\begin{itemize}

\item \powhegbox{}: We generate results using the \powhegbox{}
  implementation presented in this paper by performing the
  parton-shower matching to \pythia{}~\cite{Sjostrand:2014zea}
  ($v.~8.303$) using the \powheg{} method as described in
  sections~\ref{sec:powheg} and
  ~\ref{subsec:ttW-nlo-powheg-box}, and modelling the decay of the
  $t\tb W^\pm$ final state as discussed in
  section~\ref{subsec:ttW-decays}.  We employ by default the
  \textit{damping} parameters shown in Eq.~\eqref{eqn:dampdefault},
  that is:
\begin{equation*}
 h_\mathrm{damp} = \frac{H_T}{2}\;, \qquad h_\mathrm{bornzero} = 5\;,
\end{equation*}
where the dynamic damping parameter $h_\mathrm{damp}$ is evaluated on the 
underlying born kinematics. The impact of different choices of damping parameters
is estimated by the 5-point envelope that corresponds to choosing the values:
\begin{equation}
 \left( h_\mathrm{damp}, h_\mathrm{bornzero}\right) = \left\{
 \left(\frac{H_T}{2},5\right), \left(\frac{H_T}{2},2\right), 
 \left(\frac{H_T}{2},10\right), \left(\frac{H_T}{4},5\right),
 \left( H_T, 5\right) \right\}\;.
 \label{eqn:std_damp}
\end{equation}
For \pythia{} we use the $A14$ shower tune and turn off all its matrix element
corrections (MEC) to decay processes.

\item \mgfive{}:
We obtain results using \mgfive{} (v2.7.2) by performing the parton shower 
matching to \pythia{} with the \mcnlo{} method. Here the initial shower scale is 
set to $\muq = H_T/2$ and decays of unstable particles are taken into account via 
the \madspin{} framework. Also for \mgfive{} we asses the matching uncertainties 
related to the initial shower scale by making a 3-point envelope which 
corresponds to the values:
\begin{equation}
 \muq = \left\{ \frac{H_T}{4}, \frac{H_T}{2}, H_T \right\}\;.
\end{equation}
For \pythia{} we use the $A14$ shower tune with the standard \mgfive{} parameter
settings, which also do not include MEC to decay processes.

\item \sherpa{}:
Finally, we use the \sherpa{} (v2.2.10) parton-shower event generator to produce
a third set of independent results. For $t\tb W^\pm$ QCD production we use the 
\mcnlo{} matching procedure and \sherpa's parton shower, which is based on 
Catani-Seymour dipole factorization~\cite{Schumann:2007mg}. The corresponding
one-loop matrix elements for these studies are taken from the 
\openloops{}~\cite{Cascioli:2011va,
Buccioni:2017yxi,Buccioni:2019sur} program. To produce predictions with
\sherpa's public version for $t\tb W^\pm$ EW production we employed a truncated 
shower merging procedure (MEPS/CKKW) setup~\cite{Catani:2001cc,Hoeche:2009rj}, 
including LO matrix elements for the processes $t\tb W^\pm$ and $t\tb W^\pm j$ 
(at $\mathcal{O}(\alpha^3)$ and $\mathcal{O}(\alpha_s\alpha^3)$ respectively).  
The latter setup catches key contributions for $t\tb W^\pm$ EW production, in 
particular the large hard real contributions. We do not show scale variations or 
matching uncertainties for \sherpa, as we have checked that variations of its 
resummation scale produces similar results to those obtained with \mgfive{} 
shower scale variations, in agreement with the findings of 
Ref.~\cite{ATLAS:2020esn}. In \sherpa{} spin-correlated decays are implemented 
via the method presented in Ref.~\cite{Richardson:2001df}.
\end{itemize}

We do not include in any of the previous setups non-perturbative corrections like 
those from hadronization or multiple parton interactions. Notice that, even 
though the hard NLO computation is performed in a massless $5$ flavor scheme the 
used parton showers in this study adopt by default a non-zero bottom mass.

Combining the effect of renormalization and factorization scale variation with 
the variations induced by different choices of matching schemes and parton-shower 
codes enables us to address the numerical impact of higher-order corrections that
are inherent to each approach. We want to emphasize however that the variation of
damping parameters and the initial shower $\muq$ have different effects as 
explained at the end of section~\ref{sec:powheg}. Nonetheless, we will refer to 
the sensitivity of the predictions on these parameters as an 
\textit{uncertainty}.

The showered events are finally passed via the 
\hepmc{}~\cite{Dobbs:684090,Buckley:2019xhk} interface to an analysis routine
written in the \rivet{} framework~\cite{Buckley:2010ar,Bierlich:2019rhm}. 

In ancillary material that we make available with this document we provide the 
necessary files to reproduce our results. They include run cards for \powhegbox{}
and \sherpa{} as well as our \rivet{} analyses. 
\subsection{Inclusive NLO+PS observables}
\label{sec:ttW-inclusive}
In this section we study the on-shell inclusive production of $t\tb W^\pm$ and 
the impact of the parton shower evolution on inclusive observables. To this end 
we focus on differential distributions that can also be computed at fixed-order 
and compare the NLO fixed-order results to the corresponding results obtained by 
parton-shower matching in different frameworks.

For this analysis we keep the top quarks and $W^\pm$ bosons stable. No cuts are 
applied on the top quarks and $W^\pm$ bosons.  Jets are formed using the 
anti-$k_T$ jet algorithm~\cite{Cacciari:2008gp} with a resolution parameter of 
$R=0.4$ as implemented in \textsc{FastJet}~\cite{Cacciari:2005hq,
Cacciari:2011ma}.  We require a minimal transverse momentum of $p_T > 25~\GeV$ 
for all jets. Note that we do not distinguish between light and $b$-flavored jets
in obtaining the results presented in this section.

In the following we will discuss first the $t\tb W^\pm$ QCD and $t\tb W^\pm$ EW
contributions to the $t\tb W^\pm$ inclusive cross section separately, as they 
show different overall features. Afterwards, we will study the impact of the EW 
contribution on the QCD+EW combined prediction for a few representative 
observables.
\subsubsection*{$t\tb W^\pm$ QCD contribution}
%
\begin{figure}[h!]
 \includegraphics[width=0.49\textwidth]{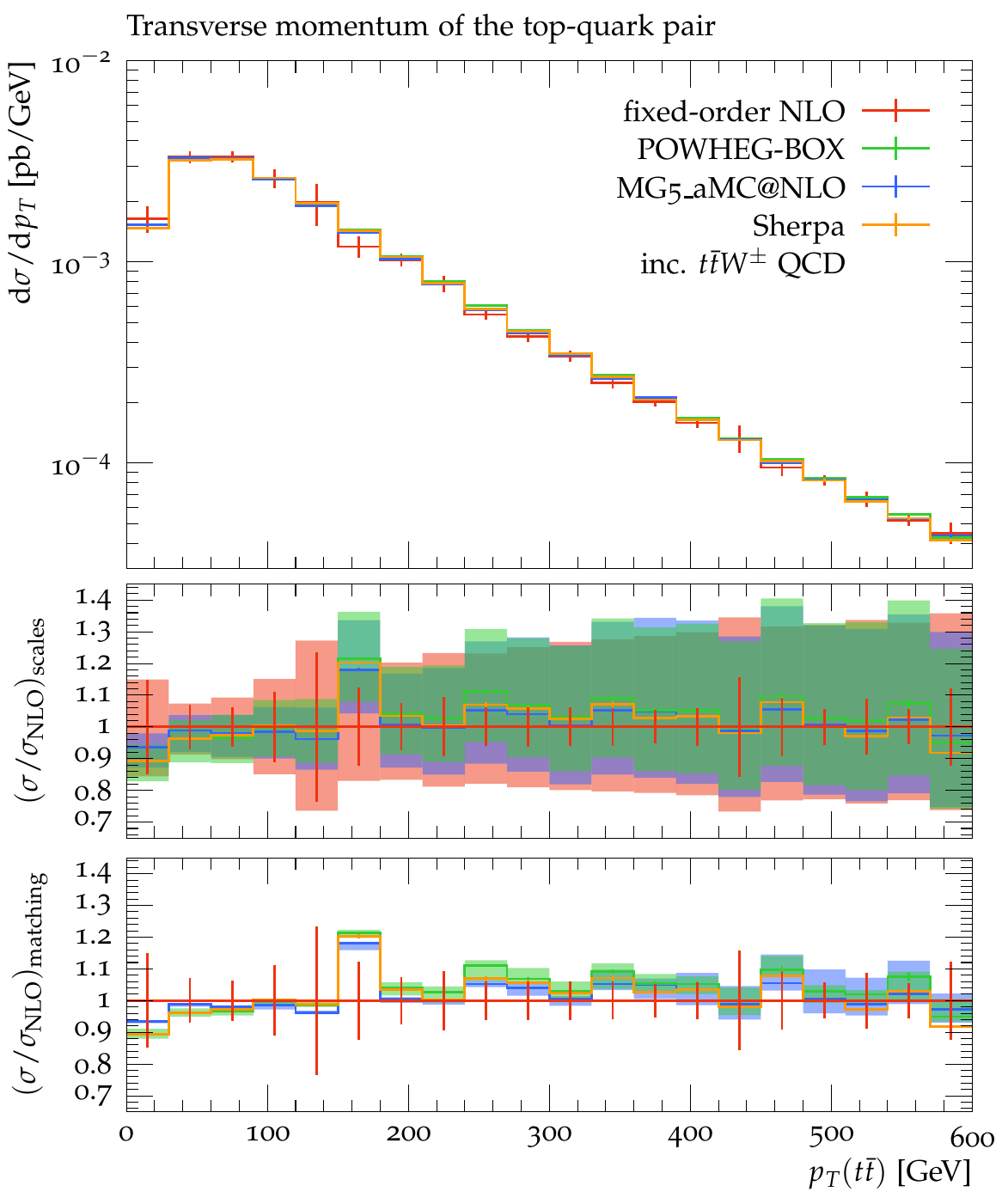}
 \includegraphics[width=0.49\textwidth]{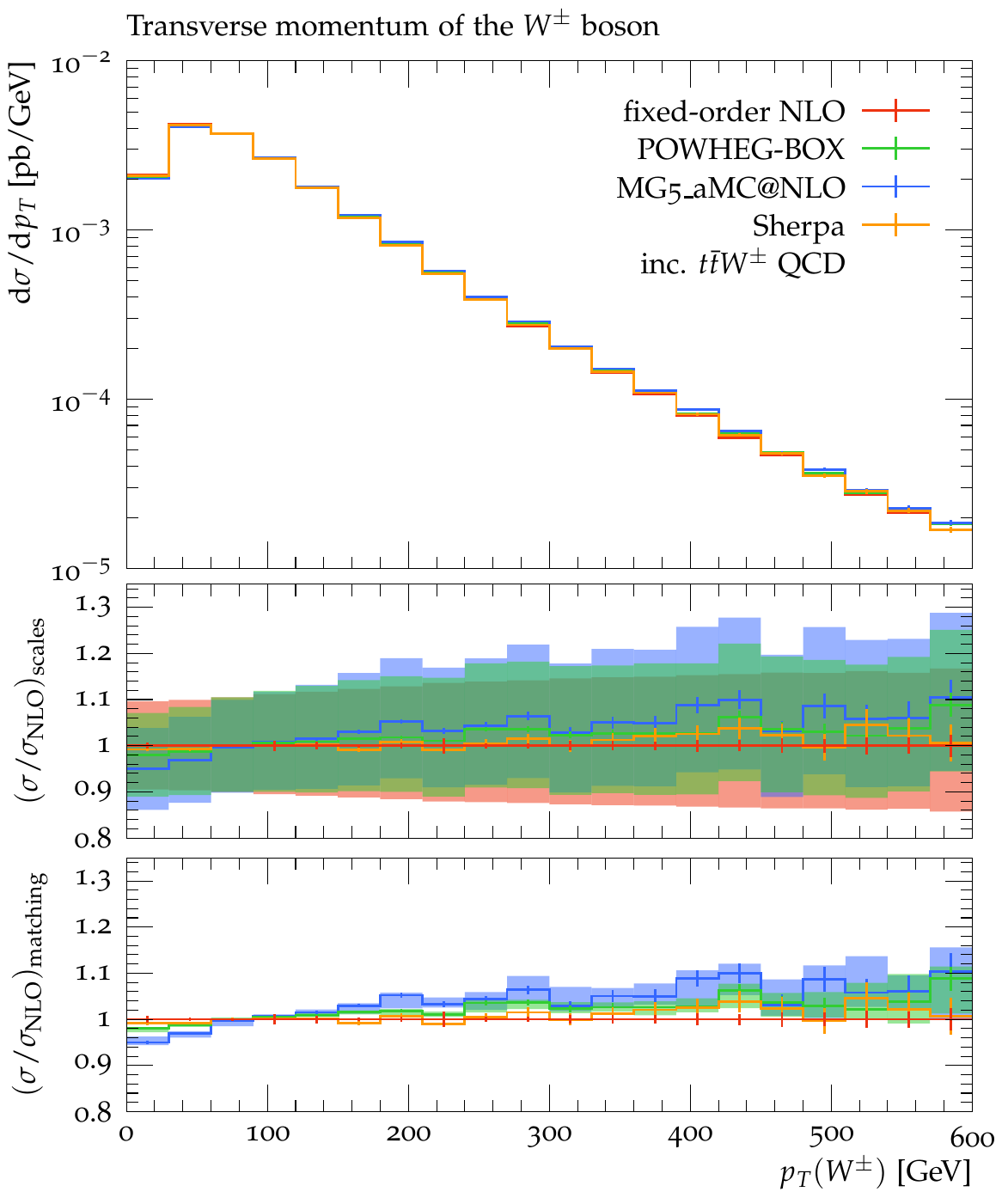}
 \caption{Inclusive $t\tb W^\pm$ QCD production as a function of the
   transverse momentum of the top-quark pair (l.h.s.) and of the $W^\pm$
   boson (r.h.s.) for various event generators. The uncertainty bands (not shown for \sherpa{})
   correspond to independent variations of renormalization and
   factorization scales (middle panel) and of the matching parameters
   (bottom panel).}
 \label{fig:inc_QCD_1}
\end{figure}
We start our discussion by focusing on the dominant $t\tb W^\pm$ QCD 
contribution. Fig.~\ref{fig:inc_QCD_1} shows the transverse momentum of the 
top-quark pair ($p_T(t\tb)$) as well as of the $W^\pm$ boson ($p_T(W^\pm)$) at 
fixed order and for various predictions including parton showers. The middle 
panel depicts the ratio with respect to the fixed-order NLO results and the 
uncertainty bands correspond to independent variations of renormalization and 
factorization scales in the hard matrix elements. The bottom panel shows the 
matching uncertainties as estimated by the variation of damping parameters and 
the initial shower scale $\muq$, as stated in section~\ref{sec:setup}, normalized
to the fixed-order NLO result.

In the case of the transverse momentum of the top-quark pair all three 
predictions agree well with the fixed-order NLO prediction over the whole plotted
range. The theoretical uncertainties due to scale variation are of the order of 
$10\%$ at the beginning and grow up to $30\%$ at the end of the spectrum. The 
matching uncertainties are smaller than the scale uncertainties in the whole 
plotted range and reach at most $7\%$ in the tail of the distribution.
Also for the transverse momentum of the $W^\pm$ boson, predictions agree well with
the fixed-order result. Only results obtained from \mgfive{} show a small shape 
difference with respect to the fixed-order NLO prediction. However, within the 
estimated theoretical uncertainties of $10-15\%$, predictions from all generators
considered agree well with each other. Again, the matching uncertainties are
small compared to the scale uncertainties, as it is expected for inclusive NLO 
observables, and are below $5\%$.

\begin{figure}[h!]
 \includegraphics[width=0.49\textwidth]{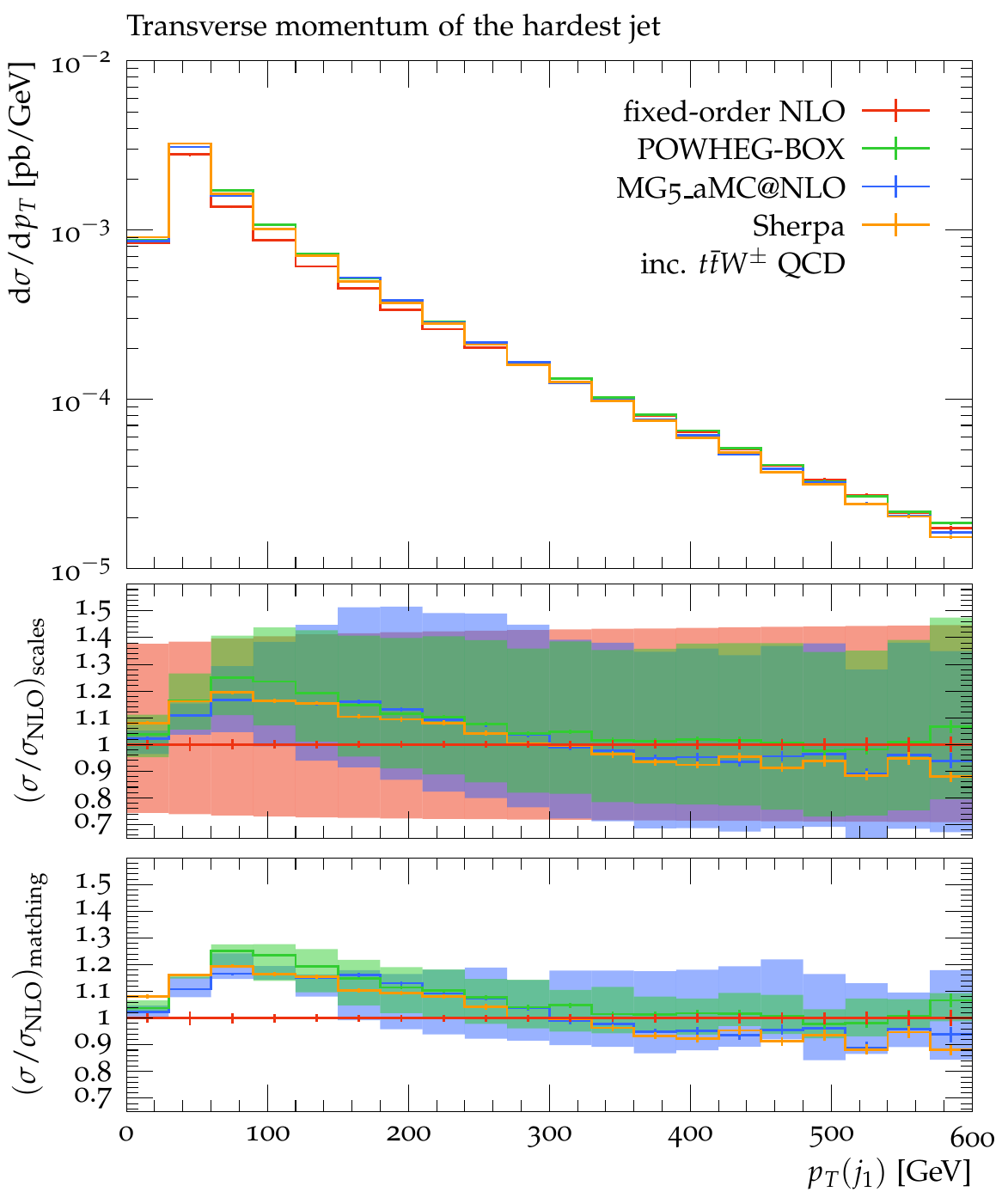}
 \includegraphics[width=0.49\textwidth]{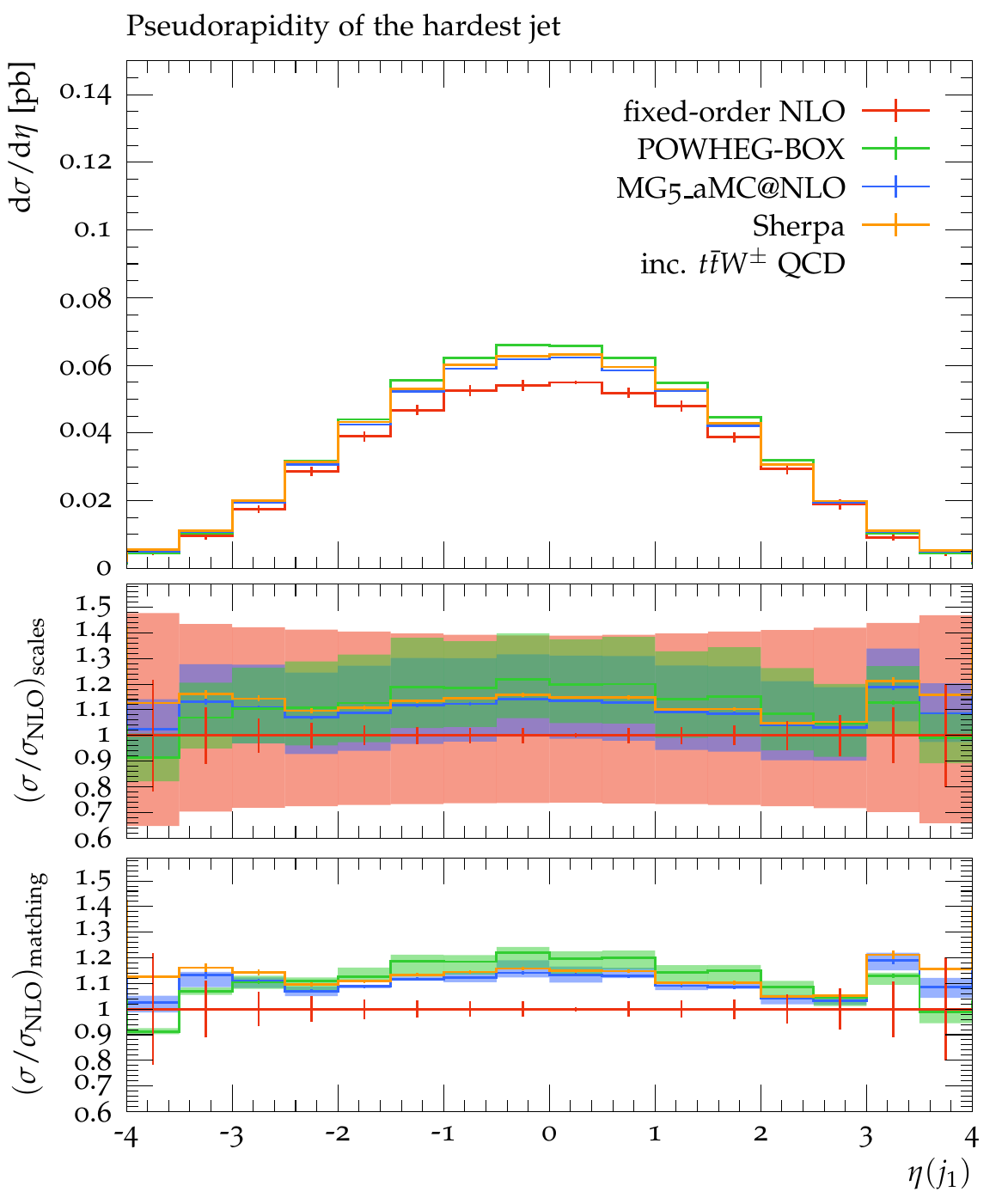}
 \caption{Inclusive $t\tb W^\pm$ QCD production as a function of the
   transverse momentum (l.h.s.) and of the pseudorapidity (r.h.s.) of the hardest
   jet for various event generators. The uncertainty bands (not shown for \sherpa{})
   correspond to independent variations of renormalization and
   factorization scales (middle panel) and of the matching parameters
   (bottom panel).}
 \label{fig:inc_QCD_2}
\end{figure}
Next we turn to less inclusive observables such as the transverse momentum 
($p_T(j_1)$) and the pseudorapidity ($\eta(j_1)$) of the hardest jet, as shown in
Fig.~\ref{fig:inc_QCD_2}. These observables are only accurate to LO since they 
do not receive one-loop corrections and can thus be more affected by the parton 
shower evolution. For example, in the case of the transverse momentum of the 
hardest jet, modifications to the spectrum are expected for small transverse 
momenta due to the Sudakov resummation in the parton shower, while the 
high-energy tail should be well described by using fixed-order matrix elements. 
Indeed all predictions including parton-shower effects differ from the 
fixed-order curve by up to $21-30\%$ for small transverse momenta, where 
\powhegbox{} predictions differ the most. On the other hand, in the high-energy 
tail of the distribution all event generators recover the fixed-order prediction 
starting from $p_T \gtrsim 300~\GeV$, with \powhegbox{} showing the best 
agreement whereas \mgfive{} and \sherpa{} give a slightly softer spectrum than 
the fixed-order curve by roughly $10\%$ at the end of the plotted spectrum.  
In addition to the shape differences we also notice a severe reduction of the
theoretical uncertainties in the beginning of the spectrum where the 
parton-shower is expected to dominate. While the fixed order has nearly constant 
uncertainties of the order of $30-35\%$ over the whole range, the parton-shower 
based predictions show only a variation of below $10\%$ at the beginning of the 
spectrum while the uncertainty grows up to $35\%$ at high $p_T$ once the real 
matrix elements dominate the spectrum again. Over the whole plotted range the 
scale uncertainties dominate over the matching related ones. However, the latter 
are larger for \mgfive{} as compared to the \powhegbox{} and amount to roughly 
$15\%$.

The plot on the right-hand side of Fig.~\ref{fig:inc_QCD_2} illustrates the
pseudorapidity of the hardest jet. Here, we observe that including parton-shower 
effects generates overall positive corrections of the order of $13-20\%$. These
corrections can be simply attributed to the fact that after the parton shower 
evolution the number of events with at least one jet is higher than in the 
corresponding fixed-order NLO computation. To be precise, \mgfive{} predicts 
$11\%$, \sherpa{} $13\%$, and the \powhegbox{} $15\%$ more events with at least 
one hard jet. The small differences between \mgfive{} and \sherpa{} can be 
attributed to different shower dynamics, while the difference between the 
\powhegbox{} and \mgfive{} are related to the parton shower matching scheme 
(as shower differences are minor between these two). Also notable is the 
reduction of the theoretical uncertainties by a factor of two, from $\pm 30\%$ at
fixed-order to $\pm 15\%$ in the central rapidity bins once the predictions are 
matched to a parton shower. As we can see from the bottom panel the matching 
uncertainties are negligible over the whole spectrum.

\subsubsection*{$t\tb W^\pm$ EW contribution}
\begin{figure}[h!]
 \includegraphics[width=0.49\textwidth]{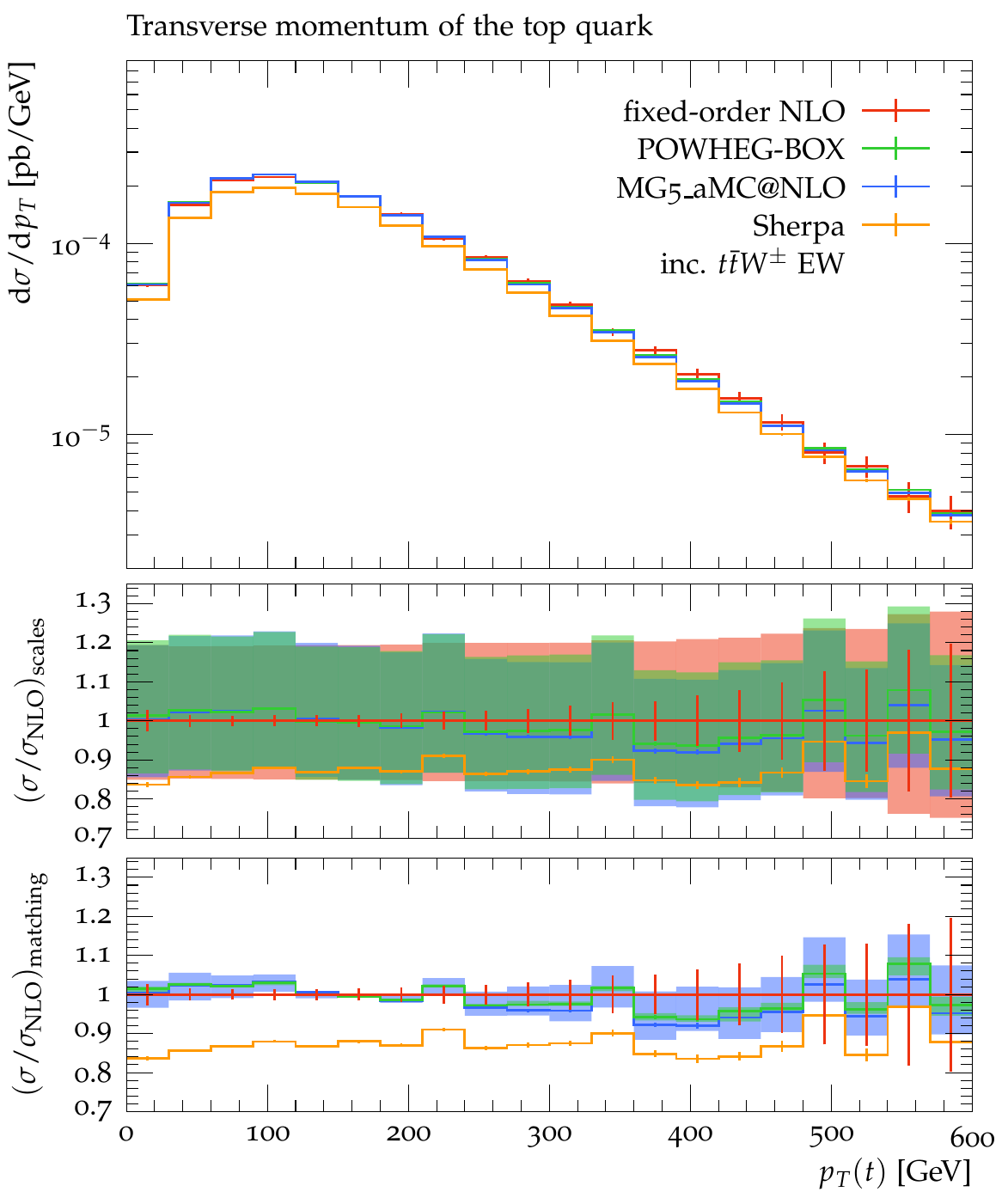}
 \includegraphics[width=0.49\textwidth]{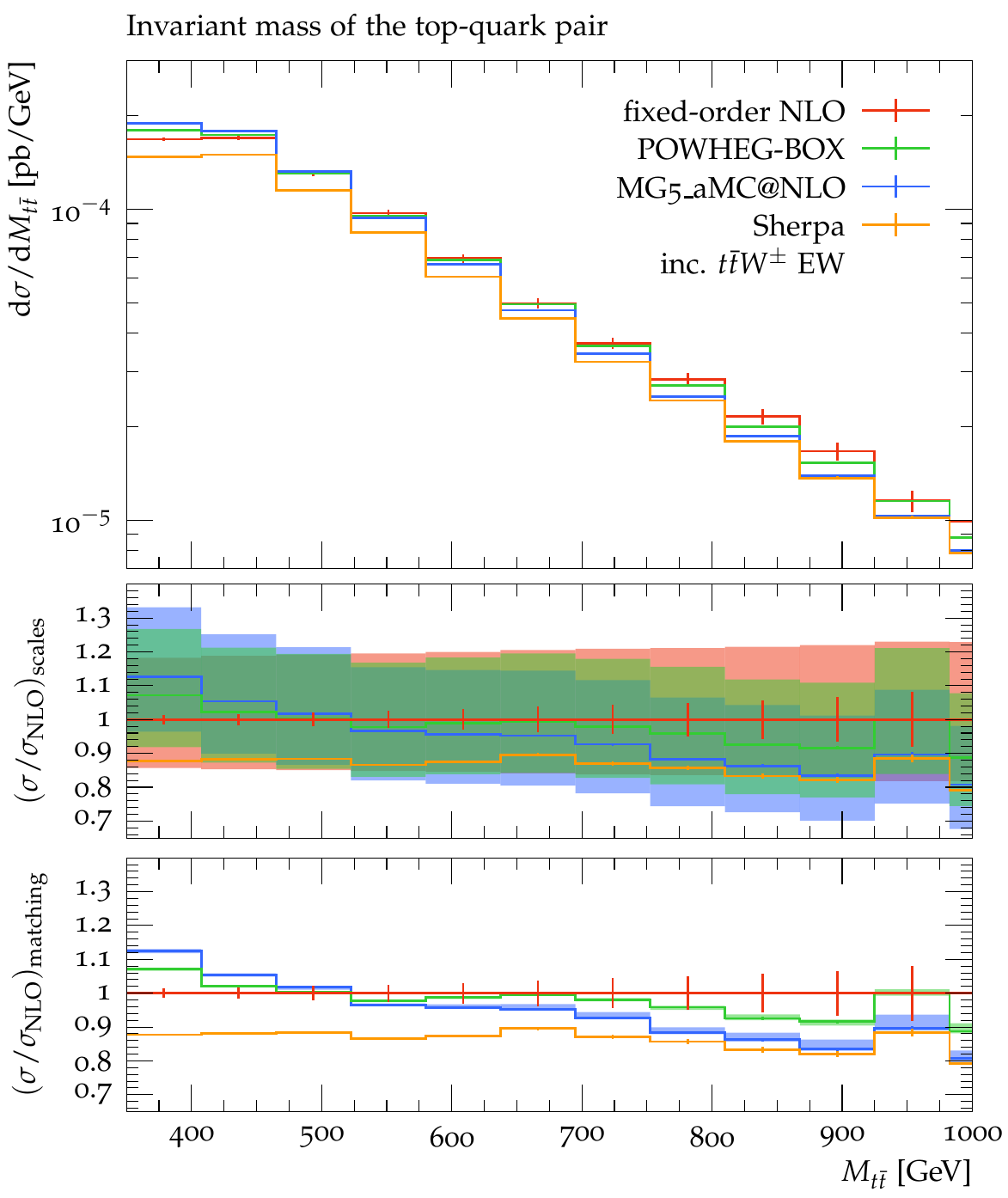}
 \caption{Inclusive $t\tb W^\pm$ EW production as a function of the
   transverse momentum of the top quark (l.h.s.) and the invariant mass
   of the top-quark pair (r.h.s.) for various event generators.
   The uncertainty bands (not shown for \sherpa{}) correspond to independent variations of renormalization 
   and factorization scales (middle panel) and of the matching parameters
   (bottom panel).}
 \label{fig:inc_EW_1}
\end{figure}

Moving on to $t\tb W^\pm$ EW contributions to the inclusive $t\tb W^\pm$ 
signature, we illustrate in the left-hand side plot of Fig.~\ref{fig:inc_EW_1} 
the transverse momentum of the top quark ($p_T(t)$). As this observable is 
already accurate to NLO we find the expected good agreement between fixed-order 
prediction and the \powhegbox{} as well as \mgfive{}. On the other hand, the
\sherpa{} prediction captures the shape of the distribution well but 
underestimates the normalization by $13\%$ as can be seen from the corresponding 
bottom panel. The reason for the discrepancy is that the \sherpa{} result is 
based on merging of tree-level matrix elements for $t\tb W^\pm$ and 
$t\tb W^\pm j$ and therefore misses higher-order loop corrections, as well as 
parts of the dominant real radiative corrections below the merging scale.  The 
estimated theoretical uncertainties from scale variations is around 
$\pm 20-25\%$ for fixed-order, \powhegbox{}, and \mgfive{} result. As the 
distribution is NLO accurate we only observe a mild dependence on the matching 
scheme of at most $10\%$. 
A slightly different situation is encountered when considering the invariant mass 
of the top-quark pair ($M_{t\tb}$) shown on the right-hand side of 
Fig.~\ref{fig:inc_EW_1}. In this case, only the \powhegbox{} agrees with the 
fixed-order result within the statistical uncertainties nearly over the whole 
plotted range. Results obtained with \mgfive{} show shape differences with 
respect to the fixed-order result. At the beginning of the spectrum \mgfive{} is 
$13\%$ higher, while it is lower in the high-energy tail of the fixed-order 
distribution at $M_{t\tb} \approx 1~\TeV$ by nearly $20\%$. As in the previous 
case \sherpa{} underestimates the normalization by about $-12\%$. Nevertheless,
within the scale uncertainties of each prediction, which amount to $16\%$ at the 
beginning of the spectrum and $21\%$ at the end of the plotted range, all 
distributions agree well with each other. Moreover, we find that the observable 
is very stable with respect to  matching related parameters, as these 
uncertainties are almost negligible.

\begin{figure}[h!]
 \includegraphics[width=0.49\textwidth]{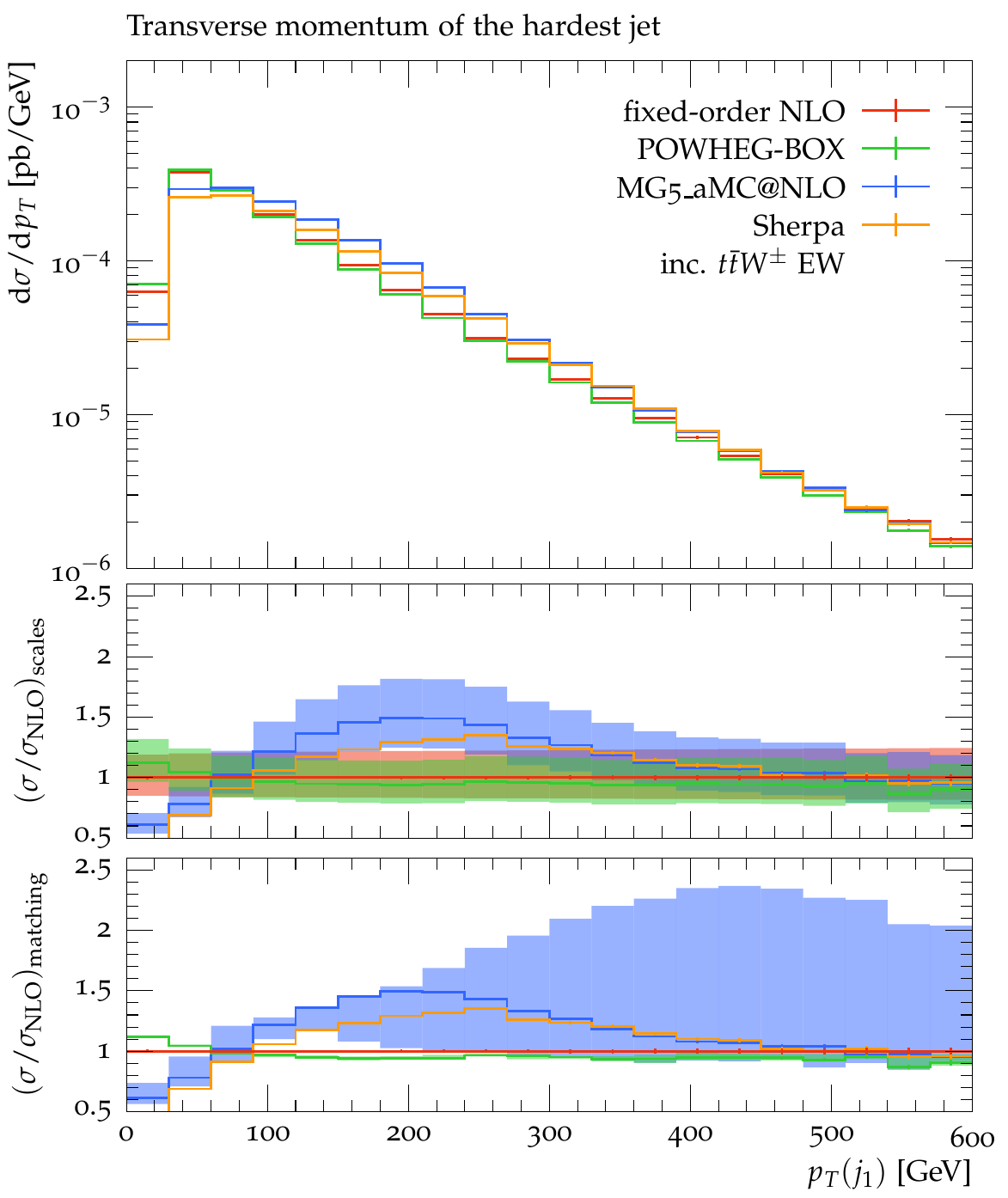}
 \includegraphics[width=0.49\textwidth]{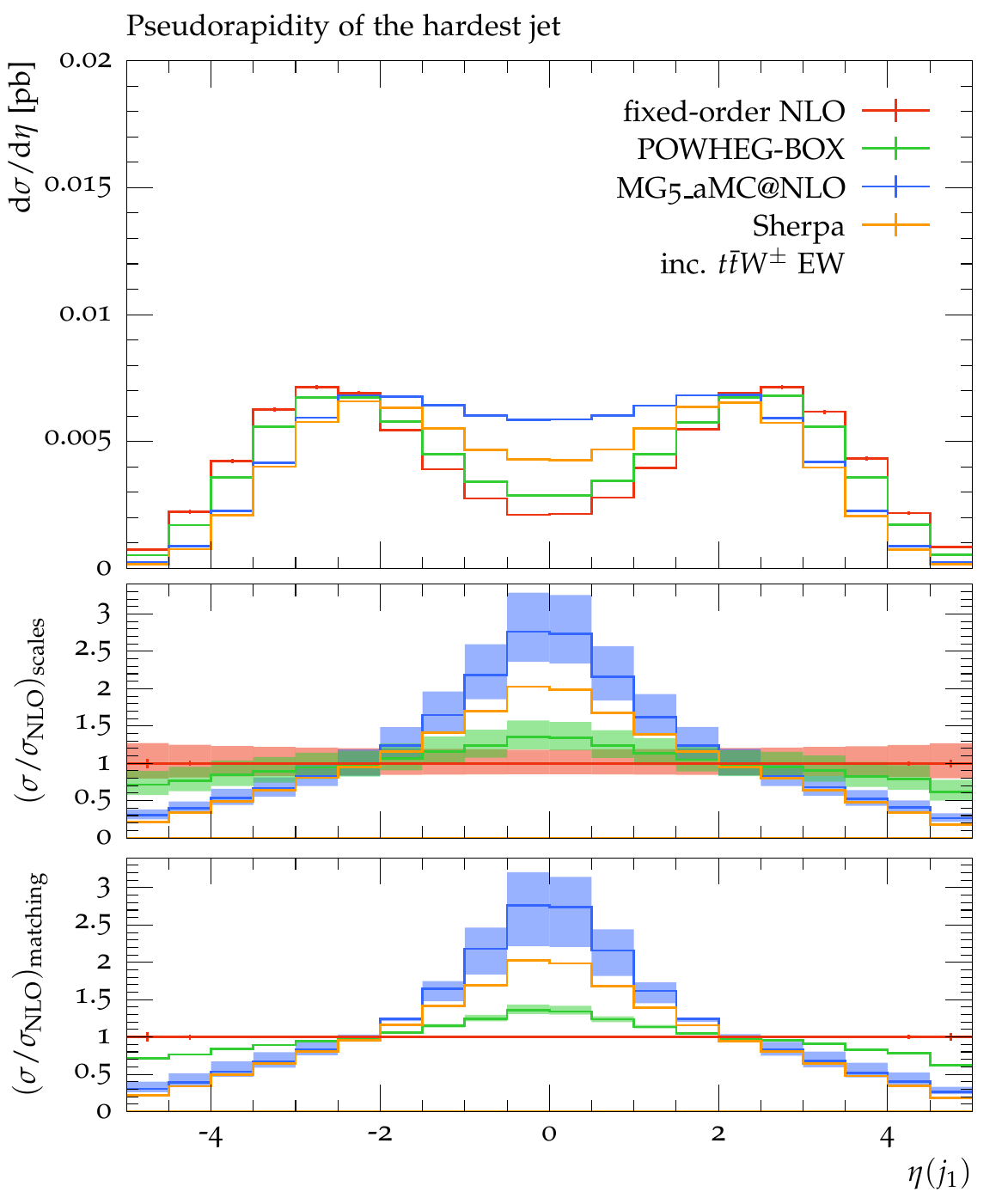}
 \caption{Inclusive $t\tb W^\pm$ EW production as a function of the
   transverse momentum (l.h.s.) and of the pseudorapidity (r.h.s.) of the leading
   jet for various event generators. The uncertainty bands (not shown for \sherpa{})
   correspond to independent variations of renormalization and
   factorization scales (middle panel) and of the matching parameters
   (bottom panel).}
 \label{fig:inc_EW_2}
\end{figure}

In Fig.~\ref{fig:inc_EW_2} we further consider the transverse momentum (l.h.s.) 
and the pseudorapidity (r.h.s.) of the hardest jet. Contrary to the $t\tb W^\pm$
QCD predictions we observe large differences between the various predictions.
For the transverse momentum distribution of the leading jet we find that the 
\powhegbox{} predicts a slightly softer spectrum than the fixed-order result with 
$+12\%$ corrections in the first bin, while for the remaining spectrum the curve 
is nearly a constant $-5\%$ below the fixed-order one. It is the parton shower 
evolution that softens the spectrum slightly and thus these corrections are of 
higher order. \mgfive{} generates a much harder intermediate spectrum with 
corrections up to $+50\%$ around $p_T\approx 200~\GeV$, and falls back to the 
fixed-order result towards the end of the plotted range. Last, the \sherpa{} 
prediction shows similar features to \mgfive{} but less pronounced. The 
theoretical uncertainties are of the order of $\pm 20\%$ for the fixed-order as 
well as the parton shower matched predictions. However, because of the severe 
shape differences the uncertainty band of \mgfive{} barely overlaps with the 
fixed-order or the \powhegbox{} result for intermediate values of the transverse 
momentum. Furthermore, we see that the \powhegbox{} results are very stable with 
respect to the damping parameters as variations at the level of only $5\%$ are 
visible. Contrary, the \mgfive{} result depends strongly on the initial shower 
scale which dominates the theoretical uncertainties from $p_T \gtrsim 150~\GeV$.

Finally, also in the case of the hardest-jet pseudorapidity distribution we 
observe quite different predictions. While the \powhegbox{} prediction is still 
the closest to the fixed-order NLO result we note shape differences, with 
deviations of nearly $+30\%$ in the central region as well as $-30\%$ in the 
forward rapidity regions. The \mgfive{} prediction generates even larger 
corrections, which reach up to a factor of $2.8$ in the central region and are 
suppressed by nearly $-70\%$ in the forward regions, while the \sherpa{} 
prediction lies in between \mgfive{} and the \powhegbox{}. Indeed, \sherpa{} 
agrees with the \mgfive{} prediction in the very forward region, while it is 
closer to the \powhegbox{} for the central rapidity bins. The significant shape 
differences also lie mostly outside of the estimated uncertainty bands. Let us 
remind the reader that, as illustrated in Fig.~\ref{fig:damp}, the \powhegbox{}
reproduces the fixed-order distribution well when no shower evolution is taken 
into account. Therefore, the significant corrections in the central rapidity 
region can be attributed to formally higher-order corrections generated by the 
parton shower. This can be also deduced from the dependence of the \mgfive{} 
results on the initial shower scale. We observe a large shape differences of  
around $20\%$ in the central rapidity bins by varying the shower scale, which is 
comparable to the scale uncertainties. 
\subsubsection*{Combined $t\tb W^\pm$ QCD+EW contribution}
As we have seen the $t\tb W^\pm$ QCD and EW contributions differ quite 
substantially in size, in their associated uncertainties, and in their 
sensitivity to the parton-shower matching schemes, where the strongest modelling 
dependence has been found in the EW production channel. Thus the question arises 
of how much of these effects are visible in the total predictions once EW and QCD
contributions are combined.
\begin{table}[h!]
\begin{tabular}{ccccc}
\hline
$\sigma$~[fb] & QCD & EW & QCD+EW & $\frac{\text{QCD+EW}}{\text{QCD}}$ \\
\hline
$\sigma^\nlo$ & $554.8(3)^{+61.5~(11\%)}_{-57.4~(10\%)}$ & 
$49.0(2)^{+9.5~(19\%)}_{-7.3~(15\%)}$ & $603.8(3)^{+71.0~(12\%)}_{-65.2~(11\%)}$ 
& $1.09$\\[0.2cm]
$\sigma^\nlops_\mathrm{PWG}$ & $555.6(4)^{+61.5~(11\%)}_{-57.8~(10\%)}$ & 
$49.2(1)^{+9.5~(19\%)}_{-7.4~(15\%)}$ & $604.8(4)^{+71.0~(12\%)}_{-65.2~(11\%)}$ & 
$1.09$\\[0.2cm]
$\sigma^\nlops_\mathrm{MG5}$ & $554.0(6)^{+60.8~(11\%)}_{-57.5~(10\%)}$ & 
$49.1(1)^{+9.5~(19\%)}_{-7.4~(15\%)}$ & $603.1(6)^{+70.4~(12\%)}_{-64.8~(11\%)}$ & 
$1.09$\\[0.2cm]
$\sigma^\nlops_\mathrm{Sherpa}$ & $553.7(8)$ & $42.7(1)$ & $596.4(8)$ & $1.08$\\
\hline
\end{tabular}
\caption{Inclusive $t\tb W^\pm$ cross section contributions at NLO accuracy for 
fixed-order and parton shower matched results including the theoretical 
uncertainty estimated from independent scale variations. In parenthesis we 
include the statistical error.}
\label{tab:xsecs}
\end{table}
To highlight the overall size of the EW contribution we list in  
table~\ref{tab:xsecs} the total cross sections for the QCD and EW contributions
separately as well as combined, both for the fixed-order NLO computation and for 
the three generators considered in our study~\footnote{We remind the reader that 
in our setup \sherpa{} employs tree-level merging of matrix elements for the EW 
contribution and thus does not have to agree with the other generators.}. We 
observe that the inclusion of the EW contribution is a $8-9\%$ correction on top 
of the dominant QCD contribution, while the theoretical uncertainties estimated 
via scale variations of the QCD contribution amounts to already $\pm 11\%$. 
Therefore, in order to have a visible effect of the EW contribution on 
differential distributions one has to focus on phase space regions where the EW 
production mode is enhanced with respect to the QCD one. In the following we show
two representative observables that illustrate the impact of the EW contribution 
on combined differential distributions using our \powhegbox{} implementation.
\begin{figure}[h!]
 \includegraphics[valign=t,width=0.49\textwidth]{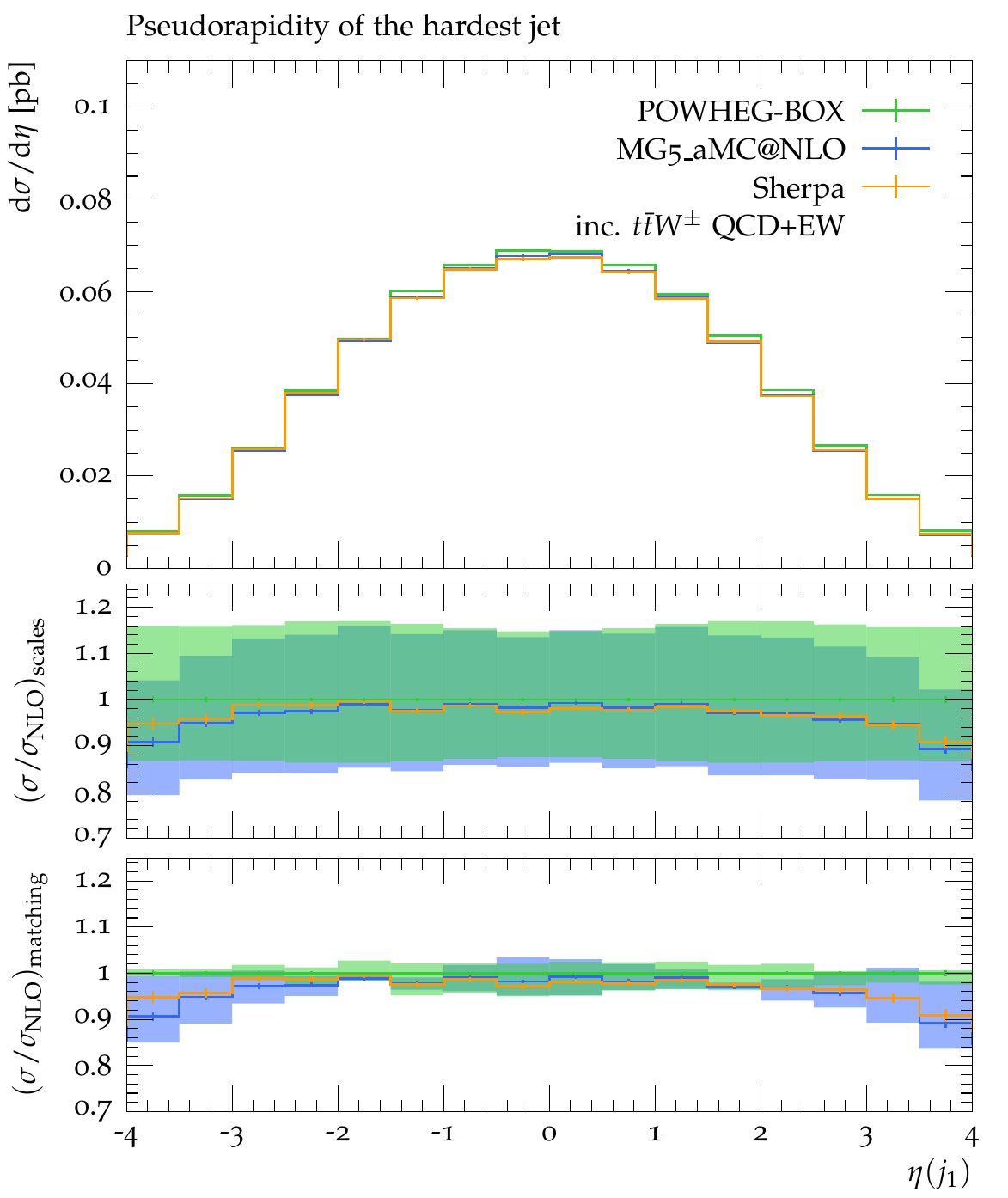}
 \includegraphics[valign=t,width=0.49\textwidth]{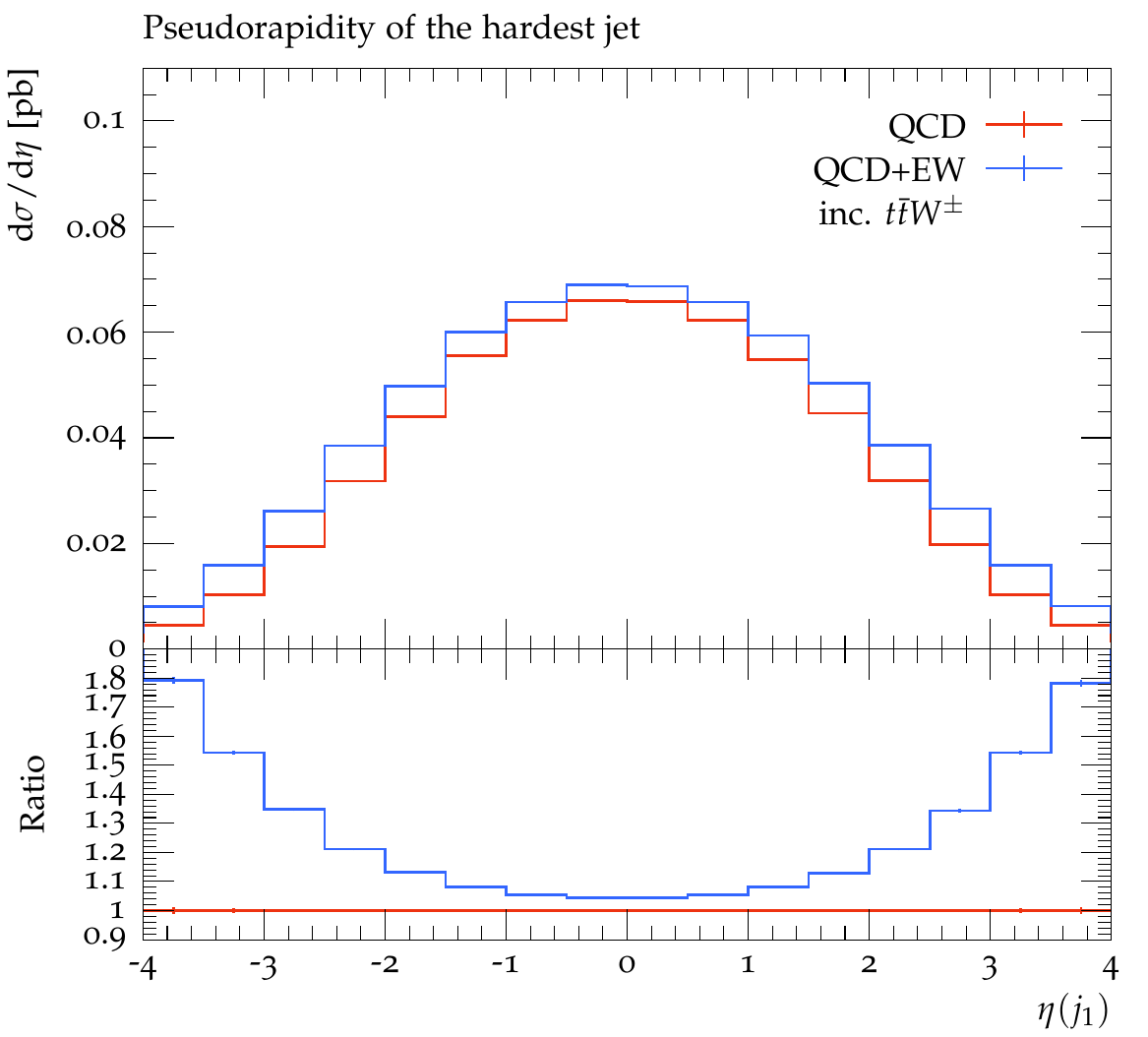}
 \caption{Inclusive $t\tb W^\pm$ EW$+$QCD production as a function of the
   pseudorapidity of the leading jet shown for various event generators (l.h.s.)
   and split into QCD and QCD+EW contribution (r.h.s.). The uncertainty bands
   in the l.h.s. plot (not shown for \sherpa{}) correspond to independent variations of renormalization and
   factorization scales (middle panel) and of the matching parameters
   (bottom panel).}
 \label{fig:inc_QCDEW_1}
\end{figure}
In Fig.~\ref{fig:inc_QCDEW_1} we show again the pseudorapidity of the hardest 
jet, for the three considered Monte Carlo generators on the left-hand side while 
on the right-hand side the same distribution is shown as predicted by the 
$t\tb W^\pm$ QCD contribution only and the combined $t\tb W^\pm$ QCD+EW 
contribution. The overall agreement between the different predictions is good 
with shape differences of only $10\%$ in the very forward region. As can be 
deduced from the right plot of Fig.~\ref{fig:inc_QCDEW_1} the EW contribution 
becomes sizable in the forward region by modifying the shape of the distribution 
by nearly $80\%$ and thus the observed shape differences between the various 
predictions are related to the modelling discrepancies in the EW contribution. 
Even though the theoretical uncertainties are dominated by missing higher-order 
corrections and amount to roughly $\pm 15\%$ the inclusion of the EW contribution
represents a systematic shift of at least $+5\%$. 
\begin{figure}[h!]
 \includegraphics[valign=t,width=0.49\textwidth]{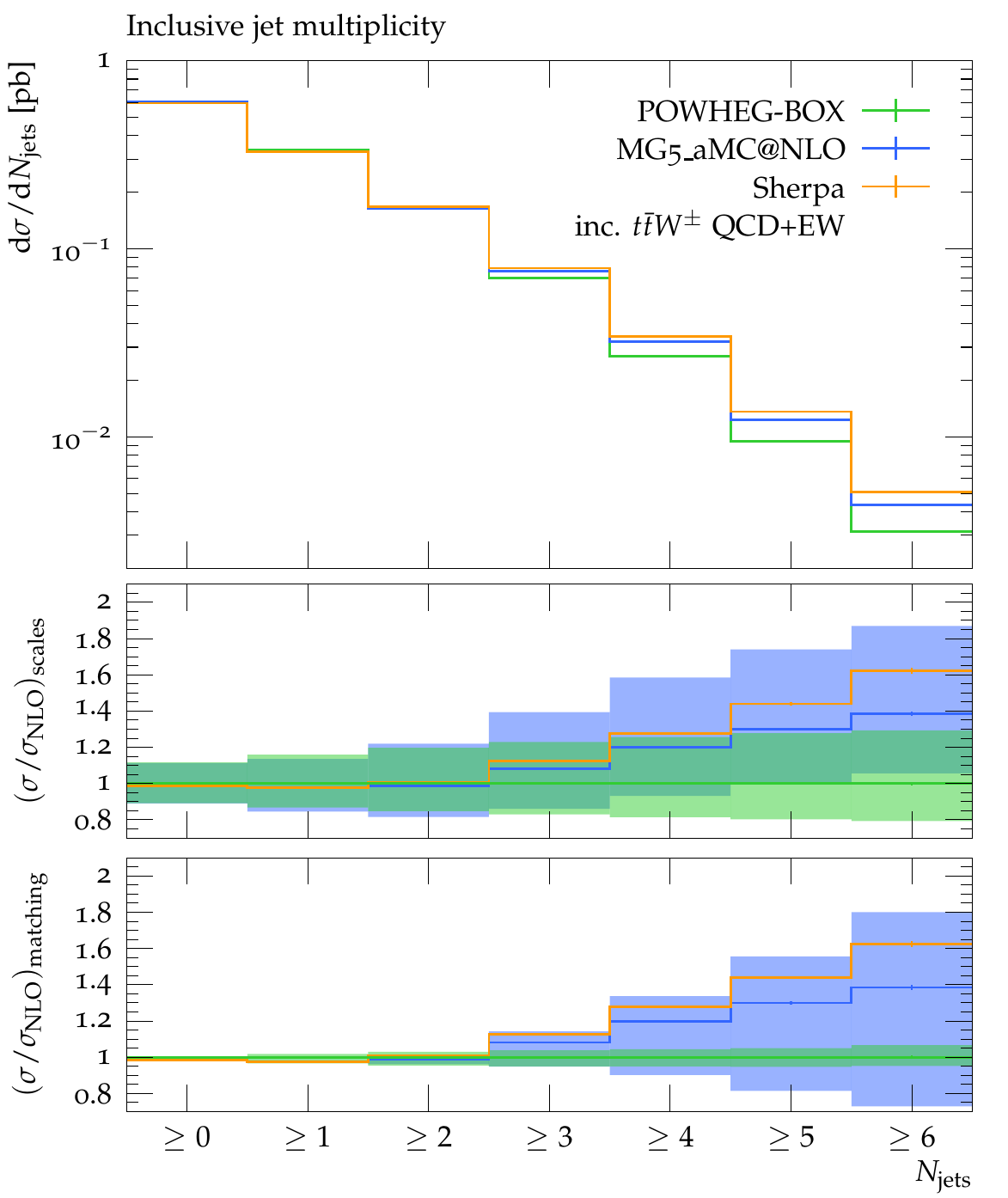}
 \includegraphics[valign=t,width=0.49\textwidth]{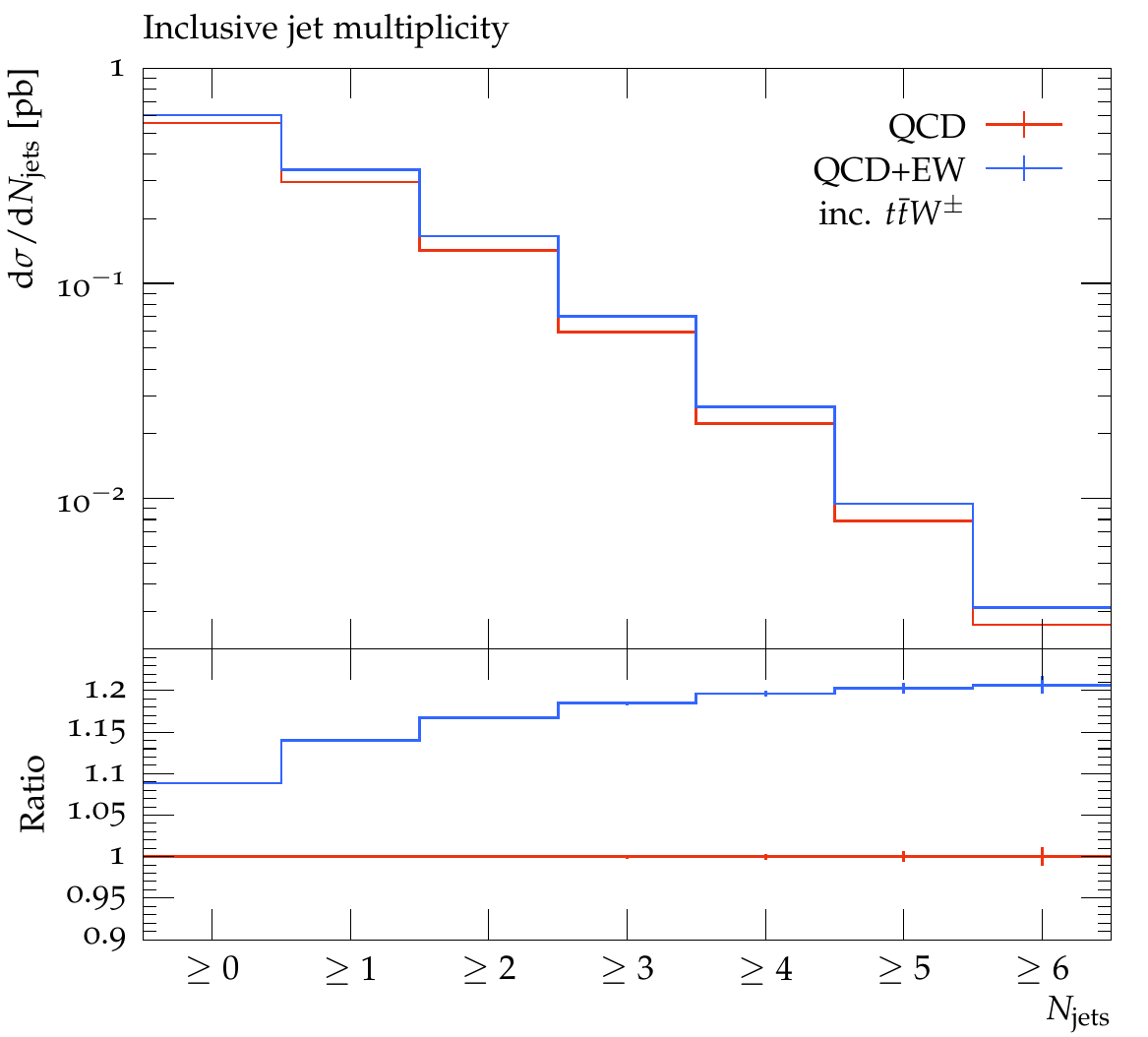}
 \caption{Inclusive $t\tb W^\pm$ EW$+$QCD cross section as a function of the
   the number of jets shown for various event generators (l.h.s.)
   and split into QCD and QCD+EW contribution (r.h.s.). The uncertainty bands
   in the l.h.s. plot (not shown for \sherpa{}) correspond to independent variations of renormalization and
   factorization scales (middle panel) and of the matching parameters
   (bottom panel).}
 \label{fig:inc_QCDEW_2}
\end{figure}
As a second example we show in Fig.~\ref{fig:inc_QCDEW_2} the inclusive cross 
section as a function of the number of jets ($N_{\mathrm{jets}}$), both light and
$b$ jets. For the first three bins we find excellent agreement between the three 
generators. Afterwards the predictions diverge and the \powhegbox{} predicts the 
smallest while \sherpa{} the largest cross section for high jet multiplicities 
with differences as large as $60\%$ for the jet bin with $6$ or more jets.
However, while the first three bins are dominated by uncertainties stemming from 
missing higher-order corrections the remaining bins are mostly affected by 
parton-shower effects whose uncertainty can be large. Indeed, while in the case 
of \powhegbox{} the matching uncertainty does not exceed 5\%,  in the \mgfive{} 
case the spectrum is dominated by the dependence on the initial shower scale 
$\muq$ that can easily account for the aforementioned shape differences. Finally, 
we want to notice that even if the impact of the inclusive $t\tb W^\pm$ EW 
contribution starts at $+9\%$ and increases with the number of jets to about a 
$+20\%$ correction on top of the QCD prediction for six and more jets, in this 
region the uncertainty of the QCD contribution overshadows this correction.

To summarize our findings of this section, we can say that $t\tb W^\pm$ QCD
inclusive production is rather robust with respect to matching uncertainties and 
different parton-shower algorithms. On the other hand, the $t\tb W^\pm$ EW 
contribution is very sensitive to different matching procedures as sizable 
higher-order corrections can be generated by the parton shower even for inclusive
observables. This however, might not be surprising all together in this 
particular case, since the NLO corrections in $t\tb W^\pm$ EW production are 
highly dominated by real radiation matrix elements and therefore the description 
of this process is essentially only at LO accuracy. This can also be seen from 
the fact that the \sherpa{} prediction, which in the EW case is based only on 
tree-level matrix elements, still recovers the main features of many observables.
In addition, the EW production channel is mediated by t-channel exchanges of 
color singlets and could be thus also sensitive to radiation patterns of parton 
shower implementations~\cite{Rainwater:1996ud}. At last, we have to note that 
even though we observe large differences in the modelling of the $t\tb W^\pm$ EW 
contribution these differences are much less visible once the QCD and EW 
contributions are combined.

\subsection{Two same-sign leptons signature}
\label{sec:2lSS}
In this section we focus on the experimental signature of two same-sign leptons 
in association with additional jets, usually denoted as ``\SSL{}''. The final 
state is selected by requiring exactly two same-sign leptons\footnote{We exclude 
$\tau$ leptons here, since these typically form different signatures at hadron 
collider detectors. Thus we focus on: $e^\pm e^\pm$, $e^\pm\mu^\pm$, 
$\mu^\pm\mu^\pm$.}  with $p_T(\ell) > 15$ GeV and $|\eta(\ell)| < 2.5$. Jets are 
formed using the anti-$k_T$ jet algorithm with a separation parameter of $R=0.4$.
We further require that jets (light as well as $b$ jets) fulfill $p_T(j) > 25$ 
GeV and $|\eta(j)| < 2.5$. Finally, we require to have at least two light jets as
well as two tagged $b$ jets. 

Results shown in this section correspond to the sum of $t\tb W^\pm$ QCD and 
$t\tb W^\pm$ EW production modes which contribute to the \SSL{} signature defined
above. Emphasis will be placed on identifying the major sources of theoretical 
uncertainty from scale variation and parton-shower matching following the 
procedure discussed in section~\ref{sec:setup}.

For the range of parameters described in section~\ref{sec:setup} and the 
selection cuts above, we obtain fully consistent fiducial cross sections in all 
the frameworks considered in our study, namely:
\begin{equation}
\begin{split}
 \sigma^\nlops_\mathrm{PWG} &= 6.79(1)~^{+0.84~(12\%)}_{-0.75~(11\%)}~
 \text{[scales]}~^{+0.05~(1\%)}_{-0.09~(1\%)}~\text{[matching]}~\text{fb}\;, \\
 \sigma^\nlops_\mathrm{MG5} &= 6.80(1)~^{+0.86~(13\%)}_{-0.76~(12\%)}~
 \text{[scales]}~^{+0.07~(1\%)}_{-0.05~(1\%)}~\text{[matching]}~\text{fb}\;, \\
 \sigma^\nlops_\mathrm{Sherpa} &= 6.80(1)~\text{fb}\;,
 \label{eqn:xsec_fid}
\end{split}
\end{equation}
where the residual uncertainty as estimated from the variation of renormalization
and factorization scales (here and in the plots labeled as ``\textit{scales}'') 
are of the order of $12\%$, while the impact of varying matching related 
parameters, i.e. the damping factors in the case of the \powhegbox{} and the 
initial shower scale in the case of \mgfive{} (here and in the plots labeled as 
``\textit{matching}''), is a $1\%$ effect in both cases.  The central result of 
\sherpa{} agrees perfectly with the other generators. 
Notice that the scale uncertainties of the fiducial cross sections in 
Eq.~\eqref{eqn:xsec_fid} amount to $\pm 12\%$ and are consistent with the 
corresponding ones of the total inclusive cross sections shown in 
Tab.~\ref{tab:xsecs}. This emphasizes the fact that even though the fiducial 
cross section is tremendously reduced by branching ratios and phase space cuts on
the decay products the signature is still inclusive with respect to the 
production of the heavy $t\tb W^\pm$ final state.

\subsubsection*{Hadronic observables}
\begin{figure}[h!]
 \includegraphics[width=0.49\textwidth]{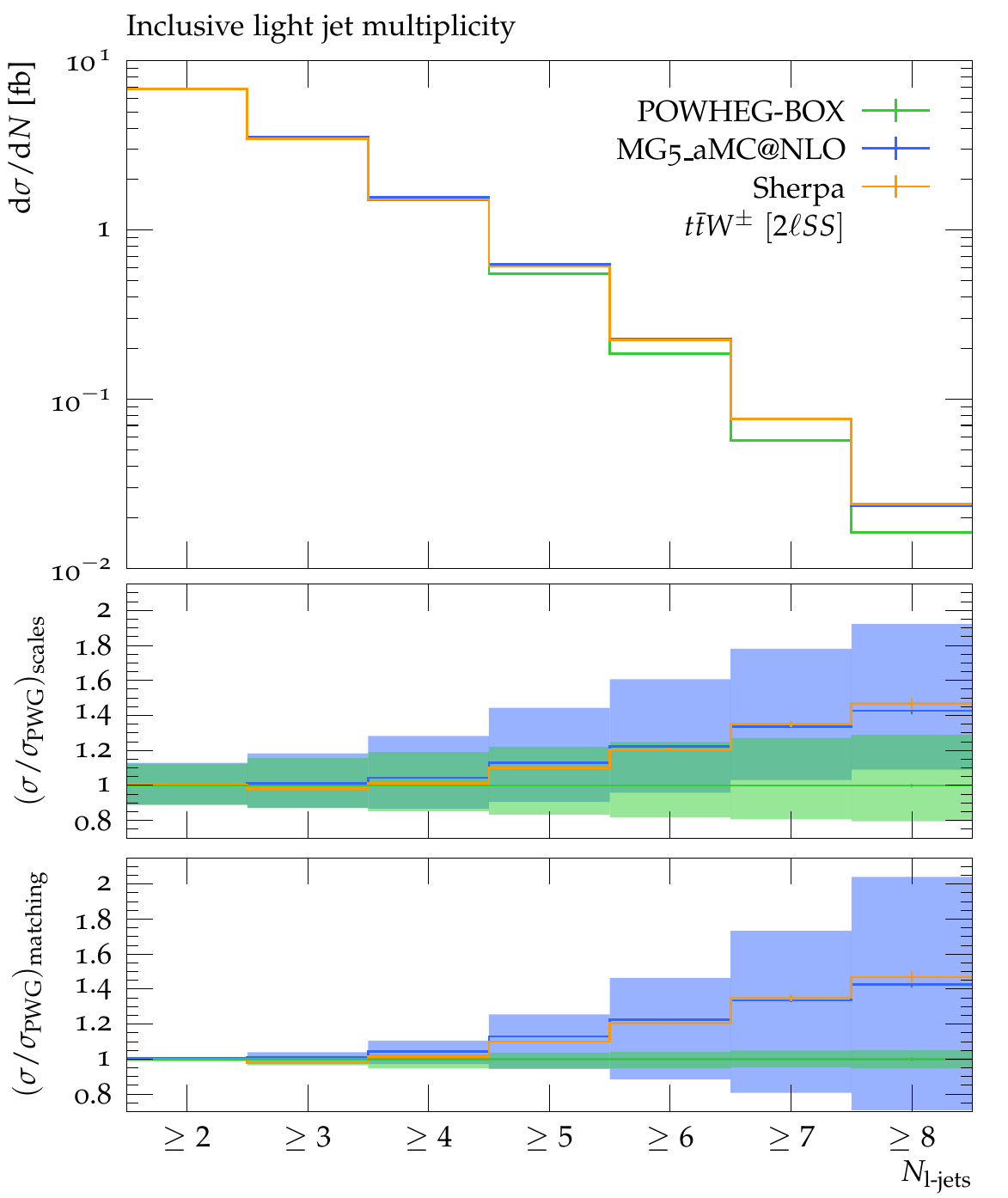}
 \includegraphics[width=0.49\textwidth]{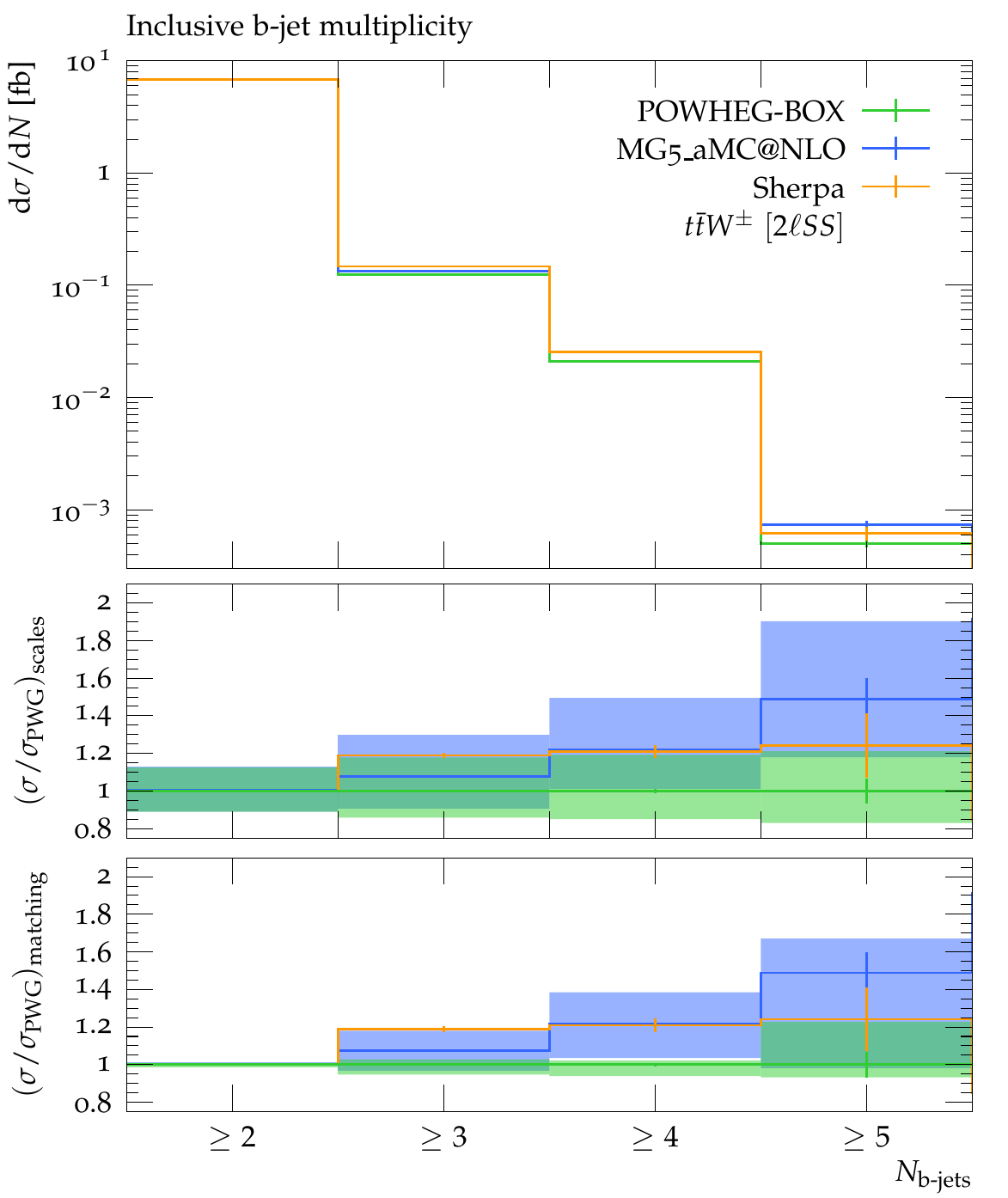}
 \caption{Inclusive cross section in the \SSL{} fiducial region as a 
 function of the number of light jets (l.h.s.) and the number of $b$ jets 
 (r.h.s.) for various event generators. 
 The uncertainty bands (not shown for \sherpa{}) correspond to independent 
 variations of renormalization and factorization scales (middle panel) and of 
 the matching parameters (bottom panel).}
 \label{fig:2ssl_1}
\end{figure}

\begin{figure}[h!]
 \includegraphics[width=0.49\textwidth]{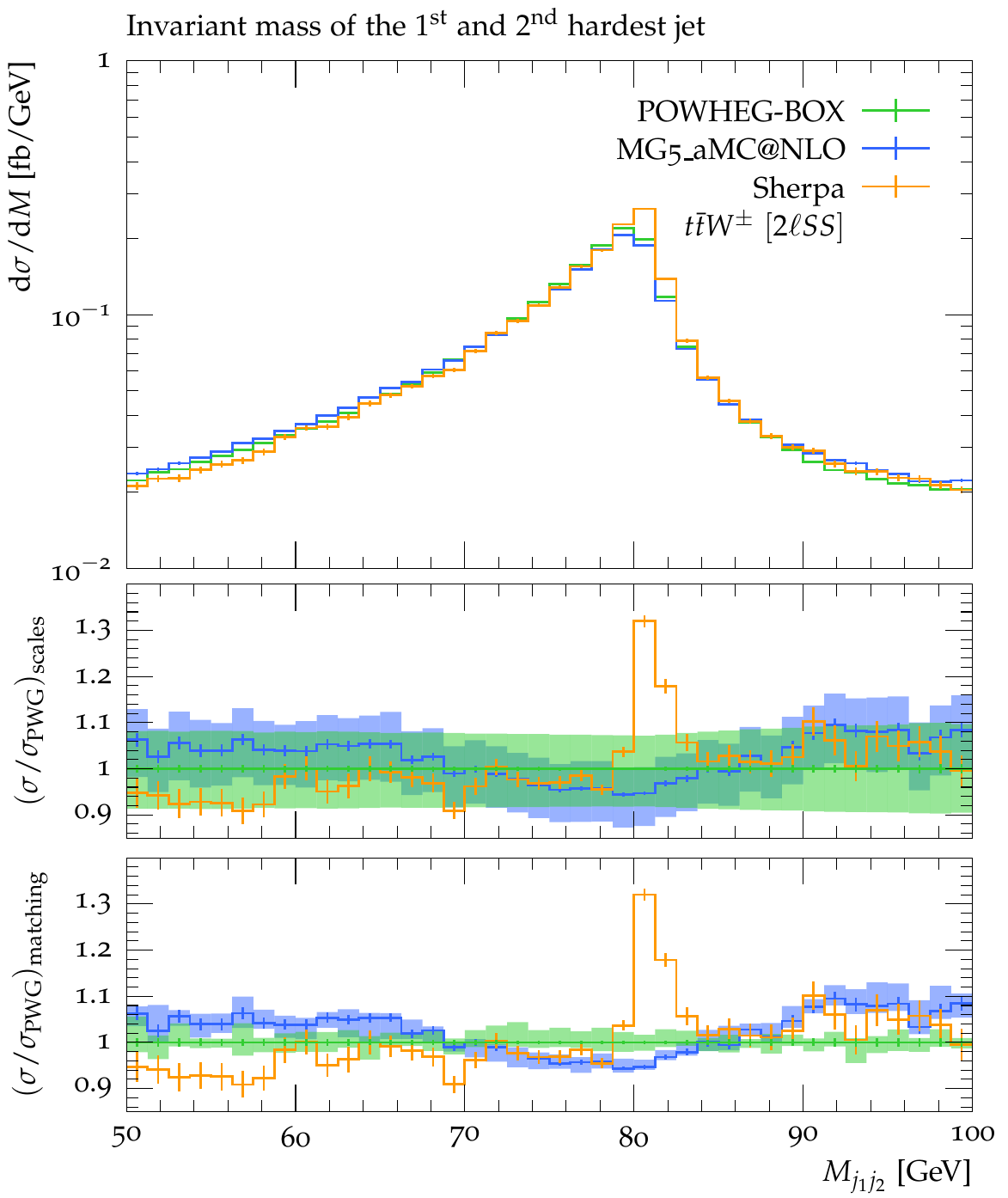}
 \includegraphics[width=0.49\textwidth]{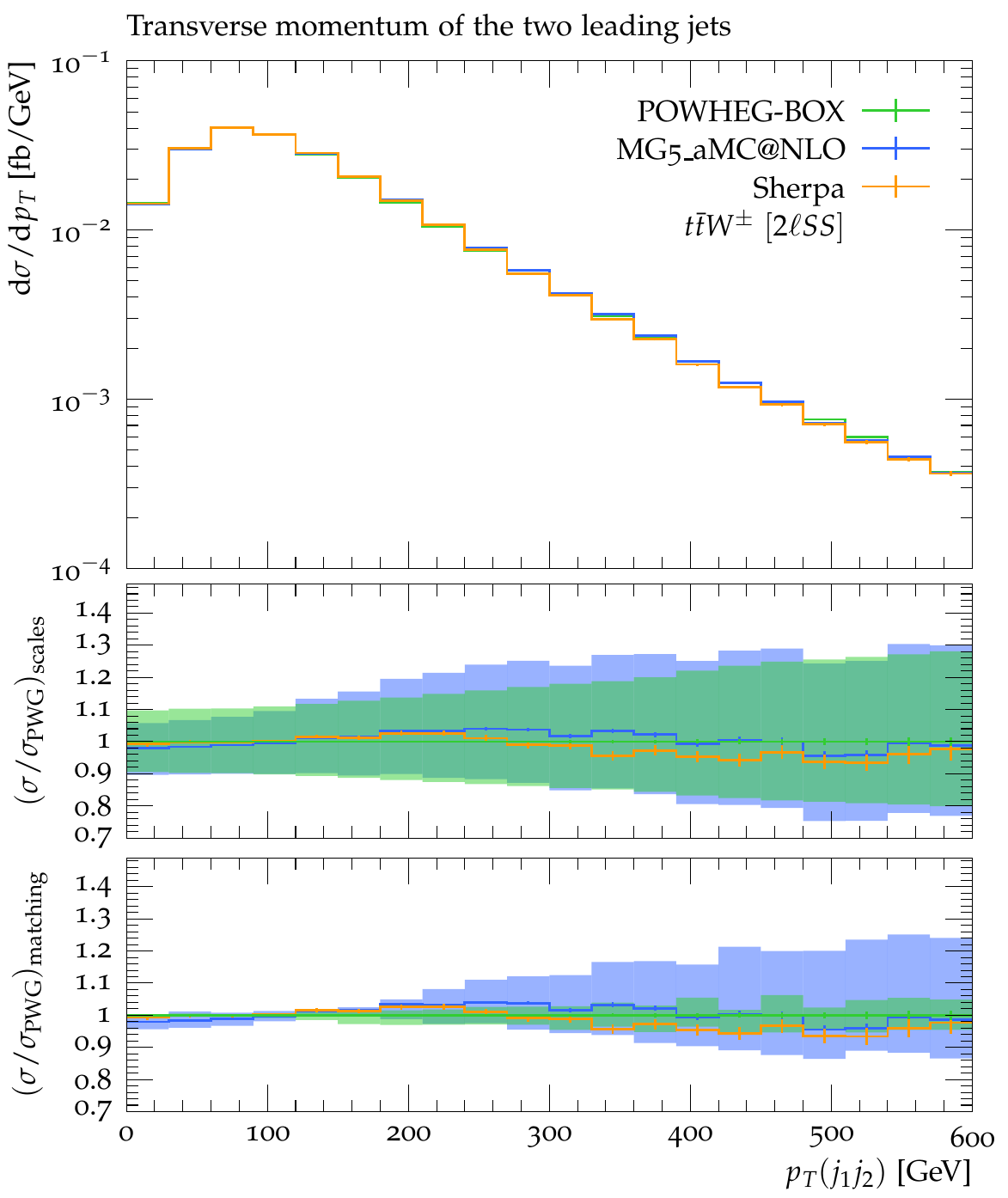}
 \caption{Differential cross section in the \SSL{} fiducial region as a function 
   of the invariant mass (l.h.s.) and the transverse momentum (r.h.s.) of the two 
   leading light jets. 
 The uncertainty bands (not shown for \sherpa{}) correspond to independent 
 variations of renormalization and factorization scales (middle panel) and of 
 the matching parameters (bottom panel).}
 \label{fig:2ssl_2}
\end{figure}

\begin{figure}[h!]
 \includegraphics[width=0.49\textwidth]{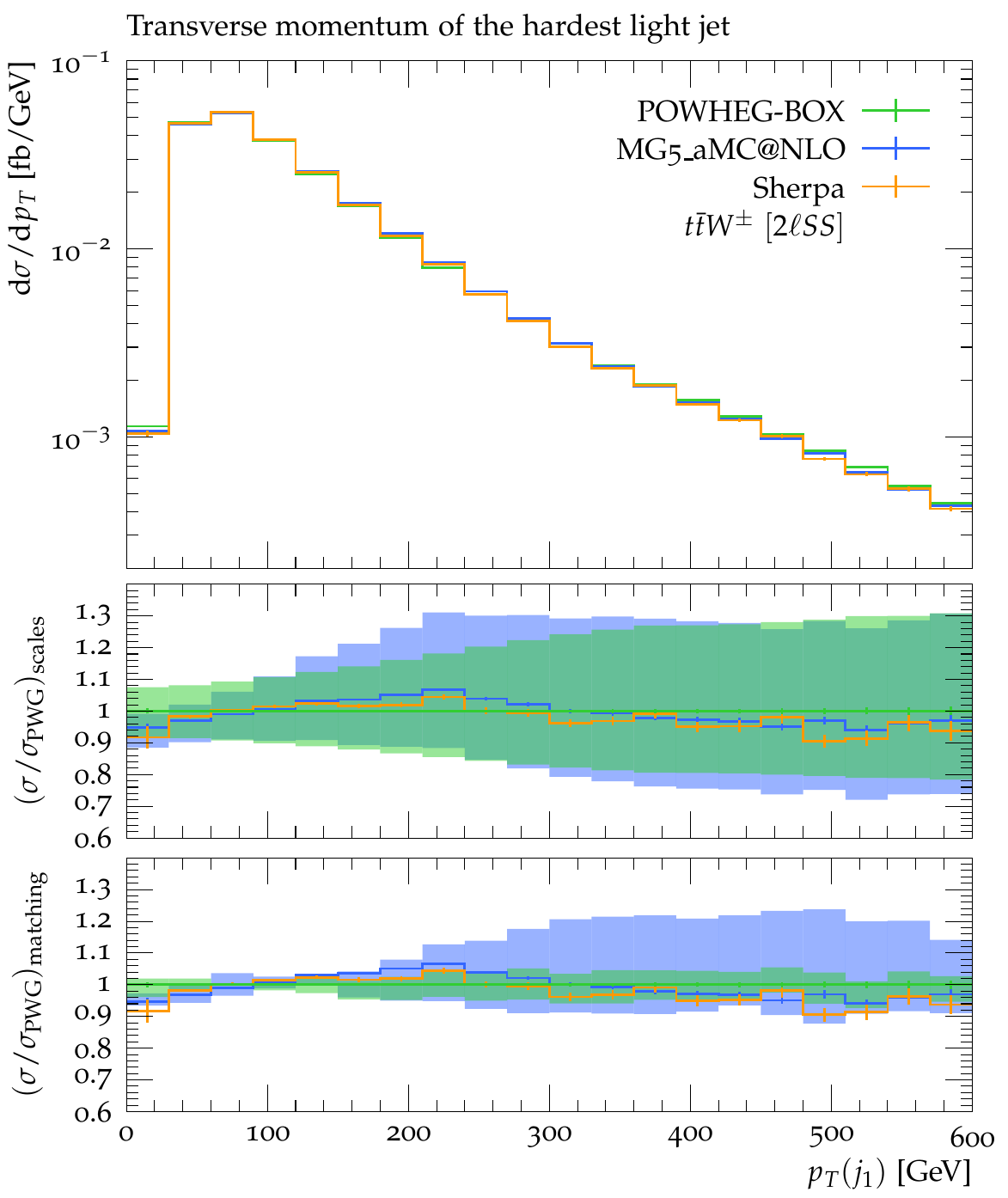}
 \includegraphics[width=0.49\textwidth]{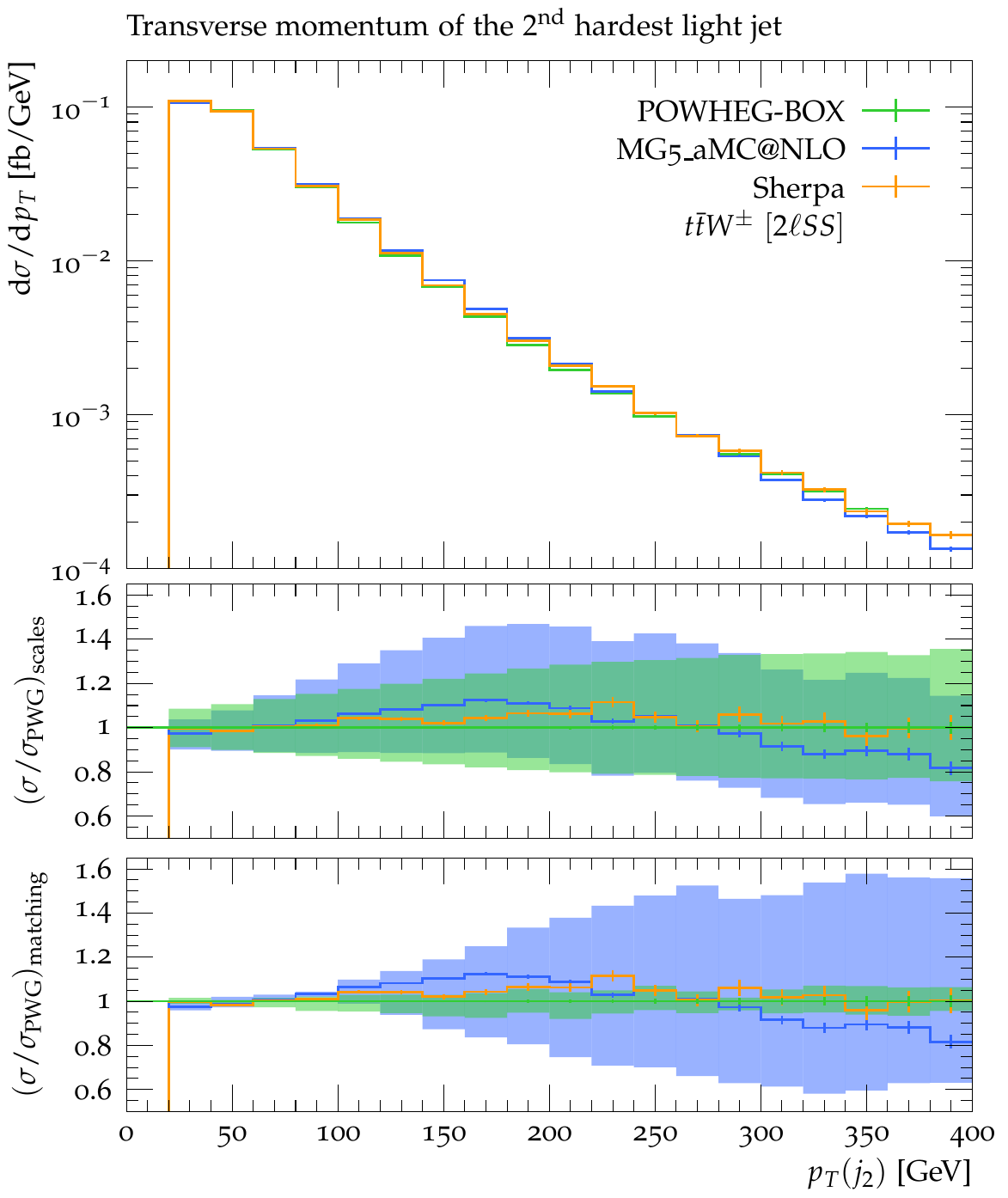}
 \caption{Differential cross section for the \SSL{} fiducial region as a function 
   of the transverse momentum of the hardest (l.h.s.) and second hardest (r.h.s.) 
   light jet. 
 The uncertainty bands (not shown for \sherpa{}) correspond to independent 
 variations of renormalization and factorization scales (middle panel) and of 
 the matching parameters (bottom panel).}
 \label{fig:2ssl_3}
\end{figure}

\begin{figure}[h!]
 \includegraphics[width=0.49\textwidth]{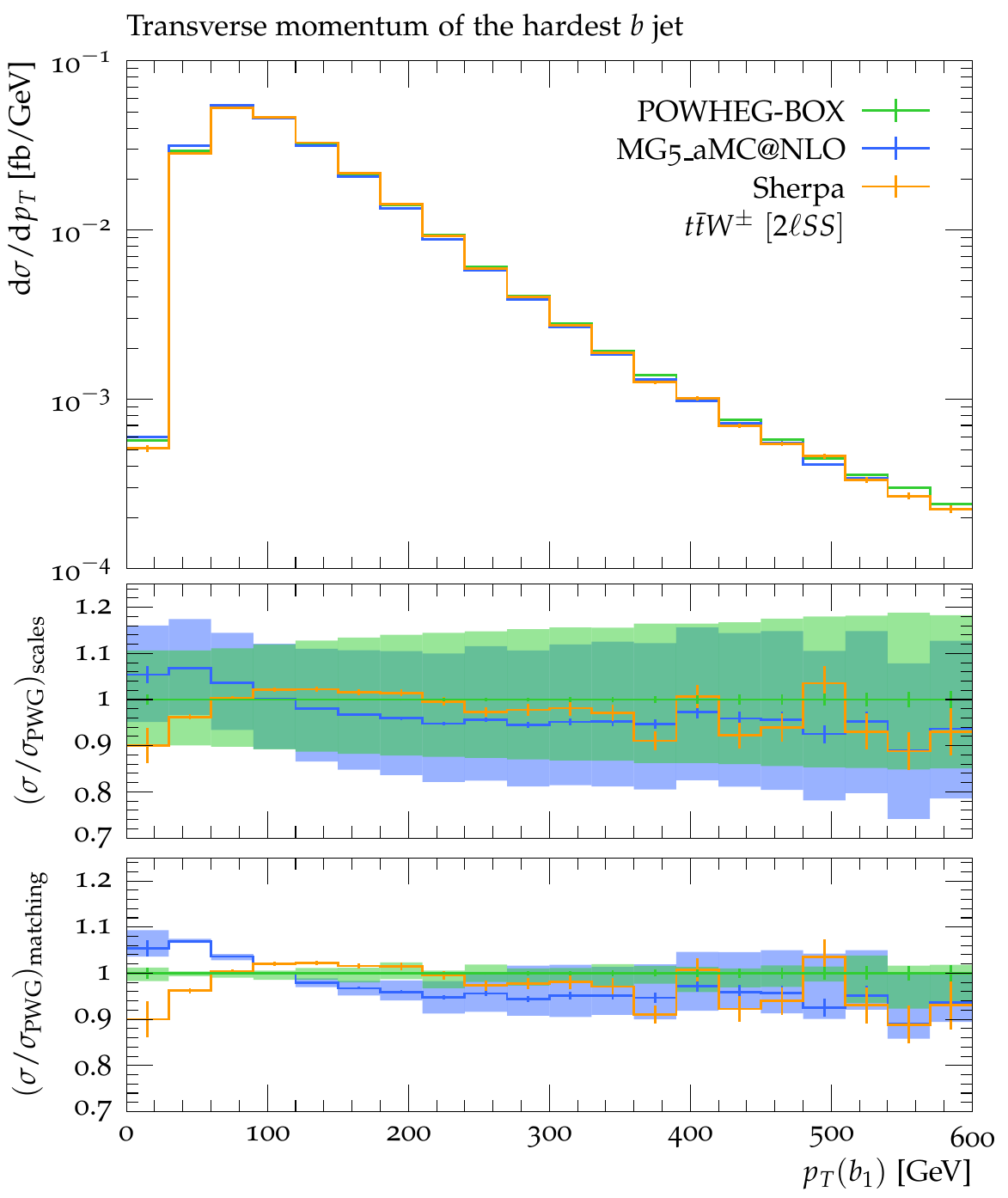}
 \includegraphics[width=0.49\textwidth]{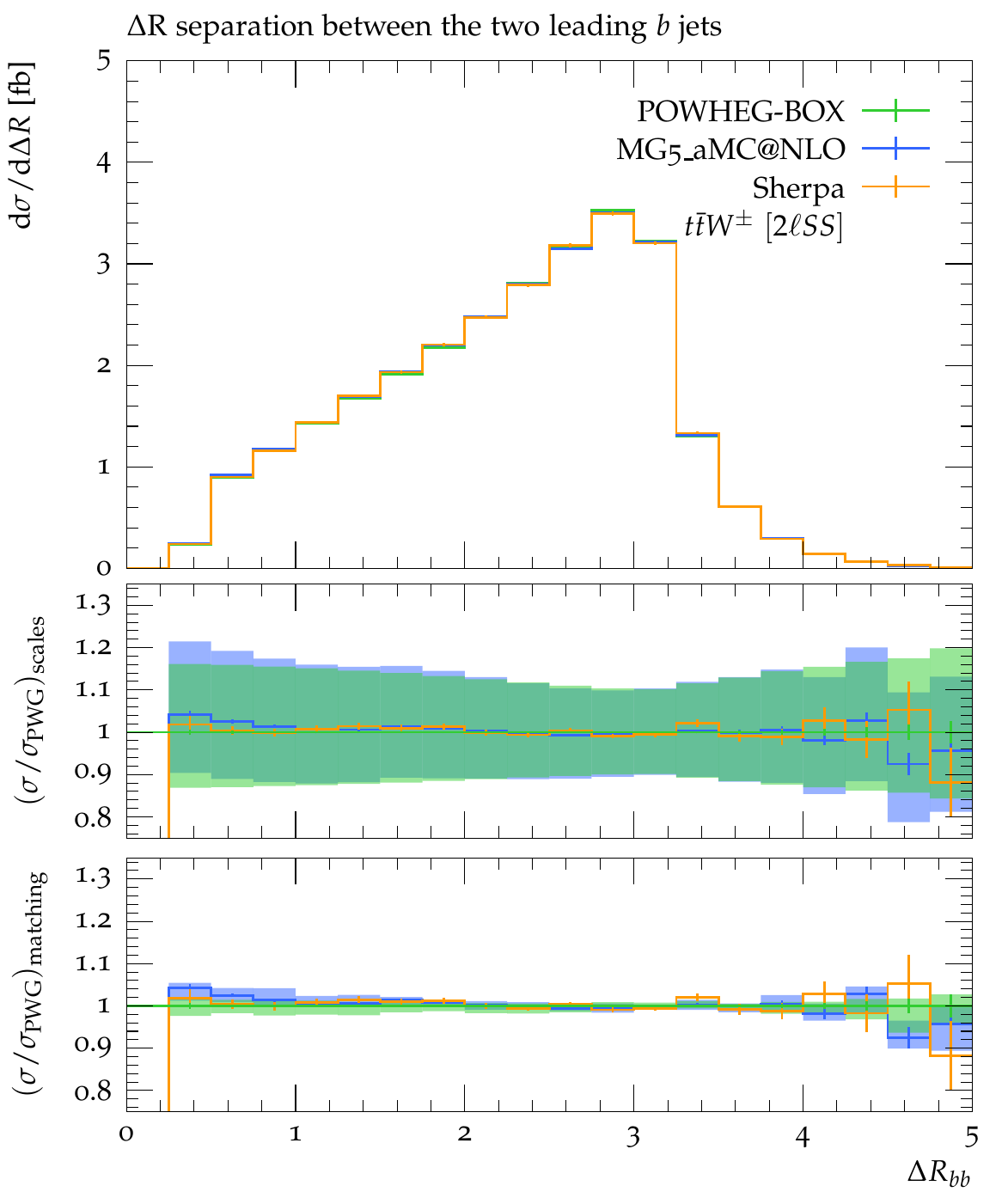}
 \caption{Differential cross section for the \SSL{} fiducial region as
   a function of the transverse momentum of the hardest $b$ jet (l.h.s.)
   and the $\Delta R$ separation between the two leading $b$ jets
   (r.h.s.). 
 The uncertainty bands (not shown for \sherpa{}) correspond to independent 
 variations of renormalization and factorization scales (middle panel) and of 
 the matching parameters (bottom panel).}
 \label{fig:2ssl_4}
\end{figure}

Theoretical uncertainties on the total cross sections are of course not 
representative of the actual uncertainties affecting differential distributions 
in different regions of the corresponding kinematic observables. With this in 
mind we will discuss in the following several distributions of phenomenological 
interest and emphasize how from the analysis of the residual scale and matching 
uncertainties we can derive indications of how to improve the corresponding 
theoretical predictions. All plots presented in 
Figs.~\ref{fig:2ssl_1}-\ref{fig:2ssl_6} consist of three panels where the upper 
panel shows the central predictions for the three considered generators, the 
middle one illustrates the scale uncertainty band based on the variation of 
$\mu_R$ and $\mu_F$, and the bottom one the matching uncertainty band obtained 
from the variation of damping factors and the initial shower scale $\muq$ . Both 
scales and matching uncertainties are normalized to the central 
\powhegbox{} prediction.

We start by considering the fiducial cross section as a function of the number of
light ($N_{\mathrm{l-jets}}$) or bottom  ($N_{\mathrm{b-jets}}$) jets, as shown 
in Fig.~\ref{fig:2ssl_1}. In the case of light jets,  shown on the left side of
Fig.~\ref{fig:2ssl_1}, we observe that \mgfive{} and \sherpa{} predictions align 
really well in all jet bins. Comparing to the \powhegbox{} we note that 
predictions for at least $2$, $3$, and $4$ light jets agree well among all 
generators, while starting from at least $5$ light jets onward the \powhegbox{} 
generates a softer spectrum and the deviation grows from $10\%$ up to $47\%$ for 
events with at least $8$ light jets, although the shape differences are  within 
the estimated theoretical uncertainties. Scale uncertainties are slightly 
asymmetric and start, after symmetrization, at $12\%$ for at least two jets and
increase to $25\%$ for the \powhegbox{} and $29\%$ for \mgfive{}. On the other 
hand, the matching uncertainties are rather different between the predictions. 
Varying the damping parameters in the \powhegbox{} only leads to modifications of 
the cross sections of at most $\pm5\%$ in the last jet bin, while variations of 
the initial shower scale in \mgfive{} lead to variations of up to $\pm 47\%$.
Thus, in the case of \mgfive{} the initial shower scale uncertainties become the 
dominant starting from at least $6$ light jets.

Similar conclusions can be drawn for the distribution in the number of $b$ jets. 
Beyond the first bin all three generators give slightly different predictions, 
where the \powhegbox{} shows the softest spectrum and \mgfive{} the hardest, with
nearly $50\%$ deviations with respect to the \powhegbox{} distribution for events
with 5 $b$ jets. However, Monte Carlo errors become sizable in this region as
well. \sherpa{} resembles more closely the \powhegbox{} spectrum over the whole 
range with differences of at most $25\%$. As for the case of light jets, the 
large deviations between the generators are compatible with the estimated 
theoretical uncertainties. While scale variations in the first bin still amount 
to $12\%$ uncertainties they increase rapidly to $19-24\%$ for each prediction. 
Also matching uncertainties increase tremendously for more than $3$ $b$ jets and 
show that these regions are clearly dominated by the parton shower, as expected 
for this particular observable. In general, also for other less obvious 
observables, a sudden strong dependence on the initial shower conditions indicate 
that these regions are dominated by the parton shower instead of the underlying
NLO computation.

Turning to other exclusive observables we compare the modelling of fiducial 
differential distributions for the considered event generators starting with the 
invariant mass spectrum of the two hardest jets $M_{j_1j_2}$ depicted on the left
of Fig.~\ref{fig:2ssl_2}. It is remarkable to see the pronounced Breit-Wigner 
shape of the $W$ boson, which suggests how the hardest jets in an event are 
predominantly generated by the hadronic decay of one of the $W$ bosons. More 
information can be derived from a closer inspection of the $p_T$ distribution of 
these jets, as we will discuss in the following. With respect to the \powhegbox{}
curve, the \mgfive{} prediction for the $M_{j_1j_2}$ distribution shows a 
slightly different shape with deviations of the order of $5\%$. \mgfive{} is 
slightly more off-shell as can be seen by the depletion of events in the 
resonance region by about $5\%$. \sherpa{} on the other hand is considerable more 
on-shell with $32\%$ more events on the $W$ resonance. The scale uncertainties
are nearly constant for \mgfive{} and of the order of $7\%$ in the plotted range. 
For the \powhegbox{} prediction scale uncertainties are constant and of the order
of $8\%$ below $M_{j_1j_2} \approx M_W$, and increase mildly to $10\%$ at the end
of the spectrum. The matching uncertainties are for the \mcnlo{} or \powheg{} 
scheme both at most $4\%$. The small impact of the shower scale variation can be
understood from the fact that \pythia{} preserves the momentum of resonant 
decaying particles. Therefore, the available phase space for further radiation 
is naturally limited by a shower scale $\muq \lesssim M_W$, which is considerable
smaller than $\muq = H_T/2$ and thus the radiation pattern of a hadronic decaying
$W$ can to a large extent be independent of the variation of the initial shower
scale we have chosen.

On the right-hand side of Fig.~\ref{fig:2ssl_2} we also show the overall 
transverse momentum distribution of the two hardest light jets $p_T(j_1j_2)$. 
All three predictions agree remarkably well over the whole plotted range with 
only minor shape differences, with deviations of up to $5\%$. The dominant 
contribution to the theoretical uncertainties originates from missing 
higher-order corrections. They amount to at least $\pm 10\%$ and increase with 
growing transverse momentum to $-27\%$ and $+30\%$ at $p_T \approx 600~\GeV$. 
Below $p_T \lesssim 180~\GeV$ the matching uncertainties are about $2-3\%$ for 
both the \mcnlo{} and \powheg{} schemes. Above, the initial shower scale 
dependence grows steadily to $20\%$, while the dependence of damping factors in 
the \powhegbox{} stays below $5\%$. The increase in the theoretical 
uncertainties with $p_T$ and in particular above $p_T \ge 180~\GeV$ suggests that
the spectrum is sensitive to the real radiation and thus one of the leading jets 
actually originates from an emission in the production of the $t\tb W^\pm$ final 
state. Below a transverse momentum value of $180~\GeV$ the spectrum is dominated 
by radiative top-quark decays. This explains the reduced scale dependence of the 
\powhegbox{} and \mgfive{} predictions in this region, since this part of the
distribution is inclusive with respect to the production of the $t\tb W^\pm$ 
system. Radiative top decays also account for the insensitivity on the initial 
shower scale of the predictions because \pythia{} preserves the top-quark 
resonance and thus a natural shower scale of $\muq \approx m_t \ll H_T/2$ is in 
place. 
To investigate the modelling of the first and second hardest light jets in more 
detail, we show in Fig.~\ref{fig:2ssl_3} the distributions in their individual
transverse momentum ($p_T(j_1)$ and $p_T(j_2)$ respectively). For both 
observables the shapes of the different generator's predictions agree well with 
each other. In the case of the hardest light jet, differences do not exceed 
$10\%$, while for the transverse momentum of the second hardest light jet 
deviations at the level of $10\%$ are visible and can become as large as $20\%$ 
in the tail of the distribution for \mgfive{}. The two transverse momentum
spectra however exhibit rather different theoretical uncertainties. For the 
hardest light jet the uncertainties are dominated by scale uncertainties that 
start around $6-8\%$ and increase up to $26-29\%$ in the tail of the 
distribution. Note that the \powhegbox{} scale uncertainties start slightly 
larger but grows lower than those of \mgfive{}. On the other hand, the matching 
uncertainties for the \powhegbox{} are below $\pm 5\%$ variation over the whole 
spectrum, while for \mgfive{} the uncertainties are of comparable size only below
$p_T \approx 180~\GeV$, but grow to about $+27\%$ and $-9\%$ for higher $p_T$. 
The still modest impact of the shower scale can be attributed to the fact that 
the tail of the spectrum is stabilized by the NLO real radiation and thus depends 
less on the shower evolution. The scale dependence for the transverse momentum 
distribution of the second hardest light jet behaves identically to the previous 
case. At the beginning there are uncertainties of the order of $7-9\%$ that rise 
to $30-33\%$ in the tail. However, the matching uncertainties dominate quickly 
the theoretical uncertainty of the \mgfive{} prediction, while, as in the 
previous cases, the \powhegbox{} predictions show damping uncertainties below 
$5\%$ over the full $p_T$ range. While the \mgfive{} uncertainties are compatible 
with the \powhegbox{} only up to $120~\GeV$ the corresponding shower scale 
dependence increases dramatically and generates corrections between $-23\%$ and 
$+91\%$. Clearly, all shape differences of the three different generators are 
within that uncertainty.

After discussing light-jet distributions we turn now to $b$ jets that are 
predominantly generated in the decay of top quarks and are thus a good testing 
ground to compare different top-quark modelling strategies. Starting with the 
transverse momentum of the hardest $b$ jet $p_T(b_1)$ shown on the left-hand side
of Fig.~\ref{fig:2ssl_4}, we observe small shape differences between the various 
predictions especially in the beginning of the distribution. The \powhegbox{} 
prediction is slightly harder than \mgfive{} and \sherpa{}, however the shape 
difference are well below $10\%$ and the prediction of the $b$ jet spectrum is 
sensitive to the modelling of top-quark decays~\cite{Jezo:2016ujg} as well as to 
the modelling of radiation from $b$-quarks in the parton shower. However, the 
dominant source of uncertainty remains the missing higher-order corrections which
amount to $10-18\%$, as shown by the scale dependence of the high-$p_T$ spectrum.
Regarding the matching uncertainties one can note that the \powhegbox{} 
predictions show in general a very reduced dependence on the damping parameters, 
while \mgfive{} obtains uncertainties as large as $8\%$ in the tail of the 
distribution. 
 
Additionally, on the right-hand side of Fig.~\ref{fig:2ssl_4} we show the 
$\Delta R_{bb}$ separation of the two hardest $b$ jets. As can be seen from the 
shoulder in the distribution, the two leading $b$ jets tend to be generated most 
of the time in a back-to-back configuration, which is typical for top-quark 
decays, whereas jets originating from a collinear $g\to b\bar{b}$ splitting in 
the parton shower would peak for small values of $\Delta R_{bb}$. The predictions 
of all three generators are identical, and uncertainties are largely compatible, 
and dominated by scale sensitivity. Scale uncertainties are the smallest around 
the peak at $\Delta R_{bb} \approx 3$ where they amount to $\pm 10\%$, while
towards the beginning and the end of the spectrum they increase to about $18\%$. 
Matching uncertainties are for the \powhegbox{} as well as \mgfive{} below $5\%$ 
over the whole spectrum.

\subsubsection*{Leptonic observables}
%
\begin{figure}[h!]
 \includegraphics[width=0.49\textwidth]{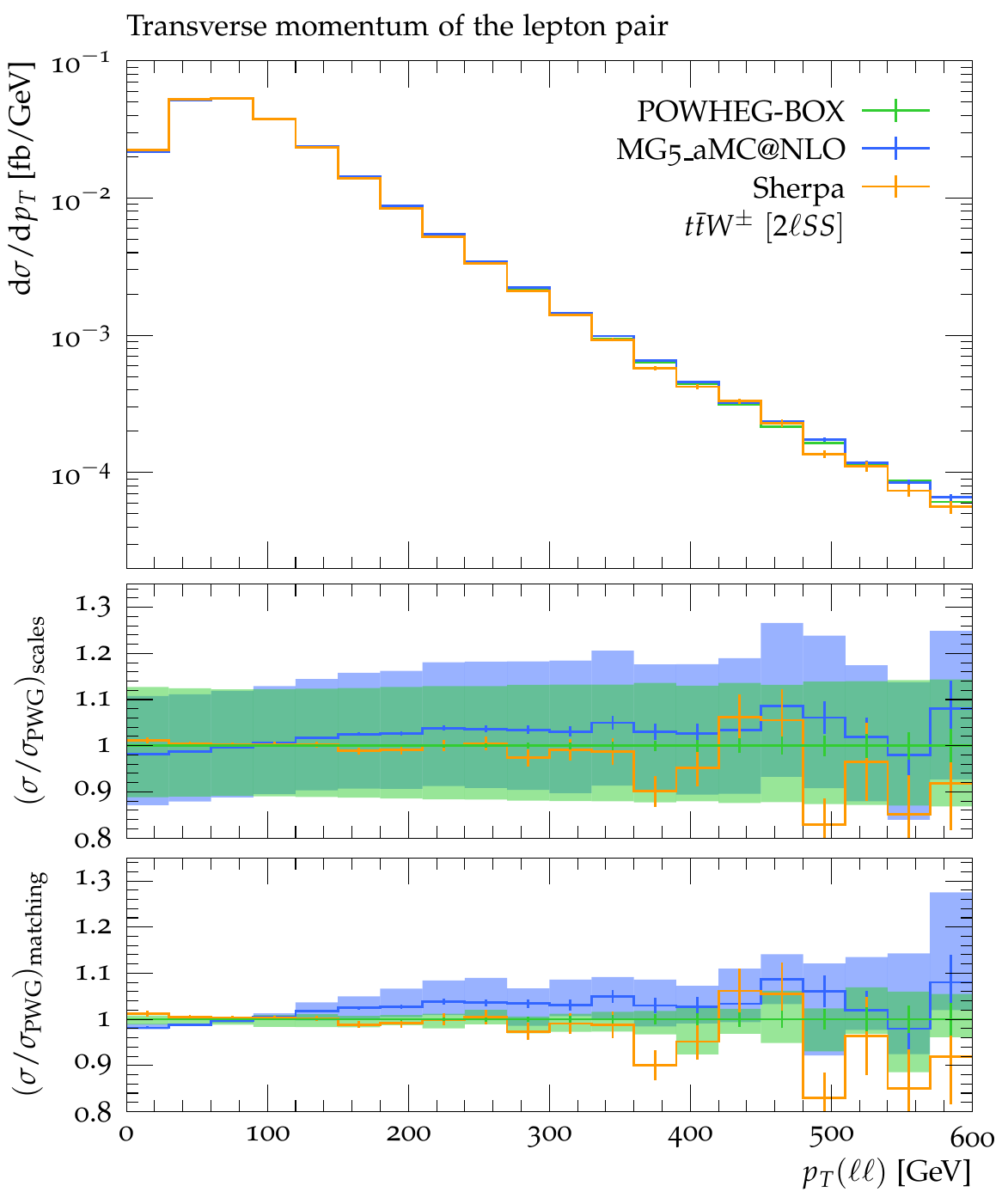}
 \includegraphics[width=0.49\textwidth]{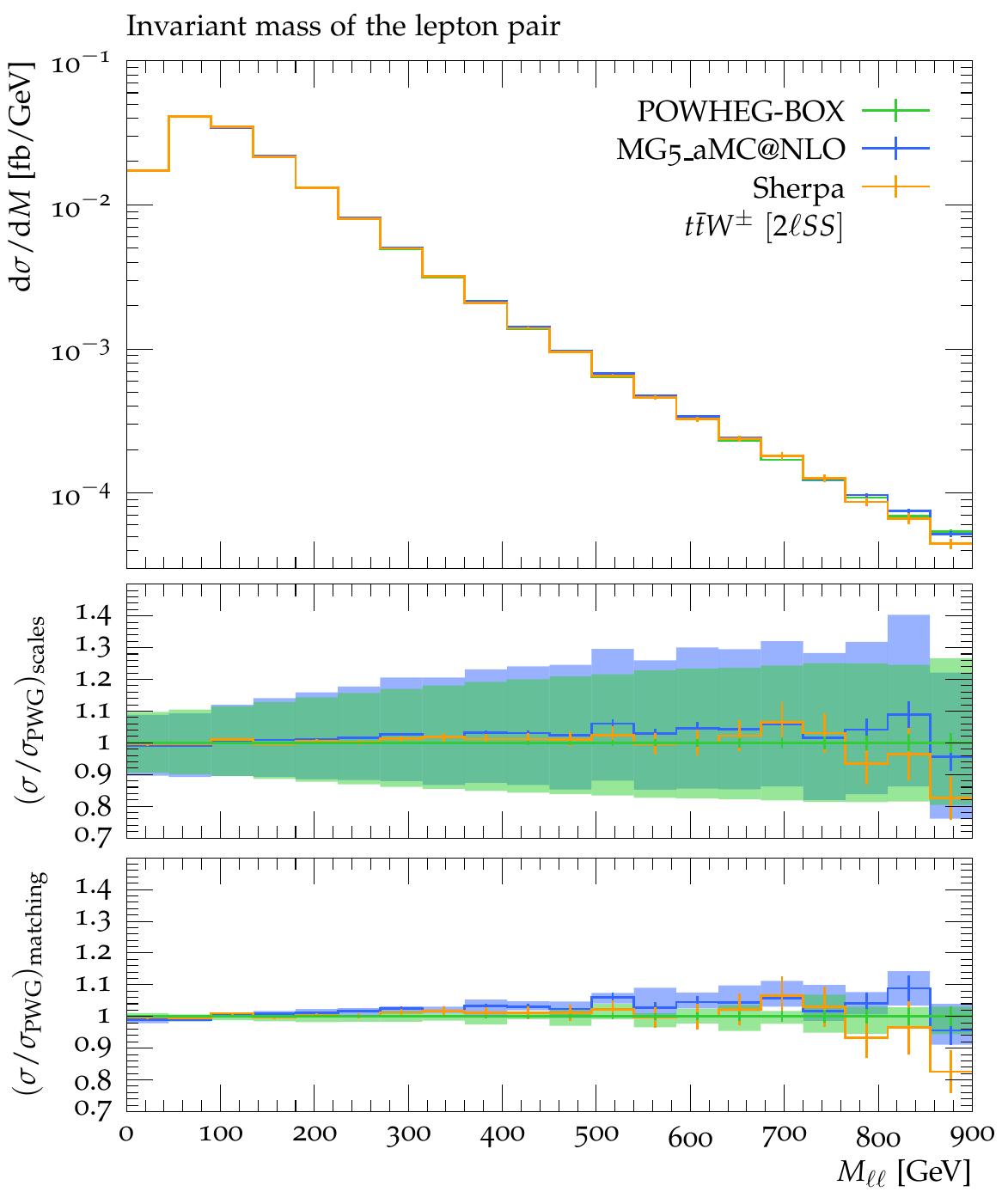}
 \caption{Differential cross section for the \SSL{} fiducial region as
   a function of the transverse momentum (l.h.s.) and the invariant mass
   (r.h.s.) of the same-sign lepton pair. 
 The uncertainty bands (not shown for \sherpa{}) correspond to independent 
 variations of renormalization and factorization scales (middle panel) and of 
 the matching parameters (bottom panel).}
 \label{fig:2ssl_5}
\end{figure}

We now focus on leptonic observables, such as the transverse momentum 
$p_T(\ell\ell)$ and the invariant mass $M_{\ell\ell}$ spectrum of the two 
same-sign lepton pair that are shown in Fig.~\ref{fig:2ssl_5}. These observables 
are only indirectly affected by QCD corrections, because the leptons will only 
recoil against further emissions in the parton shower evolution. Nonetheless, 
these observables are crucial to illustrate the dynamical correlations between
the decay of the prompt $W$ boson and of the top quark, and therefore might be 
sensitive to the overall shower evolution. For the transverse momentum 
distribution shown on the left of Fig.~\ref{fig:2ssl_5} we observe that \sherpa{} 
and \powhegbox{} predictions are essentially identical within the Monte Carlo 
uncertainty. In contrast, \mgfive{} shows a clear but very small shape difference 
with respect to the \powhegbox{}. The distribution is clearly affected mostly by 
scale variations at a level of $10\%$ over the whole range. Matching 
uncertainties on the other hand are negligible at the beginning and increase 
steadily towards the end of the spectrum where they become comparable in size 
with the scale uncertainties.
For the case of the invariant mass spectrum of the lepton pair, depicted on the
right of Fig.~\ref{fig:2ssl_5}, we find again very good agreement between all
generators over the whole range. The theoretical uncertainties are larger as 
compared to the transverse momentum distribution. Here, scale uncertainties start
at $10\%$ and increase up to $25\%$ at the end of the spectrum. In addition, the 
spectrum is not very sensitive to the parton shower evolution as the 
corresponding uncertainties are below $5\%$ for both \mgfive{} and the 
\powhegbox{}.

\begin{figure}[h!]
 \includegraphics[valign=t,width=0.49\textwidth]{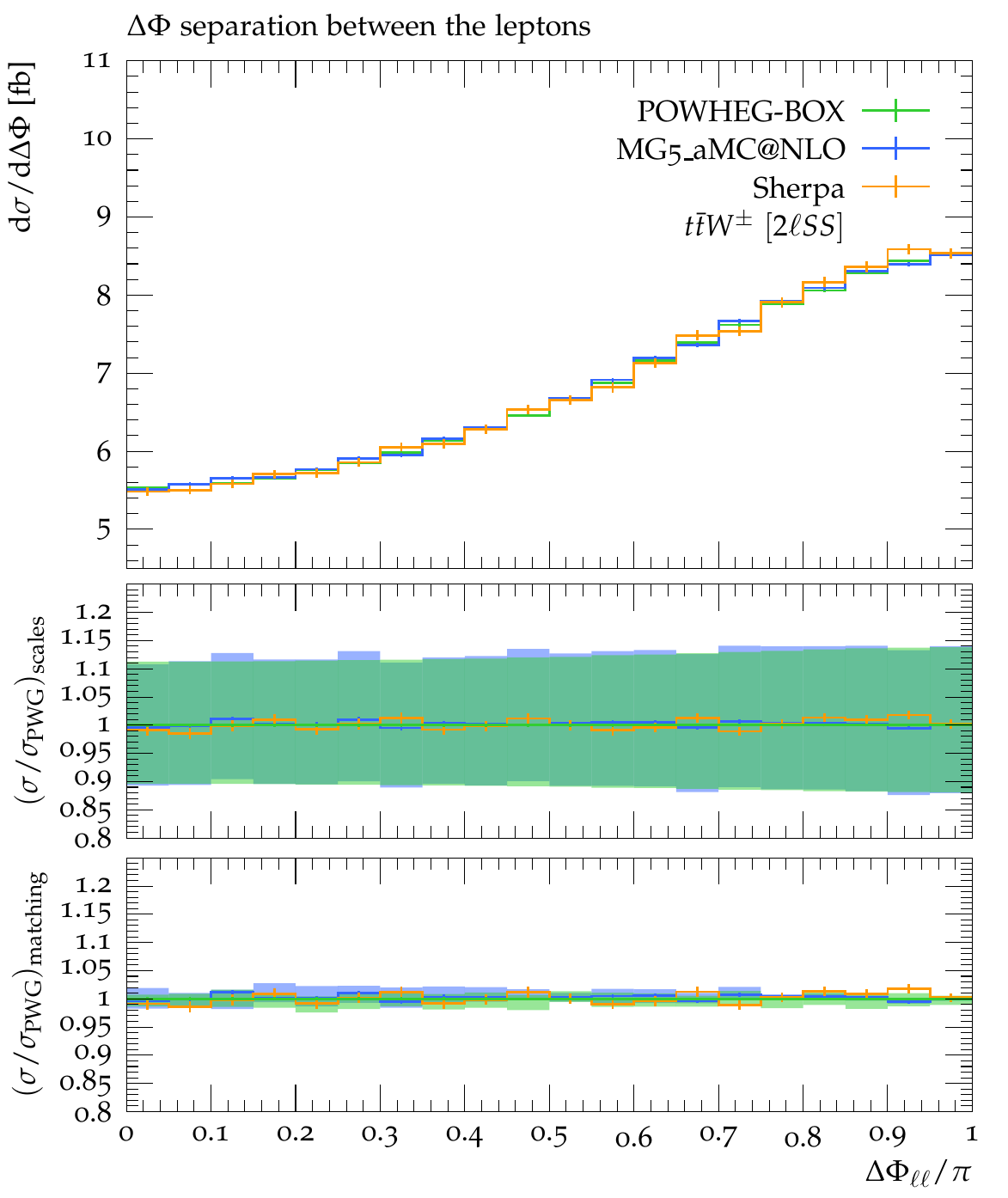}
 \includegraphics[valign=t,width=0.49\textwidth]{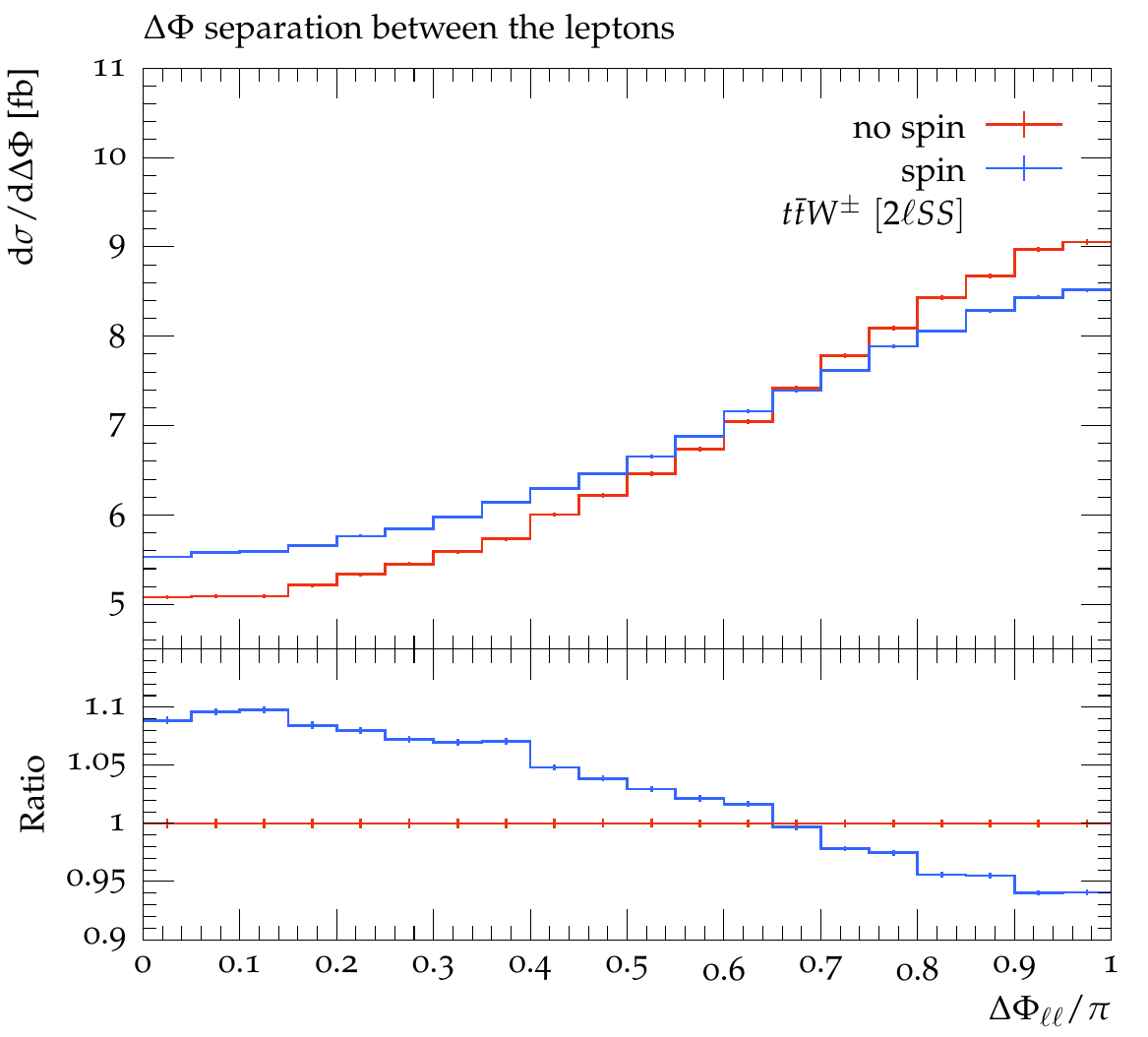}
 \caption{Differential cross section for the \SSL{} fiducial region as
   a function of the azimuthal angle $\Delta \Phi_{ll}$ between the
   two same-sign leptons (l.h.s.) and the impact of spin correlations in
   the decay modelling (r.h.s.). 
 The uncertainty bands in the l.h.s. plot (not shown for \sherpa{}) correspond to independent 
 variations of renormalization and factorization scales (middle panel) and of 
 the matching parameters (bottom panel).}
 \label{fig:2ssl_6}
\end{figure}
Additionally, we show in Fig.~\ref{fig:2ssl_6} the azimuthal angle between the 
two same-sign leptons $\Delta\phi_{\ell\ell}$ which is sensitive to spin
correlations. Let us note at this point that the observed spin correlations in 
our case have a different origin than those typically studied in top-quark pair 
production~\cite{Bigi:1986jk,Barger:1988jj,Bernreuther:1995cx,Mahlon:1995zn,
Behring:2019iiv,Czakon:2020qbd}. In top-pair production the dilepton pair 
originating from the top-quark decays are spin correlated. In our case however, 
since we are considering signatures with same-sign leptons, only one lepton 
emerges from a top-quark decay, while the other one is produced by the decay of 
the prompt $W$ boson. At the same time, it is exactly the presence of the prompt 
$W$ boson that fully polarizes the top quarks~\cite{Maltoni:2014zpa}, and as a
consequence $\Delta \Phi_{\ell\ell}$ depends on the spin correlation between 
them. To highlight the importance of spin correlations, in the right-hand side 
plot of Fig.~\ref{fig:2ssl_6} we show \powhegbox{} predictions that either 
include or neglect spin correlations in the decay modelling for the \SSL{} final 
state. We observe that spin correlations lead to a shift in the spectrum, where 
events are shifted from the upper end of the spectrum at $\Delta\Phi_{\ell\ell} 
\approx \pi$ to the opposite end. The induced corrections to the distribution can 
reach up to $10\%$ for $\Delta\Phi_{\ell\ell}/\pi \lesssim 0.15$. Similar effects 
are found in other leptonic observables (either dimensionless or dimensionful).

\subsubsection*{Impact of $t\tb W^\pm$ EW contributions}
We conclude this section by analyzing the impact of the EW production mode of the 
$t\tb W^\pm$ final state on the \SSL{} signature under consideration using our 
\powhegbox{} implementation. In this context, we would like to stress that, while 
we highlight in the following the few special cases where the $t\tb W^\pm$ EW 
contribution becomes significant and cause visible shape modifications, for most 
observables the inclusion of the $t\tb W^\pm$ EW process amounts to a rather flat 
$+10\%$ correction at the differential level. Since the EW production mode 
enhances the production of light jets in the forward region, as can be seen from 
the pseudorapidity distribution in Fig.~\ref{fig:inc_QCDEW_1}, we expect the most 
severe impact in light-jet observables.

\begin{figure}[h!]
 \includegraphics[width=0.49\textwidth]{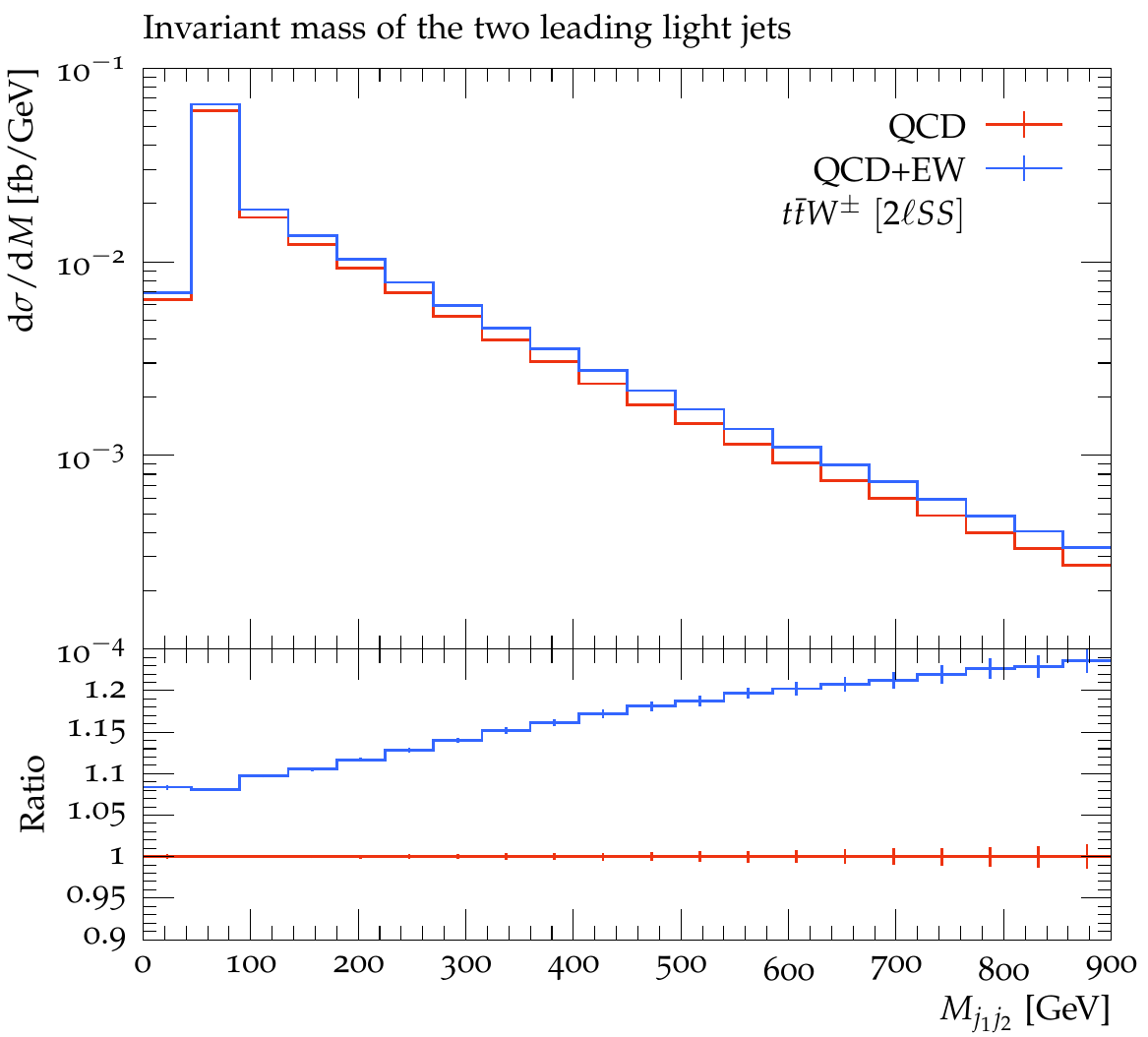}
 \includegraphics[width=0.49\textwidth]{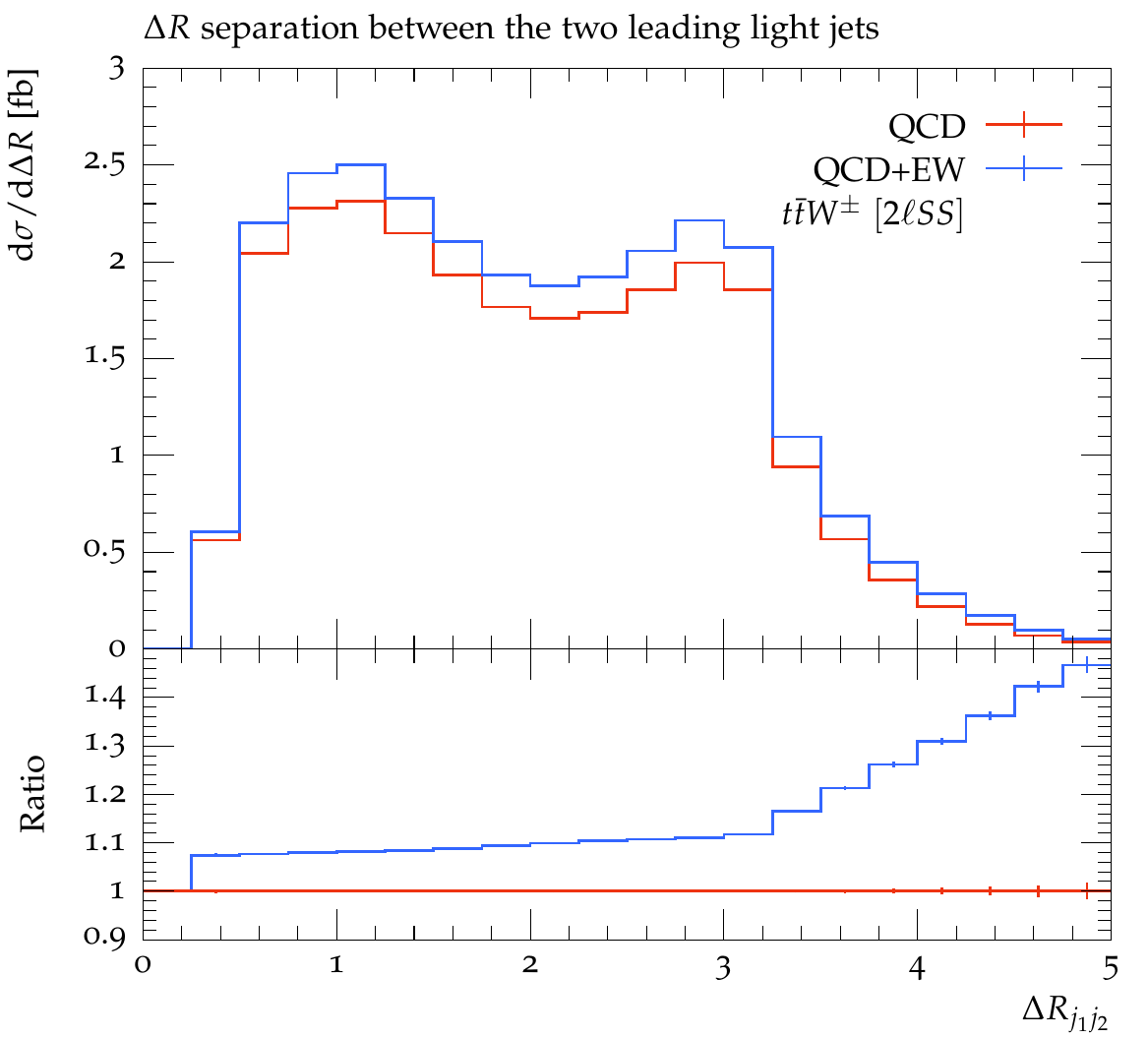}
 \caption{Differential cross section for the \SSL{} fiducial region as a 
 function of the invariant mass (l.h.s.) and $\Delta R$ separation between the 
 two hardest lights jets (r.h.s.).  The predictions based only on $t\tb W^\pm$ 
 QCD production are given in red while the total $t\tb W^\pm$ QCD+EW 
 contributions are shown in blue. The bottom panel shows the percentage change in 
 the shape of the distribution.}
 \label{fig:2ssl_8}
\end{figure}
To illustrate this point, we show in the left-hand side plot of 
Fig.~\ref{fig:2ssl_8} the invariant mass distribution of the leading two light 
jets $M_{j_1j_2}$ as predicted by the $t\tb W^\pm$ QCD contribution only and the 
total $t\tb W^\pm$ QCD+EW contribution. The bottom panel shows the ratio with 
respect to only the $t\tb W^\pm$ QCD contribution. Below $M_{j_1j_2} \lesssim 
100~\GeV$ we see that the $t\tb W^\pm$ EW contribution is small, about a $+8\%$ 
correction, which can be understood from the fact that this region should be 
dominated by jets originating from the hadronically decaying $W$ boson instead of 
jets emitted in the production process. However, above $100~\GeV$ the 
$t\tb W^\pm$ EW contribution starts to grow until it reaches a quite significant 
$+25\%$ correction at the end of the plotted range. Also in the 
$\Delta R_{j_1j_2}$ separation between the leading light jets, shown on the 
right-hand side of Fig.~\ref{fig:2ssl_8}, we can see a strong impact of the 
$t\tb W^\pm$ EW contribution. Below $\Delta R_{j_1j_2} \approx \pi$ the 
additional contribution is small between $8-12\%$ and starts growing rapidly 
beyond that point. In fact, the $t\tb W^\pm$ EW contribution generates 
corrections of the order of $50\%$ at $\Delta R_{j_1j_2} \approx 5$ which 
highlights the fact that this contribution preferably populates the forward 
regions.

To further highlight the impact of the EW contribution we study jets in the 
forward region. To this end we look for events that have passed the selection 
cuts specified in section~\ref{sec:setup} and have additional jets in the forward 
region defined by:
\begin{equation}
 p_T(j) > 25~\GeV\;, \qquad 2.5 \le |\eta(j)| \le 5.0\;,
\end{equation}
where we do not distinguish between light or $b$ jets.

\begin{figure}[h!]
 \includegraphics[width=0.49\textwidth]{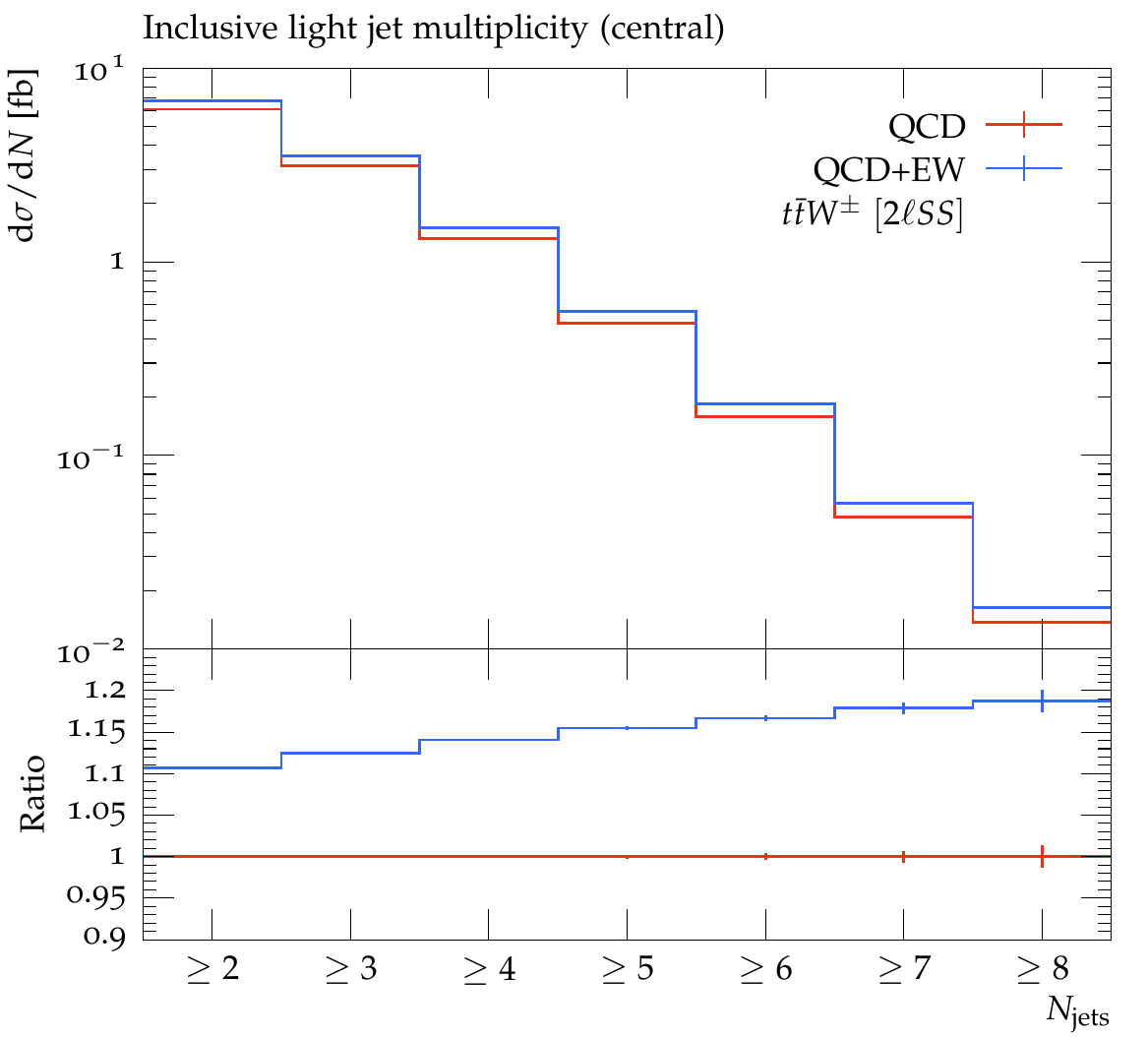}
 \includegraphics[width=0.49\textwidth]{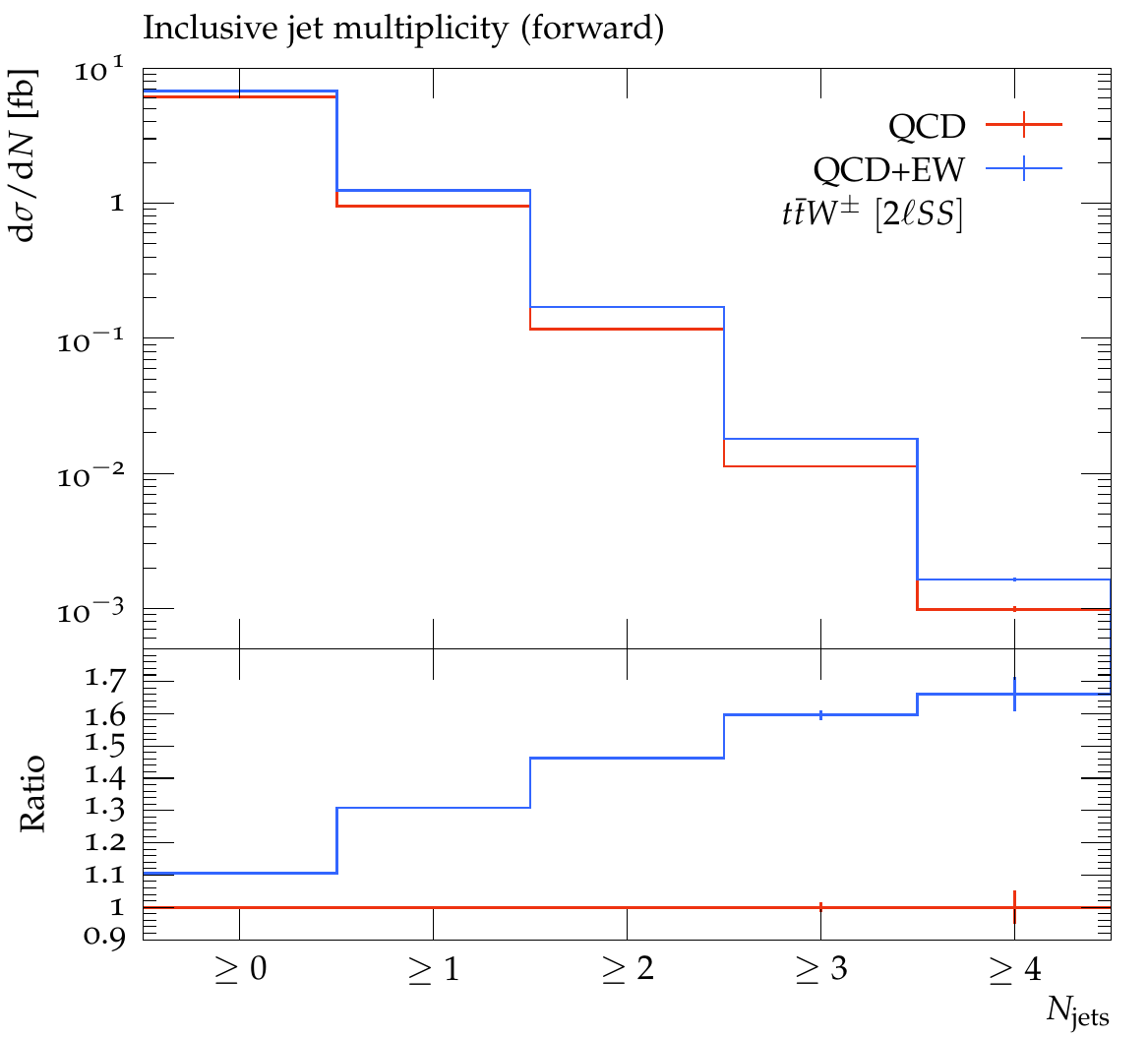}
 \caption{The inclusive cross section for the \SSL{} fiducial region as a 
 function of the number of hard jets in
 the central phase space volume (l.h.s.) and in the forward region (r.h.s.). 
 The predictions based only on $t\tb W^\pm$ QCD production are given in red
 while the total $t\tb W^\pm$ QCD+EW contributions are
 shown in blue. The bottom panel shows the
 percentage change in the shape of the distribution. }
 \label{fig:2ssl_9}
\end{figure}
In Fig.~\ref{fig:2ssl_9} we show the inclusive cross section as a function of the
number of jets in the forward region (r.h.s.) and contrast it with the 
corresponding plot in the central region (l.h.s.). In the central phase space 
volume the $t\tb W^\pm$ EW contribution only has a very mild impact, and, for 
events with at least two light jets, this contribution amounts to $11\%$, while 
for events with at least eight light jets it grows to $19\%$. On the contrary, if 
we look at the inclusive jet multiplicities in the forward region as shown on the 
right plot in Fig.~\ref{fig:2ssl_9} then we see that for the first bin a $11\%$ 
corrections is visible that then quickly increases up to $66\%$ for events that
have at least four additional forward jets.

\begin{figure}[h!]
 \includegraphics[width=0.49\textwidth]{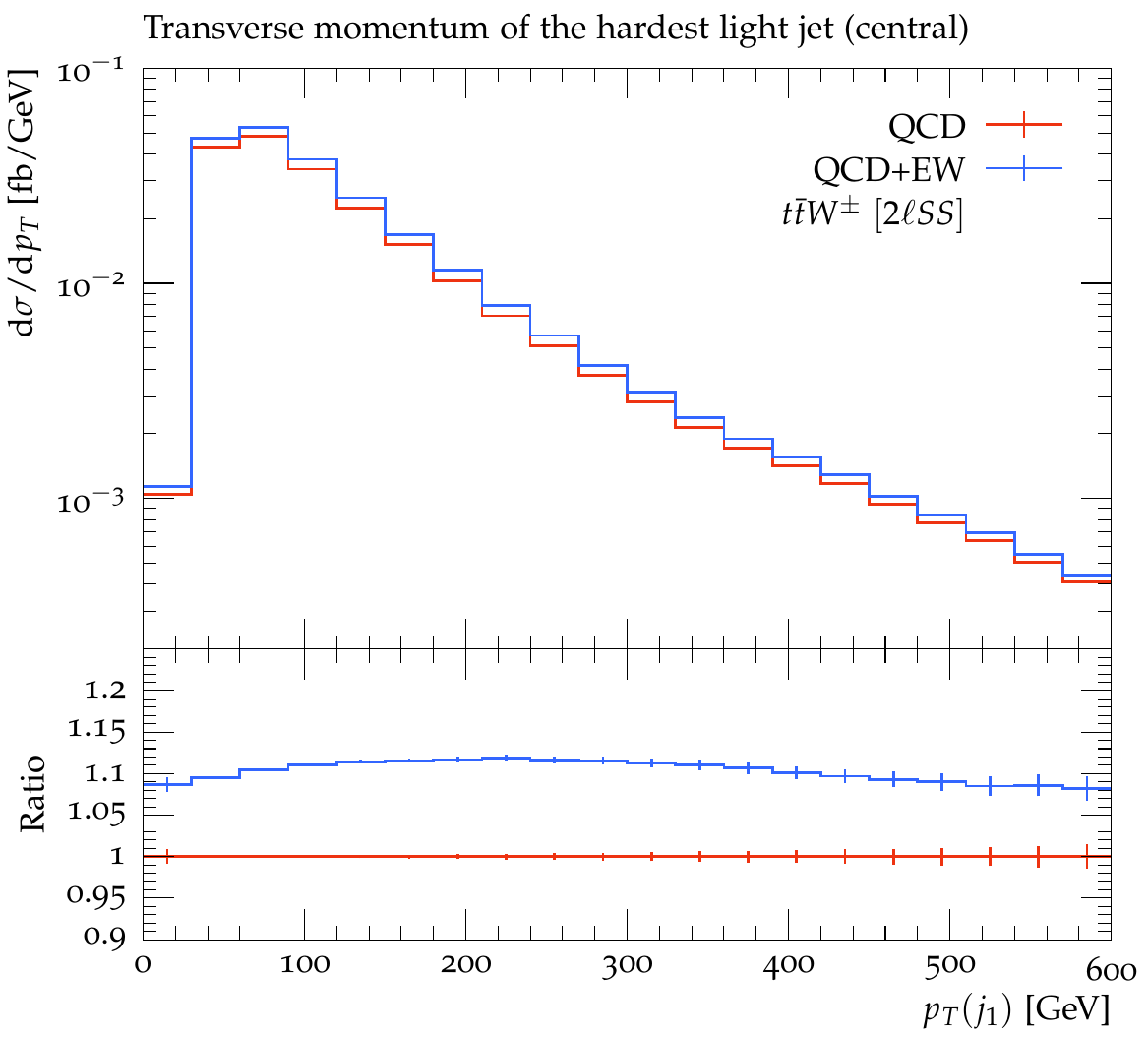}
 \includegraphics[width=0.49\textwidth]{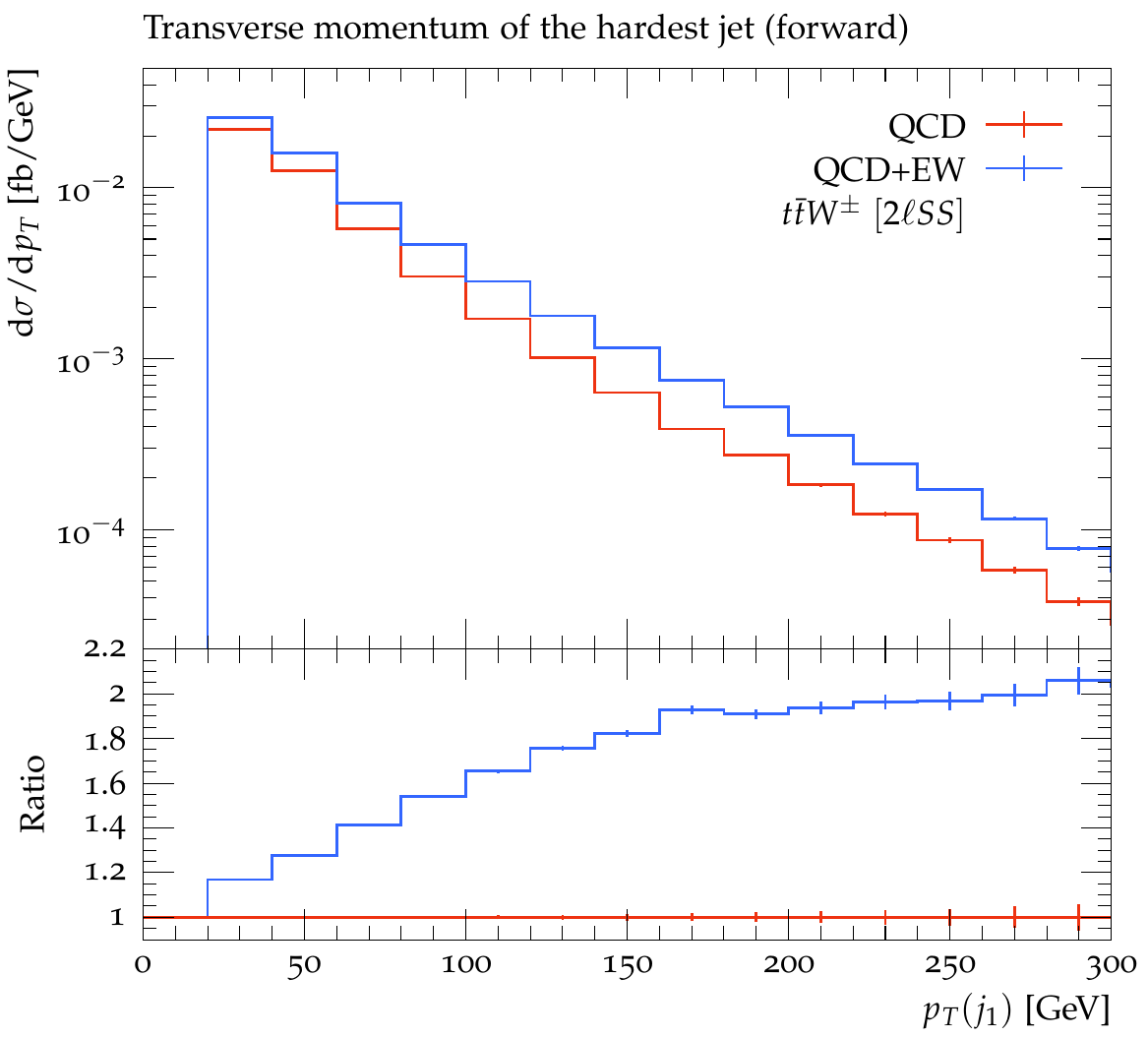}
 \caption{Differential cross section for the \SSL{} fiducial region as a 
 function of the transverse momentum of
 the leading jet in central phase space volume (l.h.s.) and in the forward 
 region (r.h.s.). The predictions based only on $t\tb W^\pm$ QCD production 
 are given in red while the total $t\tb W^\pm$ QCD+EW contributions
 are shown in blue. The bottom 
 panel shows the percentage change in the shape of the
 distribution.}
 \label{fig:2ssl_10}
\end{figure}
Last we study the impact on the transverse momentum spectrum of the leading light 
jet, $p_T(j_1)$, in the central and forward regions as depicted in 
Fig.~\ref{fig:2ssl_10}. As before, in the central phase space volume we find a 
rather constant correction of $9-11\%$ over the whole range of the plotted 
spectrum. The maximal corrections of $11\%$ are obtained around $p_T \approx 
200~\GeV$. On the other hand, the hardest forward jet receives large corrections 
from $t\tb W^\pm$ EW production. Essentially starting from $17\%$ corrections at 
the beginning of the distribution the $t\tb W^\pm$ EW contribution gives rise to 
$100\%$ corrections for $p_T \gtrsim 200~\GeV$. 

\section{Conclusions}
\label{sec:conclusions}
In this paper we have presented a new NLO parton-shower Monte Carlo event 
generator for the hadronic production of a top-quark pair in association with a 
$W^\pm$ boson taking into account the dominant NLO corrections at 
$\mathcal{O}(\alpha_s^3\alpha)$ and $\mathcal{O}(\alpha_s\alpha^3)$. Decays of 
unstable particles are included at leading order retaining spin-correlation
effects. The \powheg{} event generator \texttt{Wtt\_dec} is publicly available as 
part of the \powhegbox{} repository under
\begin{center}
 \url{http://powhegbox.mib.infn.it}
\end{center}

Motivated by the current tension~\cite{ATLAS:2019nvo} between the 
state-of-the-art SM predictions for the $t\tb W^\pm$ cross section and the 
corresponding measurement derived when $t\tb W^\pm$ is extracted from a combined 
signal and background fit in $t\tb H$ analyses, we performed a detailed generator 
comparison involving the \powhegbox{}, \mgfive, and \sherpa{}. A comparison at 
the level of on-shell $t\tb W^\pm$ production has revealed good agreement for the
$\mathcal{O}(\alpha_s^3\alpha)$ production mode, while the 
$\mathcal{O}(\alpha_s\alpha^3)$ contribution is very sensitive to the details of 
a given generator setup. Furthermore, a comparison has been made at the fully 
decayed stage for a fiducial phase space volume corresponding to the \SSL{} 
signature with two same-sign leptons and both light and $b$ jets. We provide this 
as a proof of concept to estimate in a more robust way the residual theoretical 
uncertainty on the $pp\to t\tb W^\pm$ cross section from both fixed-order and 
parton-shower components, at the inclusive and fiducial level. We found good 
agreement between all three generators at the differential level. Theoretical 
uncertainties have been addressed via means of independent variations of the
renormalization and factorization scales as well as matching related parameters 
specific of each generator, such as damping factors (\powhegbox{}) or the initial
shower scale (\mgfive{}). The investigation of the latter dependence allowed us 
to explain peculiar shape differences between the generators. In addition, we 
also investigated the impact of LO accurate spin-correlated decays on 
the differential distributions and found that leptonic observables are 
particularly sensitive to spin-correlation effects. Finally, we also quantified 
the impact of the $\mathcal{O}(\alpha_s\alpha^3)$ contributions at the 
differential level. For most observables these contributions amount to a flat 
$+10\%$ correction, while for a few observables sensitive to forward jets they 
can become more sizable.

It is interesting to notice that even though the on-shell modelling of  
$t\tb W^\pm$ EW production largely depends on the matching and parton-shower 
settings of each generator, these effects are much less visible for full 
predictions once QCD and EW contributions are combined. Furthermore, giving the 
fact that the two same-sign lepton signature is dominated by jets emerging from a 
hadronic $W$ decay that in all generators is modelled only at LO, for this 
particular signature it will be very important to improve on the theoretical 
description of these jets in the future. With respect to this, the modelling of 
the two same-sign lepton signature could be improved by a fixed-order full 
off-shell computation similar to Ref.~\cite{Denner:2017kzu}, as the corresponding 
matching to parton showers would be computationally challenging. The narrow-width 
approximation presents an alternative to include one-loop QCD corrections to the 
top-quark and $W$ boson decays and could be included in an event generator, as 
has been already shown in Ref.~\cite{Campbell:2014kua}.
On the other hand, other fiducial signatures, like the ones involving three 
leptons and no hadronic $W$ decay, may require the inclusion of higher-order
corrections (like $\mathcal{O}(\alpha_s^4\alpha)$ NNLO corrections) to the 
production process to reach a better control of the corresponding theoretical 
systematics. In all cases the impact of radiative top-quark decays should be 
studied carefully.
Finally, we want to emphasize the fact that even though for the current 
center-of-mass energy of the LHC of $\sqrt{s}=13~\TeV$ the modelling of the 
subleading $\mathcal{O}(\alpha_s\alpha^3)$ contribution only plays a minor role 
at the fiducial level it will be crucial to improve on its theoretical accuracy 
for higher center-of-mass energies, as its radiative contribution will 
increase~\cite{Frederix:2017wme}.

It is clear from our discussion that providing a robust theoretical prediction 
for hadronic $t\tb W^\pm$ production cannot be framed as a unique recipe and care 
must be taken to analyze the specific characteristics of different observables 
measured in experiments. Having at our disposal several well tested tools that
allow to implement state-of-the-art theoretical calculations in the modelling of 
collider events is clearly valuable and offers us the possibility of studying the
problem in its complexity and identify where improvement is most needed.

\section*{Acknowledgements}
We would like to thank Maria Moreno Llacer for very helpful discussions, Seth 
Quackenbush and Diogenes Figueroa for their support with the \nlox{} code and 
Vasily Sotnikov for his support with the \blackhat{} library of 
Ref.~\cite{Anger:2017glm}. We are grateful to Carlo Oleari for his assistance in 
making this code publicly available on the \powhegbox{} repository. The work of 
F.~F.~C. and L.~R.  is supported in part by the U.S. Department of Energy under 
grant DE-SC0010102.

\appendix
\section{Off-shell momentum mapping}
\label{app:offshellproj}
In this section we elaborate on the momentum mapping between on-shell and 
off-shell momenta used in section~\ref{subsec:ttW-decays} to include a smearing 
of the particle masses according to a Breit-Wigner distribution. The mapping 
discussed below is an adaptation of the method presented in 
Ref~\cite{Czakon:2015cla}, which itself was based on the momentum mapping of 
Ref.~\cite{Nagy:2007ty}. The momentum mapping is Lorentz invariant, preserves the
center-of-mass energy as well as the overall momentum of the momentum 
configuration. The mapping is constructed such that it generates a new off-shell 
momentum $\hat{p}_l$ with $\hat{p}_l^2 = v_l^2$ starting from an on-shell 
momentum $p_l$, with $p_l^2 = m_l^2$ by borrowing some of the necessary energy 
from the remaining recoiling final-state. The new momentum $\hat{p}_l$ is 
parametrized by
\begin{equation}
 \hat{p}_l = \lambda p_l + \frac{1-\lambda+y}{2a_l}Q\;, \qquad 
 \hat{p}_l^2 = v_l^2\;,
\end{equation}
where $Q$ is the total final state momentum and $v_l^2$ is the new virtuality of 
the off-shell momentum. Furthermore, we introduce the following dimensionless 
variables
\begin{equation}
  a_l = \frac{Q^2}{2(Q\cdot p_l)}\;, \qquad b_l = \frac{m_l^2}{2(Q\cdot p_l)}\;,
\qquad c_l = \frac{v_l^2}{2(Q\cdot p_l)}\;.
\end{equation}
The parameter $\lambda$ can be determined by requiring that the invariant mass 
of the recoiling momenta before and after the mapping are preserved, which 
are given by
\begin{equation}
\begin{split}
 K &= Q - p_l\;, \qquad K^2 = 2(Q\cdot p_l)\Big[a_l+b_l-1 \Big]\;, \\
 \hat{K} &= Q - \hat{p}_l\;, \qquad \hat{K}^2 = \frac{2(Q\cdot p_l)}{4a_l}
 \Big[ (2a_l-1-y)^2 - (1-4a_lb_l)\lambda^2\Big]\;.
\end{split}
\end{equation}
Equating $K^2 = \hat{K}^2$ then yields 
\begin{equation}
 \lambda = \sqrt{\frac{(1+y)^2-4a_l(y+b_l)}{1-4a_lb_l}}\;.
\end{equation}
The remaining parameter $y$ can be fixed from the condition
\begin{equation}
 \hat{p}_l^2 = 2(Q\cdot p_l)(y+b_l) \stackrel{!}{=} v_l^2 = 2(Q\cdot p_l)~c_l\;,
\end{equation}
which defines $y$ as a measure of the virtuality
\begin{equation}
 y = c_l - b_l = \frac{v_l^2 - m_l^2}{2(Q\cdot p_l)}\;.
\end{equation}
The upper boundary on the virtuality is given by the value $y_{max}$ for which
$\lambda$ vanishes and is given by 
\begin{equation}
  y_{max} = (\sqrt{a_l} - \sqrt{a_l+b_l-1})^2 - b_l\;.
\end{equation}
In order to preserve momentum conservation the recoiling momenta have to be 
boosted
\begin{equation}
  \hat{p}_i^{~\mu} = \Lambda^\mu_{~\nu} p_i^\nu\;, \qquad i \neq l\;,
\end{equation}
with the Lorentz transformation
\begin{equation}
 \Lambda^\mu_{~\nu} = g^\mu_{~\nu} - 
  \frac{2(K+\hat{K})^\mu(K+\hat{K})_\nu}{(K+\hat{K})^2}
  + \frac{2\hat{K}^\mu K_\nu}{K^2} \;.
\end{equation}
%
\bibliography{ttW}
\end{document}